\documentclass[aps, prd, reprint,10pt, notitlepage, a4paper, floatfix,amsmath, amssymb, amsfonts,superscriptaddress,nofootinbib,longbibliography]{revtex4-2}

\pdfoutput=1

\usepackage{graphics}
\usepackage{graphicx}
\usepackage{amsmath,amsfonts,amssymb,amsthm}
\usepackage[table,usenames,dvipsnames]{xcolor}
\usepackage{gensymb}
\usepackage{appendix}
\usepackage{listings}
\usepackage{url}
\usepackage[normalem]{ulem}
\usepackage{alltt}
\usepackage[colorlinks]{hyperref}
\usepackage{cleveref}
\usepackage{longtable}
\usepackage{enumitem}
\setlist{nosep}
\usepackage{color}
\usepackage{calc}
\usepackage{tensor}
\usepackage{bm}
\usepackage{multirow}
\usepackage[varg]{txfonts}
\usepackage{dcolumn}
\usepackage[nolist,nohyperlinks]{acronym}
\usepackage{xspace}
\usepackage[english]{babel}
\usepackage[abs]{overpic}
\usepackage{pict2e}
\usepackage[caption=false]{subfig} 
\allowdisplaybreaks[1]
\usepackage[utf8]{inputenc}
\usepackage{stackengine}
\usepackage{boldline,multirow}
\usepackage{braket}
\usepackage{makecell}
\usepackage{tabularx}
\usepackage{rotating}
\usepackage{booktabs}

\newcommand{\AEI}{Max Planck Institute for Gravitational Physics (Albert Einstein Institute), Am M\"uhlenberg 1, Potsdam 14476, Germany}
\newcommand{\Maryland}{Department of Physics, University of Maryland, College Park, MD 20742, USA}
\newcommand{\ICECSIC}{Institut de Ci\`encies de l'Espai (ICE, CSIC), Campus UAB, Carrer de Can Magrans s/n, 08193 Cerdanyola del Vall\`es, Spain}
\newcommand{\Nottingham}{Nottingham Centre of Gravity \& School of Mathematical Sciences, University of Nottingham, University Park, Nottingham NG7 2RD, United Kingdom}

\definecolor{dodgerblue}{HTML}{1E90FF}
\definecolor{viennared}{HTML}{DA0A14}

\hypersetup{citecolor=dodgerblue}

\def\mF{\mathcal{F}}
\DeclareMathOperator{\sech}{sech}
\DeclareMathOperator{\arcsinh}{arcsinh}

\begin{document}

\title{
	Pushing the accuracy of the quasi-circular baseline for the SEOBNR family of models
}
\author{Cheng Foo}\email{cheng.foo@aei.mpg.de}
\affiliation{\AEI}

\author{Alessandra Buonanno}
\affiliation{\AEI}
\affiliation{\Maryland}

\author{H\'ector Estell\'es}
\affiliation{\ICECSIC}

\author{Lorenzo Pompili}
\affiliation{\Nottingham}
\date{\today}

\begin{abstract}

	Gravitational-wave observations have opened a new era of astronomy, with accurate modeling of binary black hole signals becoming increasingly important for precise source characterization and astrophysical inference as detector sensitivities improve.
	We present \texttt{SEOBNRv5iHM}, an updated quasi-circular, aligned-spin inspiral-merger-ringdown gravitational-waveform model with higher-order modes, constructed within the effective-one-body formalism and building upon the previous model \texttt{SEOBNRv5HM}.
	The improvements mainly concern the late inspiral and the merger-ringdown: we remove the non-quasi-circular corrections used in previous models and instead introduce a phenomenological intermediate region between the inspiral and the ringdown, which allows the ringdown attachment time to be chosen independently for each mode, including at the peak of the mode's amplitude.
	The model is calibrated and validated against 591 (of the 592) quasi-circular aligned-spin waveforms from the third Simulating eXtreme Spacetimes catalog.
	For the $(2,2)$ mode, when considering the maximum unfaithfulness against numerical relativity over a total mass range of $10$--$300\,M_\odot$, \texttt{SEOBNRv5iHM} achieves a median value of $4.25\times10^{-5}$, a factor of 4.5 below \texttt{SEOBNRv5HM}, and a factor of $\sim 3-8$ below other state-of-the-art semi-analytical models, with $84\%$ of cases below $10^{-4}$.
	At higher masses of $M\ge150 M_\odot$, the median drops to $9.7\times10^{-6}$, achieving comparable accuracy to the numerical relativity surrogate \texttt{NRHybSur3dq8}.
	For polarization mismatches at inclination $\pi/3$, considering waveforms up to $\ell_\mathrm{max} = 5$, the median maximum unfaithfulness is $4.2\times10^{-4}$, a factor of 2.6 below \texttt{SEOBNRv5HM}, a factor of 1.9 below \texttt{IMRPhenomTHM}, and a factor of 6--11 below the other semi-analytical models considered.
	Since spin-precessing and eccentric models are built on top of the quasi-circular, aligned-spin baseline, these improvements will make future \texttt{SEOBNR} models more accurate, reducing waveform systematics in the analyses of upcoming observing runs.
	
\end{abstract}

\maketitle


\section{Introduction}
\label{sec:Introduction}
The field of gravitational-wave (GW) astronomy recently marked the tenth anniversary of GW150914~\cite{LIGOScientific:2016aoc}, the first detected GW signal from a binary black hole (BBH) coalescence.
In the decade since, the number of detected events has risen steadily as detector sensitivity has improved.
Following the fourth observing run of the LIGO--Virgo--KAGRA (LVK) Collaboration, the latest catalog, GWTC-5.0~\cite{LIGOScientific:2026sit,LIGOScientific:2026ifv,LIGOScientific:2026ctl,LIGOScientific:2026wfs}, contains almost 400 
GW events from merging black holes (BHs) and neutron stars, the vast majority of which are BBHs~\cite{TheLIGOScientific:2014jea,VIRGO:2014yos,KAGRA:2020tym,LIGOScientific:2018mvr,LIGOScientific:2020ibl,LIGOScientific:2021usb,LIGOScientific:2021djp,LIGOScientific:2025slb}.

To infer the properties of the GW sources, the detector data are compared against theoretical predictions for the emitted signals, provided by waveform models. 
There are several approaches to building these inspiral--merger--ringdown (IMR) models.

Numerical-relativity (NR) surrogates~\cite{Blackman:2015pia,Blackman:2017dfb,Blackman:2017pcm,Williams:2019vub,Rifat:2019ltp,Varma:2018mmi,Varma:2019csw,Yoo:2022erv,Islam:2021mha,Islam:2022laz,Yoo:2023spi,Nee:2025nmh,Ravishankar:2026akf}
interpolate NR simulations directly across parameter space.
Where available, these are typically the most accurate waveform models. 
Their reach, however, is set by the parameter coverage and duration of the underlying simulations: extending them to the many hundreds of cycles visible to current detectors for low-mass binaries, and to the full spin and mass-ratio range, requires hybridization with an analytical inspiral.

The bulk of LVK parameter estimation therefore relies on two semi-analytical families.
The first is the phenomenological \texttt{IMRPhenom} family, constructed either in the frequency domain~\cite{Husa:2015iqa,Khan:2015jqa,Pratten:2020fqn,Garcia-Quiros:2020qpx,Pratten:2020ceb,Hamilton:2021pkf,Thompson:2023ase,Ghosh:2023mhc,Hamilton:2025xru,Ramos-Buades:2026kbq} or in the time domain~\cite{Estelles:2020osj,Estelles:2020twz,Estelles:2021gvs,Planas:2025feq}.
The second is the effective-one-body (EOB) approach~\cite{Buonanno:1998gg,Buonanno:2000ef,Damour:2000we,Damour:2001tu,Buonanno:2005xu}, of which there are two state-of-the-art families:  \texttt{SEOBNR} (e.g., see Refs.~\cite{Bohe:2016gbl,Cotesta:2018fcv,Ossokine:2020kjp,Cotesta:2020qhw,Ramos-Buades:2021adz,Mihaylov:2021bpf,Pompili:2023tna,Ramos-Buades:2023ehm,Khalil:2023kep,vandeMeent:2023ols,Mihaylov:2023bkc,Julie:2024fwy,Pompili:2024yec,Pompili:2025cdc,Leather:2025nhu,Gamboa:2024hli,Estelles:2025zah,Haberland:2025luz,Nishimura:2026nse,Buonanno:2024byg,Gamboa:2026jht,Gamboa:2026ybr})
and \texttt{TEOBResumS} (e.g., see
Refs.~\cite{Nagar:2018zoe,Nagar:2019wds,Nagar:2020pcj,Gamba:2021ydi,Riemenschneider:2021ppj,Albertini:2021tbt,Chiaramello:2020ehz,Nagar:2024dzj,Nagar:2024oyk,Albanesi:2024xus,Gamba:2023mww,Nagar:2023zxh,Gamba:2024cvy,Albanesi:2025txj}). 

It has long been anticipated~\cite{Purrer:2019jcp} that, as the sensitivity of the detectors improves and the loudest events accumulate, the systematic error of the waveform models will eventually become comparable to---and then exceed---the statistical uncertainty set by the detector noise.
Several recent studies find that this regime is already being approached: differences between state-of-the-art models translate into measurable biases in the inferred masses and spins of individual events~\cite{Dhani:2024jja,MacUilliam:2024oif,Owen:2023mid}, into spurious deviations from general relativity~\cite{Hu:2022rjq}, and, once propagated through a hierarchical analysis, into biases in the inferred population~\cite{Dhani:2024jja}. The high-mass event GW231123~\cite{LIGOScientific:2025rsn}, for which the source properties obtained with different waveform families disagree noticeably, is perhaps the clearest recent example of a detection whose interpretation is limited by the models rather than by the data~\cite{Jan:2025zcm,Bini:2026kwz,Chan:2025kyu}. 

These accuracy demands will only sharpen with the upgrades planned for the current facilities and with next-generation observatories. 
The planned ground-based third-generation (3G) detectors Einstein Telescope (ET)~\cite{Punturo:2010zz,Hild:2010id,Maggiore:2019uih,Branchesi:2023mws,ET:2025xjr} and Cosmic Explorer (CE)~\cite{Reitze:2019iox,Evans:2021gyd,Evans:2023euw,Kalogera:2021bya} will increase the strain sensitivity by roughly an order of magnitude, and extend the observing band down to a few Hz. 
As a result, the loudest BBH signals will reach signal-to-noise ratios (SNRs) of several hundred to a few thousand, with tens of thousands of events expected per year~\cite{Maggiore:2019uih,Iacovelli:2022bbs,Branchesi:2023mws,ET:2025xjr}.

The upcoming space-based LISA mission~\cite{LISA:2017pwj,LISA:2022kgy,LISA:2024hlh}, together with the proposed TianQin~\cite{TianQin:2015yph} and Taiji~\cite{Hu:2017mde,Ruan:2018tsw} missions, is likely to observe massive 
black hole binaries (MBHBs), which will have SNRs of $10^2$--$10^4$. The analysis of these signals will require a huge leap in accuracy of our waveform models~\cite{Pratten:2022kug,Toubiana:2023cwr,Garg:2024qxq,Abbondanza2026}. 
MBHB signals with lower masses will also have long inspirals.
The resulting accuracy requirements on the waveform models are correspondingly one to two orders of magnitude more stringent than those met by current models~\cite{Ferguson:2020xnm,Purrer:2019jcp,Read:2023hkv,Jan:2023raq}. 

Recently, there has been a push in the community towards extending the physics content of the models, in particular to eccentric and hyperbolic orbits~\cite{Ramos-Buades:2021adz,Nagar:2021gss,Islam:2021mha,Gamboa:2024hli,Gamboa:2024imd,Nagar:2024dzj,Nagar:2024oyk,Albanesi:2024xus,Planas:2025feq,Nee:2025nmh,Gamboa:2026jht,Gamboa:2026ybr,Ramos-Buades:2026kbq}, motivated by the prospect of constraining the formation channels of the observed population~\cite{Gupte:2024jfe,Ramos-Buades:2023yhy,Vijaykumar:2024piy,Gupte:2026whi,Breivik:2016ddj,Samsing:2018isx,Zevin:2021rtf}. 
The inclusion of additional physics such as eccentricity is also vital for controlling systematic error, since unmodeled physics can appear as biases in recovered parameters of the source~\cite{Favata:2013rwa,Ramos-Buades:2019uvh,OShea:2021faf,Divyajyoti:2023rht}, or can be confused with other effects such as spin precession~\cite{CalderonBustillo:2020xms,Romero-Shaw:2022fbf,Xu:2022zza}. 

In parallel to the efforts towards generic-orbit waveform models, we need an increasingly accurate quasi-circular (QC) description. 
Because GW emission efficiently circularizes a binary orbit during the long inspiral~\cite{Peters:1964zz, Peters:1963ux,Hinder:2007qu,Sperhake:2007gu}, most field binaries are expected to have negligible eccentricity by the time they enter the LVK band, which is why the default LVK catalog analysis has so far relied on QC, generic-spin models. 
Work on improving the QC baseline is thus still important, to understand the impact of more accurate models on the analyses, to push the accuracy of the most widely used models, and to inform the QC limit of generic models. 

Generic-orbit models are typically built on top of the QC, aligned-spin baseline, and the accuracy of this baseline sets a limit for the accuracy of the generic model.
The QC, spin-precessing models that are employed for essentially all LVK parameter estimation~\cite{Ramos-Buades:2023ehm,Estelles:2021gvs,Pratten:2020ceb,Hamilton:2025xru,Colleoni:2024knd,Varma:2019csw} 
are constructed by using the co-precessing frame~\cite{Schmidt:2010it,Schmidt:2012rh,Boyle:2011gg,OShaughnessy:2011pmr,Hannam:2013oca,Estelles:2021gvs}. 
Waveforms are first modeled in the frame whose $z$-axis tracks the orbital angular momentum---in this co-precessing frame, the waveform is similar to an aligned-spin waveform. 
Precessional effects such as equatorial asymmetries are added on top of the aligned-spin modes, before the waveform is subsequently rotated into the inertial frame of the observer. 
Existing models for eccentric and hyperbolic orbits still build upon their QC limit. 
Other models, such as those for binary neutron stars~\cite{Haberland:2025luz}, or parametrized models for testing general relativity~\cite{Pompili:2025cdc}, add physics on top of the same backbone.
The accuracy of the QC aligned-spin model thus sets a floor on the modeling error of the generic models built upon it, and improvements to it will propagate to the entire set of models.

In this paper we present \texttt{SEOBNRv5iHM}, a QC, aligned-spin, multipolar IMR waveform model for BBHs in the EOB framework.
We build upon the accurate theoretical baseline of the \texttt{SEOBNRv5HM} model, as well as insights from the time-domain phenomenological model family \texttt{IMRPhenomT}.
Here, the ``\texttt{i}'' refers to the phenomenological intermediate region described below.

Relative to \texttt{SEOBNRv5HM}, the main structural changes are to the late inspiral, merger and ringdown.
We remove the non-quasi-circular (NQC) corrections that earlier \texttt{SEOBNR} models used to improve the accuracy of the EOB modes near merger.
Instead, we introduce a phenomenological \emph{intermediate region}, following the approach of the time-domain \texttt{IMRPhenomT} models~\cite{Estelles:2020osj,Estelles:2020twz,Estelles:2021gvs,Planas:2025feq}.
Because the intermediate region is modeled up to the peak of each individual mode, the ringdown of every $(\ell,m)$ mode can be modeled from its own peak rather than from the peak of the $(2,2)$ mode. The only exceptions are the $(3,2)$ and $(4,3)$ modes, whose ringdown is still modeled from the $(2,2)$ peak onward to account for mode-mixing effects.

The model is calibrated against 591 QC, aligned-spin simulations of the third catalog of the Simulating eXtreme Spacetimes (SXS) collaboration~\cite{Scheel:2025jct}, complemented by BH-perturbation-theory waveforms in the test-mass limit (TML)~\cite{Nishimura:2026nse}.

We validate the model against the same NR catalog, finding that its dominant-mode accuracy exceeds that of other state-of-the-art semi-analytical waveform models.
In the region where \texttt{NRHybSur3dq8} can generate waveforms, we find that the \texttt{SEOBNRv5iHM} $(2, 2)$ mode gives comparable results. 
Polarization mismatches likewise improve relative to other state-of-the-art models, and are limited in part, especially for anti-aligned spins, by the mode content included.

The paper is organized as follows. Sec.~\ref{sec:Motivation} examines where the accuracy of \texttt{SEOBNRv5HM} is limited, and motivates the subsequent modeling choices. 
Sec.~\ref{sec:ModelConstruction} describes the construction of \texttt{SEOBNRv5iHM}.
Sec.~\ref{sec:Calibration_Fitting} details the procedures used to calibrate the model to NR and to fit the calibration quantities across parameter space. 
We then quantify the accuracy against NR and other state-of-the-art approximants in Sec.~\ref{sec:Results22} and Sec.~\ref{sec:ResultsPolHM}, before concluding in Sec.~\ref{sec:Conclusion}.

\section*{Notation}
Throughout the paper we use geometric units, $G = c = 1$.
For binaries with component masses $m_1, m_2$, we define the following variable combinations:
\begin{equation}
\begin{gathered}
	M \equiv m_1+m_2, \qquad q \equiv m_1/m_2, \qquad \mu \equiv m_1 m_2/M,\\
	\nu \equiv \mu/M, \qquad \delta \equiv (m_1-m_2)/M = \sqrt{1-4\nu} .
\end{gathered}
\end{equation}
These are the total mass $M$, the mass ratio $q$, the reduced mass $\mu$, the symmetric mass ratio $\nu$, and the dimensionless mass difference $\delta$. 
We use the convention that $m_1 \ge m_2$.

The spins are taken to be aligned (or anti-aligned) with the orbital angular momentum $\bm{L}$, and have magnitudes $S_1, S_2$.  
They are described by the dimensionless magnitudes $\chi_i = S_i/m_i^2$, $i=1,2$, with sign encoding the orientation relative to $\bm{L}$.
We make use of the following spin variables:
\begin{equation}
\begin{gathered}
	 a_\pm \equiv m_1\chi_1 \pm m_2\chi_2, \\
	 \chi_{\rm eff} \equiv \frac{m_1\chi_1 + m_2\chi_2}{M},\\
	\chi_S \equiv \tfrac{1}{2}(\chi_1 + \chi_2), \qquad \chi_A \equiv \tfrac{1}{2}(\chi_1 - \chi_2).
\end{gathered}
\end{equation}

In all fits, $a_{\pm}$ are expressed in units of the total mass $M$, so that they are dimensionless (and $a_{+} \equiv \chi_\mathrm{eff}$). 
The EOB dynamics are written in polar phase-space variables $(r,\phi,p_r,p_\phi)$ in the orbital plane. 
The orbital frequency is $\Omega$, from which we build the dimensionless frequency parameter $v_\Omega = (M\Omega)^{1/3}$.
The time at which the amplitude of each $(\ell, m)$ mode peaks is denoted as $t_{\ell m}^{\rm peak}$, and the amplitude and frequency of a mode are $|h_{\ell m}|$ and $\omega_{\ell m} \equiv -\dot\phi_{\ell m}$, with $h_{\ell m} = |h_{\ell m}| e^{i\phi_{\ell m}}$.
\section{Where to improve accuracy}
\label{sec:Motivation}
In setting out to improve the accuracy of the QC baseline of the \texttt{SEOBNR} family of models, we first need to understand what limits the accuracy of the current baseline, \texttt{SEOBNRv5HM}.
The standard figure of merit for waveform models is the mismatch against NR as a function of total mass, using a detector noise power spectral density (PSD)~\cite{Bohe:2016gbl,Cotesta:2018fcv,Pratten:2020fqn,Garcia-Quiros:2020qpx,Estelles:2020twz,Gamboa:2026jht}. These are given for \texttt{SEOBNRv5HM} in Figs.~5 and 7 of Ref.~\cite{Pompili:2023tna}.
These plots give some insight into the accuracy of different stages of the waveform: at low total masses, the signal within the detector band is mainly dominated by the inspiral, while at high masses the merger and ringdown dominate. 
However, it is still difficult to attribute the mismatch value to a specific part of the waveform, since all sections contribute.

We therefore use a complementary diagnostic, in which we adopt a flat PSD and vary the upper frequency cutoff of the mismatch inner product. 
This allows us to evaluate the contribution of successive stages of the evolution to the total mismatch. 
We apply it mode by mode to \texttt{SEOBNRv5HM} and, as an accuracy reference, to the NR surrogate \texttt{NRHybSur3dq8}~\cite{Varma:2018mmi}.
Before describing the test, we recall the multipolar decomposition of the waveform and the mismatch conventions used throughout this work.

The GW signal emitted by a binary system is a complex combination of GW polarizations $h(t) \equiv h_+(t) - ih_\times (t)$, and can be decomposed into $-2$ spin-weighted spherical harmonics as~\cite{Pan:2011gk}
\begin{align}
	h(t;  \bm{\lambda}, \iota, \varphi_0) = \frac{1}{d_L}\sum_{\ell \geq 2}\sum_{|m|\leq \ell} {}_{-2} Y_{\ell m} (\iota, \varphi_0) \, h_{\ell m}(t;\bm{\lambda}),
\label{eq:MultiModeDecomposition}
\end{align}
where $\bm{\lambda} = \{m_1, m_2, \bm{\chi}_1, \bm{\chi}_2\}$ are the intrinsic parameters of the binary.
The extrinsic parameters, which are the luminosity distance $d_L$, the inclination $\iota$, and the azimuthal angle $\varphi_0$, give the orientation of the binary source with respect to the observer.
This decomposition allows us to separate intrinsic parameters from extrinsic parameters, so that only the modes $h_{\ell m}$ need to be modeled. 

To quantify the agreement between two waveforms $h_1(t)$ and $h_2(t)$ we use the noise-weighted inner product~\cite{Finn:1992xs,Sathyaprakash:1991mt}
\begin{equation}
	\label{eq:overlap}
	\left( h_1 \mid h_2 \right) \equiv 4 \operatorname{Re} \int_{f_\mathrm{min}}^{f_\mathrm{max}} \frac{\tilde{h}_1(f)\ \tilde{h}_2^*(f)}{S_n(f)} \mathrm{d} f ,
\end{equation}
where $\tilde h_{1,2}(f)$ are the Fourier transforms of the waveforms and $S_n(f)$ is the one-sided PSD of the detector noise.
The faithfulness is the overlap between the normalized waveforms, maximized over the relative time and phase shift~\cite{Pompili:2023tna},
\begin{equation}
	\label{eq:match}
	\left\langle h_1 \mid h_2\right\rangle=\max _{\phi_c,\ t_c} \frac{\left(h_1\left(\phi_c, t_c\right) \mid h_2\right)}{\sqrt{\left(h_1 \mid h_1\right)\left(h_2 \mid h_2\right)}}.
\end{equation}

The mismatch, or unfaithfulness, of a model waveform to the NR waveform is then 
\begin{equation}
	\label{eq:mismatch}
	\mathcal{M} = 1 - \left\langle h_{\mathrm{model}}(\boldsymbol{\lambda}) \mid h_{\mathrm{NR}}(\boldsymbol{\lambda})\right\rangle.
\end{equation}
To use this metric on time-domain waveforms, they are first tapered with a Planck window~\cite{McKechan:2010kp}, then transformed to the frequency domain.

The optimization over time shifts is typically performed efficiently using fast Fourier transforms, which yields overlap values on a discrete grid of time shifts $\Delta t$, whose spacing is set by the sampling rate of the waveform.
The time-maximized overlap is then taken as the maximum over this grid, which means that the coarseness of the time grid can affect the resulting faithfulness. 
This is unimportant when the waveforms are very dissimilar, since such a suboptimal maximization hardly changes the result.
For very similar waveforms, however, it imposes an artificial floor on the
achievable mismatch~\cite{Ajith:2012az}.

Given the accuracy targets of
\texttt{SEOBNRv5iHM}, and in order to faithfully resolve the NR error, we therefore perform an additional subsample optimization in all our faithfulness calculations, where we locate the true overlap peak in the neighborhood of the discrete maximum.

The diagnostic introduced here replaces the sweep over total mass by a sweep over the upper cutoff $f_{\max}$ of Eq.~\eqref{eq:overlap}.
We use a flat PSD and a single reference mass $M = 30\,M_\odot$, so that the total mass only sets the frequency scale and increasing $f_{\max}$ amounts to progressively including later portions of the signal in the inner product.
The cutoffs are defined for each NR case from the instantaneous frequency of the NR $(\ell, m)$ mode. 
We consider four choices:
\begin{enumerate}
	\item up to the mid-inspiral, $f_\mathrm{max} = f_{\ell m}(t_{\text {peak }}^{22}
	      - 250M)$;
	\item up to the late inspiral, $f_\mathrm{max} = f_{\ell m}(t_{\text {peak }}^{22}
	      - 80M)$;
	\item the full inspiral up to the peak of each mode, \\$f_\mathrm{max} =
	      f_{\ell m}(t_{\text{peak }}^{\ell m})$;
	\item the standard full mismatch, $f_\mathrm{max} = 2048~\mathrm{Hz}$,
	      which includes the complete signal.
\end{enumerate}
These geometric frequencies are converted to Hz at the reference mass, while $f_\mathrm{min} = 1.35 f_\mathrm{start}$, where $f_\mathrm{start}$ is the frequency at which the Fourier amplitude of the NR waveform peaks. 
This buffer factor of 1.35 is chosen to exclude Fourier artifacts from the Planck tapering of the time-domain signal~\cite{Pompili:2023tna}. 
Near-equal-mass, equal-spin configurations, for which the odd-$m$ modes vanish (or are strongly suppressed) by symmetry, are omitted from the odd-$m$ results.

Note that the mismatch can grow as $f_\mathrm{max}$ increases, simply because the inner product then includes a longer stretch of waveform containing more cycles, over which phase errors accumulate.
The absolute value at a given cutoff is therefore not by itself informative; it is the comparison against the highly accurate \texttt{NRHybSur3dq8}, processed through the identical procedure, that isolates the stage of the evolution at which \texttt{SEOBNRv5HM} loses accuracy relative to NR.

\texttt{NRHybSur3dq8} is hybridized to PN and EOB outside its ``NR region'', which begins at roughly $t = -4000M$ to $t = -3000M$. The amplitude is built using PN expressions, and the phase uses \texttt{SEOBNRv4}. 
However, for our comparisons, we are mainly probing the NR region of the surrogate.

We compare against the recently released third SXS catalog~\cite{Scheel:2025jct}, which increased the number of QC, aligned-spin SXS simulations to 592, including simulations in more extreme regions of parameter space, with higher mass ratios and spin magnitudes.
We employ 591 of the waveforms in this work, excluding \texttt{SXS:BBH:0202}, a simulation whose merger-ringdown waveform ends prematurely.

For each simulation we compute the mismatch against NR for both \texttt{SEOBNRv5HM} and \texttt{NRHybSur3dq8}.
Since \texttt{NRHybSur3dq8} can only be generated for $q \le 10$ (trained up to $q = 8$ and extrapolated beyond), we restrict the comparison to simulations in this range, leaving 565 NR waveforms.

\begin{figure}[t]
    \centering
    \includegraphics[width=\columnwidth]{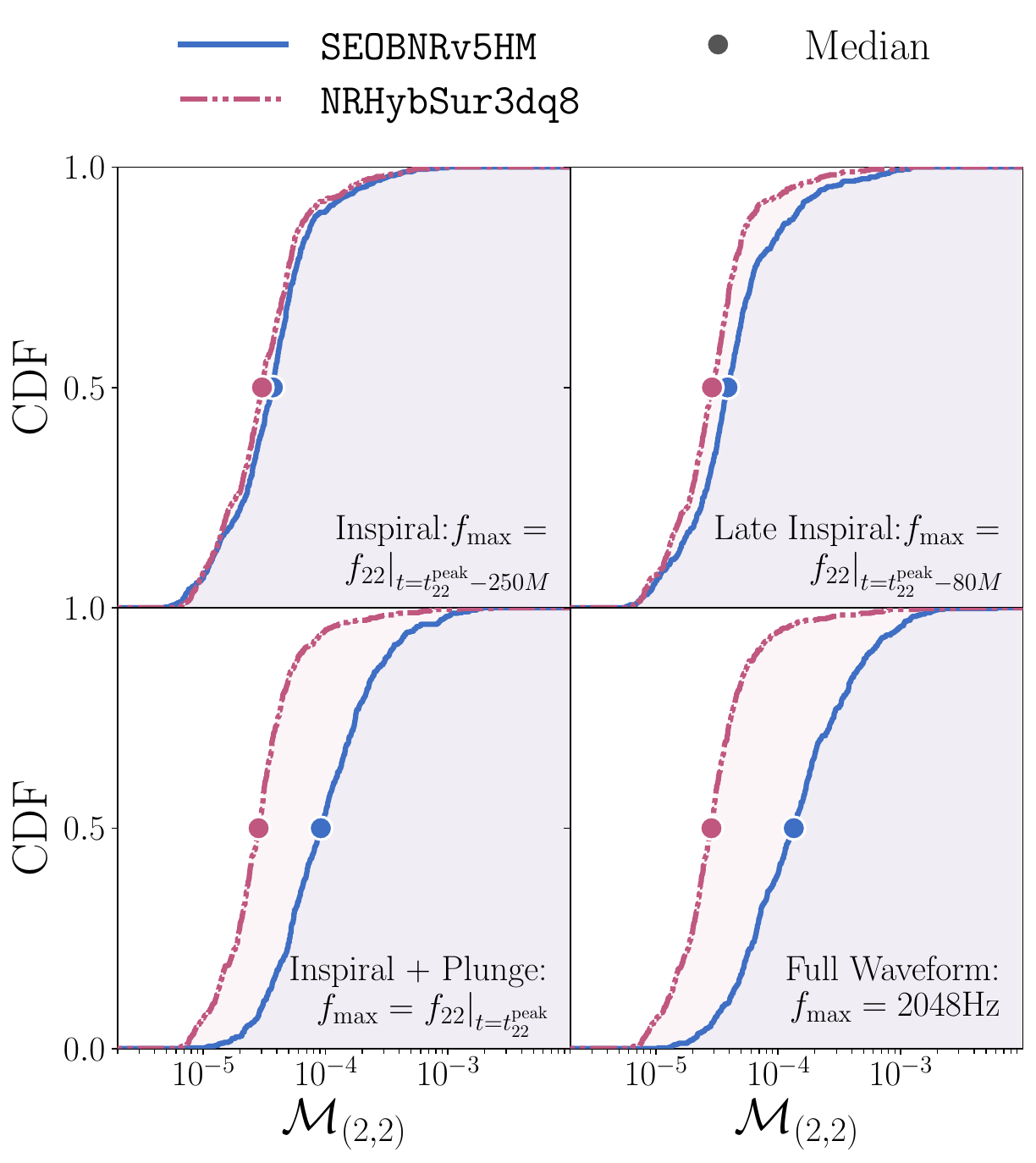}
	\caption{CDFs of the mismatch against NR for \texttt{SEOBNRv5HM} (blue-solid) and \texttt{NRHybSur3dq8} (red-dash-dotted), computed for the $(2,2)$ mode over the 565 QC, aligned-spin SXS simulations~\cite{Scheel:2025jct} with $q \le 10$.
	A flat PSD and a fixed total mass of $M = 30\,M_\odot$ are used, so that the upper cutoff $f_\mathrm{max}$ of the inner product in Eq.~\eqref{eq:overlap} controls how much of the signal enters the mismatch; the four panels progressively include later portions of the waveform. 
	From left to right, $f_\mathrm{max}$ is set to the instantaneous $(2,2)$ frequency at $250M$ and $80M$ before the peak of the $(2,2)$ mode, at the peak itself, and finally to $2048~\mathrm{Hz}$, which retains the complete signal including merger and ringdown.
	Median points are indicated by the large marker. 
	}
	\label{fig:DiagnosingError22}
\end{figure}

The results of this comparison for the $(2, 2)$ mode are shown as cumulative distribution functions (CDFs) in Fig.~\ref{fig:DiagnosingError22}. 
In the first panel, the accuracy of the \texttt{SEOBNRv5HM} $h_{22}$ mode is comparable to that of \texttt{NRHybSur3dq8} in the early to mid-inspiral, with median mismatch values of $3.7 \times 10^{-5}$ and $3.0 \times 10^{-5}$ respectively. 
As we move closer to merger, up to $80M$ before the peak of the mode, the curves in the second panel begin to separate ($3.9 \times 10^{-5}$ against $2.9 \times 10^{-5}$), though they remain close.
The third panel, which extends to the peak of the mode, shows that \texttt{SEOBNRv5HM} accumulates substantial error over the final $80M$ before the peak, degrading rapidly relative to \texttt{NRHybSur3dq8}:
the median mismatch grows to $9.1\times 10^{-5}$, while that of \texttt{NRHybSur3dq8} is unchanged at $2.8\times 10^{-5}$.
This gap widens further in the post-peak merger-ringdown region, shown in the fourth panel, where the \texttt{SEOBNRv5HM} median reaches $1.3 \times 10^{-4}$ against $2.9 \times 10^{-5}$ for \texttt{NRHybSur3dq8}.
The \texttt{NRHybSur3dq8} median is essentially constant across all four cutoffs, indicating that the additional cycles included at larger $f_\mathrm{max}$ contribute negligibly to its error budget. The growth seen for \texttt{SEOBNRv5HM} is therefore attributable to the late-inspiral and merger-ringdown portion of the waveform, rather than to the increased length of the integration.

\begin{figure}[t]
    \centering
    \includegraphics[width=\columnwidth]{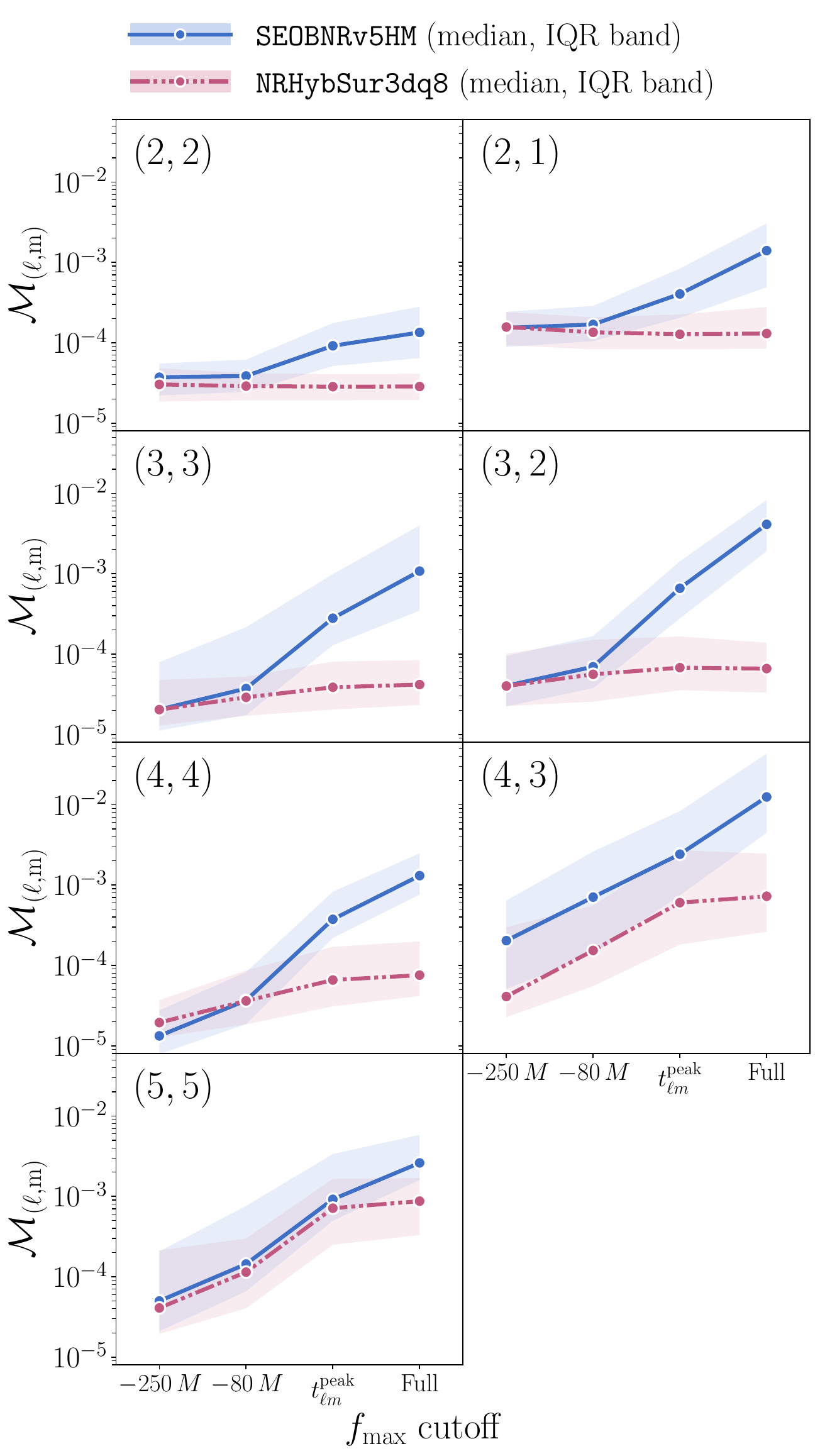}
	\caption{Mismatch against NR as a function of the upper cutoff $f_\mathrm{max}$ of the inner product in Eq.~\eqref{eq:overlap}.
	Lines give the median over the 565 QC, aligned-spin SXS simulations~\cite{Scheel:2025jct} with $q \le 10$, and shaded bands the IQR.
	The four cutoffs correspond to those in Fig.~\ref{fig:DiagnosingError22}.
	Every mode degrades over the final two cutoffs, while the $(4,3)$ mode already shows a significant loss of accuracy in the early to mid-inspiral.}
	\label{fig:DiagnosingErrorModes}
\end{figure}

To evaluate the higher modes, we instead plot only the median
for each of the four cutoffs, together with a band spanning the interquartile range (IQR), i.e.\ the 25th to 75th percentiles. 
The results are shown in Fig.~\ref{fig:DiagnosingErrorModes}.
In all cases, we see a degradation in accuracy over the final two cutoffs, i.e.\ once the last $80M$ before the peak and the merger--ringdown are included.
For some modes, however, and particularly the $(4,3)$ mode, a significant difference in accuracy is already present in the early to mid-inspiral.
In addition, for some of the higher modes, \texttt{NRHybSur3dq8} itself no longer maintains a flat accuracy across the entire waveform.

The results in this section motivate the model construction of \texttt{SEOBNRv5iHM}. 
In particular, we seek to improve the region from $t^\mathrm{peak}_{22} - 80M$ onward, so that the late inspiral and merger no longer limit the overall accuracy of the waveform. 
This will allow us to exploit the theoretical work underpinning the highly accurate long inspiral of \texttt{SEOBNRv5HM}.
This region is currently modeled using NQCs, the full construction details of which can be found in Ref.~\cite{Pompili:2023tna}.
The NQC corrections impose agreement with NR input values at a single time, the attachment point of the merger-ringdown waveform. 
They rely on their functional dependence on the EOB dynamics to correct the late inspiral, and this diagnostic shows that they do not provide sufficient control of the last $80M$ before the attachment point, where the model loses accuracy.
We instead take inspiration from the \texttt{IMRPhenomT} family of waveform models~\cite{Estelles:2020osj,Estelles:2020twz,Estelles:2021gvs,Planas:2025feq}, and introduce a phenomenological ansatz there. 
The full details of the model construction are laid out in Sec.~\ref{sec:ModelConstruction}.

\section{\texttt{SEOBNRv5\lowercase{i}HM} Model construction}
\label{sec:ModelConstruction}
In this section we describe the construction of \texttt{SEOBNRv5iHM}. The conservative and dissipative dynamics are inherited from \texttt{SEOBNRv5HM}~\cite{Pompili:2023tna,Khalil:2023kep,vandeMeent:2023ols}, up to a recalibration of the Hamiltonian parameters; the main construction change concerns the waveform modes, which are now assembled from three regions (inspiral, intermediate and merger--ringdown) instead of two.
The NQC corrections are removed, and the attachment time of the merger--ringdown can be chosen independently for each mode; for several modes, we attach it at the peak of the mode's amplitude.

\subsection{Dynamics}

The dynamics of our system are given by the EOB Hamiltonian, as in previous \texttt{SEOBNR} waveform models. 
The EOB formalism maps the dynamics of the two-body system to an effective-one-body Hamiltonian~\cite{Buonanno:1998gg,Buonanno:2000ef,Damour:2000we,Damour:2001tu,Buonanno:2005xu}. 
The energy map relating the effective Hamiltonian $H_{\mathrm{eff}}$ and the two-body EOB
Hamiltonian $H_{\mathrm{EOB}}$ is
\begin{equation}
\label{eq:HEOB}
	H_{\mathrm{EOB}}=M \sqrt{1+2 \nu \left(\frac{H_{\mathrm{eff}}}{\mu}-1\right)}\,.
\end{equation}
The aligned-spin effective Hamiltonian $H_{\mathrm{eff}}$ is the same as that used for \texttt{SEOBNRv5HM}~\cite{Khalil:2023kep,Pompili:2023tna}, which is based on that of a test mass in a deformed Kerr background~\cite{Damour:2001tu,Damour:2008qf,Pan:2009wj,Nagar:2011fx,Damour:2014sva,Balmelli:2015zsa,Khalil:2020mmr}.

We use the same choice of calibration parameters within $H_{\mathrm{eff}}$, though, given the different model construction, they have been recalibrated. 
These are $a_6$, a 5PN parameter linear in $\nu$ that enters the non-spinning sector; and $d_\mathrm{SO}$, a 4.5PN spin-orbit parameter that enters the odd-in-spin sector. 
We build new fits across parameter space for this model, as described in Sec.~\ref{sec:Calibration_Fitting}.
The fits can be found in Appendix~\ref{app:hamiltoniancalibfits}.

As in previous models, we use the momentum conjugate to the tortoise coordinate, $p_{r_*}$, instead of $p_{r}$ to improve the stability of the equations of motion during the plunge and near merger~\cite{Damour:2007xr,Pan:2009wj}. 
The definitions of $r_*$ and its conjugate momentum $p_{r_*}$ are given respectively by
\begin{equation}
	\label{eq:tortoisecoordinates}
\frac{dr_*}{dr} = \frac{1}{\xi(r)}, \qquad p_{r_*} =  p_r \,\xi(r).
\end{equation}
The choice of $\xi(r)$ is detailed in Eqs.~(7) and (14) of Ref.~\cite{Pompili:2023tna}.

The equations of motion are as follows~\cite{Pan:2011gk}:
\begin{equation}
	\label{eq:eom_prst}
	\begin{aligned}
	\dot{r} &= \xi \frac{\partial H_{\mathrm{EOB}}}{\partial p_{r_*}}, \quad
	&\dot{p}_{r_*} &= -\xi \frac{\partial H_{\mathrm{EOB}}}{\partial r} + \frac{p_{r_*}}{p_\phi} \mF_\phi, \\
	\dot{\phi} &= \frac{\partial H_{\mathrm{EOB}}}{\partial p_\phi}, \quad
	&\dot{p}_\phi &= \mF_\phi.
	\end{aligned}
\end{equation}
These equations of motion are then evolved by starting from the same QC adiabatic initial conditions derived in Ref.~\cite{Buonanno:2005xu}.

The radiation reaction (RR) force $\mF_\phi$ is obtained by summing the GW modes in factorized form~\cite{Damour:2007xr,Damour:2007yf,Damour:2008gu,Pan:2010hz}, $h_{\ell m}^{\mathrm{F}}$, which we define in Sec.~\ref{sec:inspiralmodeconstruction}.

\begin{equation}
\label{RRforce}
\mF_\phi \equiv - \frac{\Omega}{8 \pi} \sum_{\ell=2}^8 \sum_{m=1}^{\ell} m^2\left|d_L h_{\ell m}^{\mathrm{F}}\right|^2.
\end{equation}

\subsection{Waveform mode construction}

In previous EOB waveform models, the GW modes $h_{\ell m}$ have been decomposed into two parts, the inspiral--plunge and the merger--ringdown. 
In this model, we instead decompose the modes into three regions: the inspiral, an intermediate region, and the merger--ringdown. 
This is similar to the \texttt{IMRPhenomT} family of phenomenological time-domain models~\cite{Estelles:2020osj,Estelles:2020twz,Estelles:2021gvs,Planas:2025feq}. 
The same three-region split underlies the frequency-domain \texttt{IMRPhenomD} and \texttt{IMRPhenomX} models~\cite{Husa:2015iqa,Khan:2015jqa,Pratten:2020fqn,Garcia-Quiros:2020qpx,Pratten:2020ceb,Hamilton:2021pkf,Thompson:2023ase,Ghosh:2023mhc,Hamilton:2025xru,Ramos-Buades:2026kbq}.
Recent developments in the test-mass limit \texttt{SEOBNR-TML} model have followed a similar construction~\cite{Nishimura:2026nse}.

\texttt{SEOBNRv5iHM} models the same modes as \texttt{SEOBNRv5HM}: $(\ell, m) = \{(2, 2), (2, 1), (3, 3), (3, 2), (4, 4), (4, 3), (5, 5)\}$. The modes are constructed as: 
\begin{equation}
	\label{eq:h_match}
	h_{\ell m}(t)= 
	\begin{cases}
		h_{\ell m}^{\mathrm{insp}}(t), & t < t^{\mathrm{inter}}_{\ell m}, \\
		h_{\ell m}^{\mathrm{inter}}(t), & t^{\mathrm{inter}}_{\ell m}< t < t^{\mathrm{match}}_{\ell m}, \\ 
		h_{\ell m}^{\mathrm{merger-RD}}(t), & t > t^{\mathrm{match}}_{\ell m },
	\end{cases}
\end{equation}
with $t^{\mathrm{inter}}_{\ell m}$ and $t^{\mathrm{match}}_{\ell m}$ defined for each $(\ell, m)$ as
\begin{subequations}
\label{eq:tcut_tmatch}
\begin{align}
	t^{\mathrm{inter}}_{\ell m} &\equiv t^{\mathrm{peak}}_{22} - 80M, \\
	t^{\mathrm{match}}_{\ell m} &\equiv
	\begin{cases}
		t^{\mathrm{peak}}_{\ell m}, & (\ell, m) = (2,2), (3,3), (2,1), \\
		                            & \phantom{(\ell, m) = {}} (4,4), (5,5), \\
		t^{\mathrm{peak}}_{22},     & (\ell, m) = (3,2), (4,3).
	\end{cases}\end{align}
\end{subequations}
Here, $t^{\mathrm{peak }}_{\ell m}$ is the time at which the amplitude of the $(\ell, m)$ mode peaks. 

In \texttt{SEOBNRv5HM}, following \texttt{SEOBNRv4HM}, the $(5, 5)$ mode was attached at $10M$ before the peak of the $(2, 2)$ mode, due to errors in the NR input values at late times. 
In this work, we find that with the new catalog of SXS waveforms, the input values are stable enough to attach the $(5, 5)$ mode at its own peak.

Thanks to the accuracy of the \texttt{SEOBNRv5} inspiral, we are able to start the intermediate region only $80M$ before $t^{\mathrm{peak }}_{22}$, in contrast to the \texttt{IMRPhenomT} models, which start the intermediate region earlier.
A shorter intermediate region improves the accuracy there, since the ansatz has to reproduce less structure.
The choice of $80M$ balances this against the degradation of the EOB inspiral near merger, and we find it to be the optimal length for all modes. 
In addition, the diagnostic detailed in Sec.~\ref{sec:Motivation} showed a large accumulation of error in this region, which motivated replacing it here. 

\subsubsection{Inspiral modes}
\label{sec:inspiralmodeconstruction}
In the EOB formalism, inspiral modes are written in a factorized form~\cite{Damour:2007xr,Damour:2008gu,Pan:2010hz}, given by 
\begin{equation}
\label{hlmFactorised}
	h_{\ell m}^{\mathrm{F}}=h_{\ell m}^{\mathrm{N}} \hat{S}_{\mathrm{eff}} T_{\ell m} f_{\ell m} e^{i \delta_{\ell m}}.
\end{equation}

Within the factorized form of the inspiral modes, $h_{\ell m}^{\mathrm{N}}$ denotes the Newtonian (leading-order) waveform, while the factors $\hat{S}_{\mathrm{eff}}$ and $T_{\ell m}$ describe the source and tail terms, respectively. 
Explicit expressions for these can be found in Appendix~A of Ref.~\cite{Pompili:2023tna}.
$f_{\ell m}$ and $\delta_{\ell m}$ are further resummations of the residual PN amplitude and phase, and $f_{\ell m}$ can be written as 

\begin{align}
\label{frholm}
f_{\ell m} = \left\{
        \begin{array}{ll}
           \rho_{\ell m}^\ell, & \quad m \,\,\mathrm{ even}, \\\\
           (\rho_{\ell m}^\mathrm{NS})^\ell + f_{\ell m}^\mathrm{S}, & \quad m \,\,\mathrm{ odd},
        \end{array}
    \right. 
\end{align}
with $\rho_{\ell m}^\mathrm{NS}$ denoting the non-spinning part of $\rho_{\ell m}$, and $f_{\ell m}^\mathrm{S}$ denoting the spin part of $f_{\ell m}$. 
These expressions are listed in the appendices of Ref.~\cite{Pompili:2023tna}, and were derived in Refs.~\cite{Cotesta:2018fcv,Bohe:2016gbl,Pan:2010hz,Taracchini:2012ig,Henry:2022dzx,Henry:2022ccf}.

In previous EOB models, including \texttt{SEOBNRv5HM}, the final inspiral--plunge waveform modes were given by $h_{\ell m}^{\mathrm{insp-plunge }}=h_{\ell m}^{\mathrm{F}} N_{\ell m}$, where $N_{\ell m}$ are the NQC corrections. 
These are phenomenological corrections to the waveform which depend on the dynamics, as well as fitted input values from NR.
The explicit construction of these corrections can be found in Ref.~\cite{Pompili:2023tna}.

In this model, \texttt{SEOBNRv5iHM}, we remove the NQC corrections from each mode. Instead, we model the inspiral up to time $t^{\mathrm{inter}}_{\ell m}$, detailed above in Eq.~\eqref{eq:tcut_tmatch}. In this region, we thus have 
\begin{equation}
	\label{eq:hellm_insp}
	h_{\ell m}^{\mathrm{insp }}=h_{\ell m}^{\mathrm{F}}.
\end{equation}

As in \texttt{SEOBNRv5HM}, we include second-order gravitational self-force (2GSF) calibration coefficients in the non-spinning modes and RR force, determined following the method of Ref.~\cite{vandeMeent:2023ols}. Following a correction to the 2GSF fluxes, we adopt the recalibrated coefficient values computed in Ref.~\cite{Leather:2025nhu}.

The factorized modes given in Eq.~\eqref{hlmFactorised} are relatively accurate up to the chosen transition time $t^{\mathrm{inter}}_{\ell m}$, but in some cases the very last section of the inspiral deviates from NR. In order to ensure accuracy at $t^{\mathrm{inter}}_{\ell m}$ where we attach the intermediate waveform, we introduce the matching parameters $\alpha_{\ell m}$ and $\gamma_{\ell m}$ to correct the amplitude and phase respectively. These are introduced as follows:
\begin{subequations}
	\label{eq:MatchingCoefficients}
	\begin{align}
	f_{\ell m} &\rightarrow f_{\ell m} + \alpha_{\ell m} v_\Omega^{\beta_{f, \ell m}}, 	\label{eq:MatchingCoefficientsAmp}\\
	\delta_{\ell m} &\rightarrow \delta_{\ell m} + \gamma_{\ell m} v_\Omega^{\beta_{\delta, \ell m}}, \label{eq:MatchingCoefficientsPhase}
	\end{align}
\end{subequations}
where $\beta_{f, \ell m}$ and $\beta_{\delta, \ell m}$ are powers of $v_\Omega$ corresponding to higher PN orders beyond those already included in $f_{\ell m}$ and $\delta_{\ell m}$. 
The values of $\beta_{f, \ell m}$ and $\beta_{\delta, \ell m}$ used for each mode are listed in Appendix~\ref{app:modes}, which make the PN order of the matching terms explicit.

Note that these matching terms are not themselves fitted across parameter space, but are instead determined by demanding that the amplitude and frequency at $t^{\mathrm{inter}}_{\ell m}$ match NR input values, for which we have fits across parameter space. That is, if we write the factorized waveform including the matching parameters as follows
\begin{equation}
	\label{hlmFactorisedWithMatch}
	\begin{aligned}
	h_{\ell m}^{\mathrm{F}} &= h_{\ell m}^{\mathrm{N}} \hat{S}_{\mathrm{eff}} T_{\ell m} f_{\ell m}\left(\alpha_{\ell m}\right) e^{i \delta_{\ell m}\left(\gamma_{\ell m}\right)}\\
	&\equiv A^{\mathrm{F}}_{\ell m}\left(\alpha_{\ell m}\right)e^{i\phi^{\mathrm{F}}_{\ell m}\left(\gamma_{\ell m}\right)},
	\end{aligned}
\end{equation}
then $\alpha_{\ell m}$ and $\gamma_{\ell m}$ are determined by demanding
\begin{subequations}
	\begin{align}	
	\left|h_{\ell m}^{\mathrm{F}}\left(t^{\mathrm{inter }}_{\ell m}\right)\right| &\equiv \left. A^{\mathrm{F}}_{\ell m}\left(\alpha_{\ell m}\right)\right|_{t=t^{\mathrm{inter }}_{\ell m}} \nonumber\\
	&=\left|h_{\ell m}^{\mathrm{NR}}\left(t^{\mathrm{inter }}_{\ell m}\right)\right|, \quad \forall (\ell, m),\label{eq:MatchingCoefficientConditionAmp}\\
	\left. \omega_{\ell m}^\mathrm{F} \left(\gamma_{\ell m} \right) \right|_{t = t_{\ell m}^\mathrm{inter}} &\equiv \left.-\dot{\phi}^\mathrm{F} _{\ell m}\left(\gamma_{\ell m} \right) \right|_{t = t_{\ell m}^\mathrm{inter}}\nonumber \\
	&= \omega_{\ell m}^{\mathrm{NR}}\left(t^{\mathrm{inter }}_{\ell m}\right), \quad \forall (\ell, m),\label{eq:MatchingCoefficientConditionFreq}
	\end{align}
\end{subequations}
where $\left|h_{\ell m}^{\mathrm{NR}}\left(t^{\mathrm{inter}}_{\ell m}\right)\right|$ and $\omega_{\ell m}^{\mathrm{NR}}\left(t^{\mathrm{inter}}_{\ell m}\right)$ are the amplitude and frequency of the NR modes at the transition time, obtained from fits detailed in Appendix~\ref{app:ivfits}. The fitting procedure for these input values is detailed in Sec.~\ref{sec:Calibration_Fitting}. 

The $\alpha_{\ell m}$ matching terms are similar to the terms used in \texttt{SEOBNRv4HM} and \texttt{SEOBNRv5HM} to cure the minima in the $(2,1)$, $(5,5)$ and $(4,3)$ mode amplitudes. 
$\alpha_{\ell m}$ and $\gamma_{\ell m}$ are additionally similar in spirit to the ``collocation points'' used in constructing \texttt{IMRPhenom} models (e.g.~\cite{Husa:2015iqa,Khan:2015jqa,Pratten:2020fqn,Estelles:2020osj}).
However, in the earlier \texttt{SEOBNR} models they were fixed at the later time of $t^{\mathrm{match }}_{\ell m}$, and were used together with the NQC corrections to bring especially the late inspiral waveform into better agreement with NR. 
Here, we set our matching terms at an earlier time and apply them for all modes. 
Note that if we were to try to use these matching coefficients at $t^{\mathrm{match }}_{\ell m}$ without additional NQC corrections, we would obtain pathological behavior. The reason they perform well when applied at $t^{\mathrm{inter}}_{\ell m}$ is that they are correcting a minor difference.

PN predicts minima in the amplitude of the $(2,1)$, $(5,5)$ and $(4,3)$ modes. 
Previously, in Ref.~\cite{Cotesta:2018fcv}, this behavior was seen in the $(2,1)$ mode of four NR simulations, including one produced for that paper. 
No minima were found in the $(5,5)$ and $(4,3)$ modes in NR. 

In the third SXS catalog~\cite{Scheel:2025jct}, we find 15 simulations with dips in the amplitude of the $(2,1)$ mode\footnote{These are listed in the ancillary file \texttt{sxs\_simulations\_with\_dip\_in\_21.dat}.}. 
These are typically simulations where the $(2,1)$ mode amplitude is comparatively small, and in most cases, the dip occurs well before the intermediate region. 
We find that for most of these cases we accurately track the location of the zero crossing---however, the PN basis of our waveforms predicts zero crossings in more parts of the parameter space that we do not see in NR. 

In addition, we find one simulation, \texttt{SXS:BBH:1376}, where this behavior is present in the $(5, 5)$ mode, confirming in NR the PN-predicted minimum, i.e., that this feature is not merely an artifact of the PN expressions.

\subsubsection{Modeling the intermediate region}
\label{sec:intermediate_region_modelling}
To model the intermediate region waveform, defined for each mode as the region between $t^{\mathrm{inter}}_{\ell m}$ and $t^{\mathrm{match }}_{\ell m}$, we use a phenomenological ansatz. This intermediate region is motivated by that used in the \texttt{IMRPhenomT} family of models, described first in Ref.~\cite{Estelles:2020osj} and employed in Refs.~\cite{Estelles:2020twz,Estelles:2021gvs,Planas:2025feq}. 
The ansatz employed here is similar to that of \texttt{IMRPhenomTHM}, with the frequency ansatz promoted from a quartic to a quintic expression. 

In the \texttt{IMRPhenomT} framework, this intermediate region begins $150M$ before the peak of the $(2,2)$ mode ($t_{22}^\mathrm{peak} - 150M$). 
In this model, we choose to apply the intermediate region over a shorter interval, starting at $t_{22}^\mathrm{peak} - 80M$, since we find that the inspiral modes described above have a high level of accuracy until this point.
In addition, the intermediate region ansatz performs better when used on a shorter region, as it has to capture less phenomenology with a single ansatz. 

We construct the intermediate region modes as
\begin{equation}
	\label{eq:InterModes}
	h_{\ell m}^{\mathrm{inter}}(t) = A_{\ell m}^\mathrm{inter}(t)\ e^{i\phi_{\ell m}^\mathrm{inter}(t)}
\end{equation}
where the amplitude and frequency are given by the phenomenological ans\"atze. These ans\"atze follow those of the \texttt{IMRPhenomT} models, except for the promotion of the frequency ansatz from a quartic to a quintic expression:
\begin{align}
	A_{\ell m}^\mathrm{inter}(t) &= b_0 + b_1 (t - t^\mathrm{peak}_{\ell m})^2 \notag \\ 
	 &\quad + b_2 \sech^{1/7}(2\sigma_{\ell m}^\mathrm{I} (t - t^\mathrm{peak}_{\ell m})) \label{eq:inter_amp}\\ 
	&\quad + b_3 \sech(\sigma_{\ell m}^\mathrm{I} (t - t^\mathrm{peak}_{\ell m})), \notag \\ 
	\omega^\mathrm{inter}_{\ell m}(t) &= \sigma_{\ell m}^\mathrm{R} \left(1 - \sum_{k=0}^{k=5}a_k\arcsinh^k(\sigma_{\ell m}^\mathrm{I}(t - t^\mathrm{peak}_{\ell m})) \right).  \label{eq:inter_freq}
\end{align}

The term proportional to $b_3$ reproduces the $\sech$ amplitude profile derived in the Backwards One-Body (BOB) model for $\psi_4$~\cite{McWilliams:2018ztb}, with its characteristic timescale set by the damping time of the fundamental QNM. 
The remaining terms, introduced in \texttt{IMRPhenomT}, provide additional phenomenological flexibility when applying this profile directly to the strain amplitude, for which the BOB functional form is not derived.
Here, $\sigma_{\ell m}^\mathrm{R}$ and $\sigma_{\ell m}^\mathrm{I}$ are respectively the real and imaginary parts of the complex fundamental quasi-normal mode frequencies of the remnant BH, $\sigma_{\ell m 0} = \sigma_{\ell m}^\mathrm{R} - i \sigma_{\ell m}^\mathrm{I} $. 
As in \texttt{SEOBNRv5HM}, these QNM frequencies are obtained for each $(\ell, m)$ mode using the Python package \texttt{qnm}~\cite{Stein:2019mop}.
Their inputs are the final mass and spin of the remnant black hole, computed using the fitting formulas from Refs.~\cite{Jimenez-Forteza:2016oae} and \cite{Hofmann:2016yih}. 
The phase $\phi_{\ell m}^\mathrm{inter}(t)$ is then computed using 
\begin{equation}
	\label{eq:inter_phase}
	\phi_{\ell m}^\mathrm{inter}(t) = \phi_{\ell m}^\mathrm{insp}(t^{\mathrm{inter}}_{\ell m}) - \int_{t'= t^{\mathrm{inter}}_{\ell m}}^t dt' \omega_{\ell m}^\mathrm{inter}(t'),
\end{equation}
where $\phi_{\ell m}^\mathrm{insp}(t^{\mathrm{inter}}_{\ell m})$ is the phase of the inspiral mode $(\ell, m)$ at $t=t^{\mathrm{inter}}_{\ell m}$. 

The free coefficients $a_i$ and $b_i$ are fixed by conditions imposed at three times: the transition time $t^{\mathrm{inter}}_{\ell m}$, where the intermediate region begins; a collocation point inside the region, $t^{\mathrm{mid}}_{\ell m} = t^{\mathrm{peak}}_{22} - 25M$; and the peak time of each mode, $t^{\mathrm{peak}}_{\ell m}$.
For the $(3,2)$ and $(4,3)$ modes, the attachment time is $t^{\mathrm{match}}_{\ell m} = t^{\mathrm{peak}}_{22}$. Even so, the intermediate-region ansatz for these modes is still constrained at $t^{\mathrm{peak}}_{\ell m}$; the waveform is simply truncated at $t^{\mathrm{peak}}_{22}$, where the ringdown is attached.

The four amplitude ansatz constants $\{b_0, b_1, b_2, b_3\}$ are set by the following conditions, where `$\overset{!}{=}$' denotes a demanded equality:
\begin{itemize}
	\item The amplitude at the start of the intermediate region modes must match that at the end of the inspiral modes: 
	\begin{equation}
	\label{eq:InterRegCutAmpCondition1}
	\left|h_{\ell m}^{\mathrm{insp}}\left(t^{\mathrm{inter }}_{\ell m}\right)\right| \overset{!}{=} \left|h_{\ell m}^{\mathrm{inter}}\left(t^{\mathrm{inter }}_{\ell m}\right)\right| \equiv A_{\ell m}^\mathrm{inter}(t^{\mathrm{inter }}_{\ell m}).
	\end{equation}
	Since the matching terms $\alpha_{\ell m}$ fix the amplitude at this point to equal NR input values (Eq.~\eqref{eq:MatchingCoefficientConditionAmp}), this is equivalent to 
	\begin{equation}
	\label{eq:InterRegCutAmpCondition2}
	\left|h_{\ell m}^{\mathrm{inter}}\left(t^{\mathrm{inter }}_{\ell m}\right)\right| \equiv A_{\ell m}^\mathrm{inter}(t^{\mathrm{inter}}_{\ell m}) \overset{!}{=} \left|h_{\ell m}^{\mathrm{NR}}\left(t^{\mathrm{inter }}_{\ell m}\right)\right|.
	\end{equation}
	\item The first derivative of the amplitude at the start of the intermediate region modes must match that at the end of the inspiral modes: 
	\begin{align}
	\label{eq:InterRegCutDAmpCondition}
	\left. \frac{d\left| h_{\ell m}^{\mathrm{insp}}(t) \right|}{dt} \right|_{t =t^{\mathrm{inter}}_{\ell m}}
	&\overset{!}{=} \left. \frac{d\left| h_{\ell m}^{\mathrm{inter}}(t) \right|}{dt} \right|_{t =t^{\mathrm{inter}}_{\ell m}} \\
	&\equiv \left. \frac{d A_{\ell m}^\mathrm{inter}(t)}{dt} \right|_{t =t^{\mathrm{inter}}_{\ell m}}. \nonumber
	\end{align}
	Note that NR input values are not imposed for this quantity.
	\item The amplitude at the collocation point, $t^{\mathrm{mid }}_{\ell m} = t^{\mathrm{peak }}_{22} - 25M$, of the intermediate region modes must match the NR input values: 
	\begin{equation}
	\label{eq:InterRegMidAmpCondition}
	\left|h_{\ell m}^{\mathrm{inter}}\left(t^{\mathrm{mid }}_{\ell m}\right)\right| \equiv A_{\ell m}^\mathrm{inter}(t^{\mathrm{mid }}_{\ell m}) \overset{!}{=} \left|h_{\ell m}^{\mathrm{NR}}\left(t^{\mathrm{mid }}_{\ell m}\right)\right|.
	\end{equation}
	\item The amplitude at the peak of each of the intermediate region modes must match the NR input values: 
	\begin{equation}
	\label{eq:InterRegPeakAmpCondition}
	\left|h_{\ell m}^{\mathrm{inter}}\left(t^{\mathrm{peak }}_{\ell m}\right)\right| \equiv A_{\ell m}^\mathrm{inter}(t^{\mathrm{peak }}_{\ell m}) \overset{!}{=} \left|h_{\ell m}^{\mathrm{NR}}\left(t^{\mathrm{peak }}_{\ell m}\right)\right|.
	\end{equation}
\end{itemize}
Note that the amplitude ansatz is even in $t - t^{\mathrm{peak}}_{\ell m}$, and thus its amplitude has vanishing first derivative at the peak of each mode by construction.

The six frequency ansatz constants $\{a_0, a_1, a_2, a_3, a_4, a_5\}$ are fixed by requiring the following: 
\begin{itemize}
	\item The frequency at the start of the intermediate region modes must match that at the end of the inspiral modes: 
	\begin{equation}
	\label{eq:InterRegCutFreqCondition1}
	\omega_{\ell m}^{\mathrm{insp}}\left(t^{\mathrm{inter }}_{\ell m}\right) \overset{!}{=} \omega_{\ell m}^{\mathrm{inter}}\left(t^{\mathrm{inter }}_{\ell m}\right). 
 	\end{equation}
	Like the corresponding amplitude condition, since the matching terms $\gamma_{\ell m}$ fix the frequency at this point to equal NR input values (Eq.~\eqref{eq:MatchingCoefficientConditionFreq}), this is equivalent to 
	\begin{equation}
	\label{eq:InterRegCutFreqCondition2}
	\omega_{\ell m}^{\mathrm{inter}}\left(t^{\mathrm{inter }}_{\ell m}\right) \overset{!}{=} \omega_{\ell m}^{\mathrm{NR}}\left(t^{\mathrm{inter }}_{\ell m}\right). 
	\end{equation}
	\item The first derivative of the frequency at the start of the intermediate region modes must match that at the end of the inspiral modes: 
	\begin{equation}
	\label{eq:InterRegCutDFreqCondition}
	\left. \frac{d\omega_{\ell m}^{\mathrm{insp}}(t)}{dt} \right|_{t =t^{\mathrm{inter}}_{\ell m}}
	\overset{!}{=} \left. \frac{d\omega_{\ell m}^{\mathrm{inter}}(t)}{dt} \right|_{t =t^{\mathrm{inter}}_{\ell m}}.
	\end{equation}
	Note that NR input values are not imposed for this quantity.
	\item The frequency at the collocation point of the intermediate region modes must match the NR input values: 
	\begin{equation}
	\label{eq:InterRegMidFreqCondition}
	\omega_{\ell m}^{\mathrm{inter}}\left(t^{\mathrm{mid }}_{\ell m}\right) \overset{!}{=} \omega_{\ell m}^{\mathrm{NR}}\left(t^{\mathrm{mid }}_{\ell m}\right).
	\end{equation}
	\item The first derivative of the frequency at the collocation point of the intermediate region modes must match the NR input values: 
	\begin{equation}
	\label{eq:InterRegMidDFreqCondition}
	\left. \frac{d\omega_{\ell m}^{\mathrm{inter}}(t)}{dt} \right|_{t =t^{\mathrm{mid}}_{\ell m}}
	\overset{!}{=} \left. \frac{d\omega_{\ell m}^{\mathrm{NR}}(t)}{dt} \right|_{t =t^{\mathrm{mid}}_{\ell m}}. 
	\end{equation}
	This condition is not enforced in the \texttt{IMRPhenomT} models, and is included in the \texttt{SEOBNRv5iHM} version of the intermediate region by promoting the frequency ansatz to a quintic expression.  
	\item The frequency at the peak of the intermediate region modes must match the NR input values: 
	\begin{equation}
	\label{eq:InterRegPeakFreqCondition}
	\omega_{\ell m}^{\mathrm{inter}}\left(t^{\mathrm{peak }}_{\ell m}\right) \overset{!}{=} \omega_{\ell m}^{\mathrm{NR}}\left(t^{\mathrm{peak }}_{\ell m}\right).
	\end{equation}
	\item The first derivative of the frequency at the peak of the intermediate region modes must match the NR input values: 
	\begin{equation}
	\label{eq:InterRegPeakDFreqCondition}
	\left. \frac{d\omega_{\ell m}^{\mathrm{inter}}(t)}{dt} \right|_{t =t^{\mathrm{peak}}_{\ell m}}
	\overset{!}{=} \left. \frac{d\omega_{\ell m}^{\mathrm{NR}}(t)}{dt} \right|_{t =t^{\mathrm{peak}}_{\ell m}}. 
	\end{equation}
\end{itemize}

The NR input values referred to in Eqs.~\eqref{eq:InterRegCutAmpCondition2}--\eqref{eq:InterRegPeakDFreqCondition} are obtained from fits across parameter space, built on values extracted from QC, aligned-spin NR waveforms from the third SXS catalog~\cite{Scheel:2025jct}, and BH-perturbation-theory waveforms described in Ref.~\cite{Nishimura:2026nse}. 
The fitting procedure is described in Sec.~\ref{sec:Calibration_Fitting}, and the fits can be found in Appendix~\ref{app:ivfits}.

\begin{figure}[t]
    \centering
    \includegraphics[width=\columnwidth]{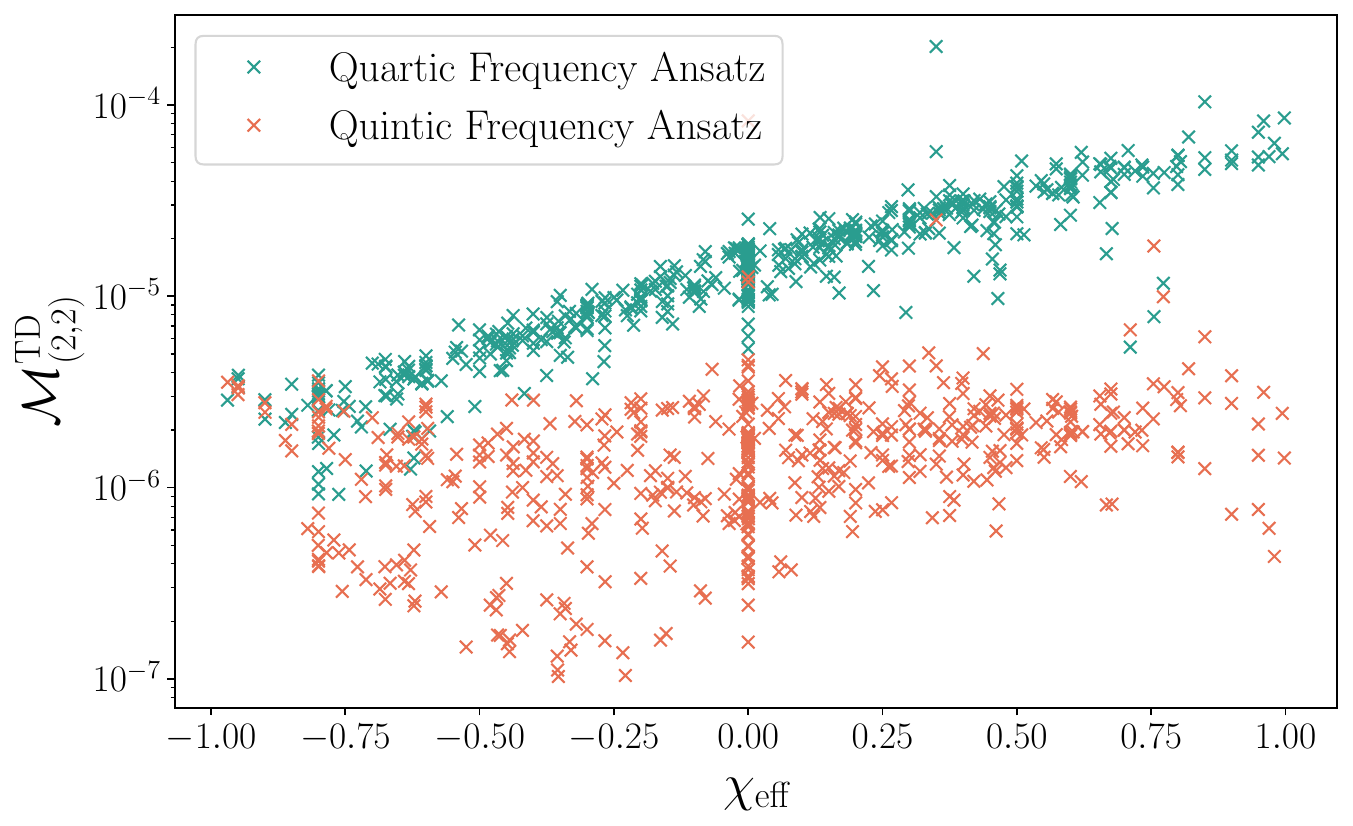}\\[\medskipamount]
    \includegraphics[width=\columnwidth]{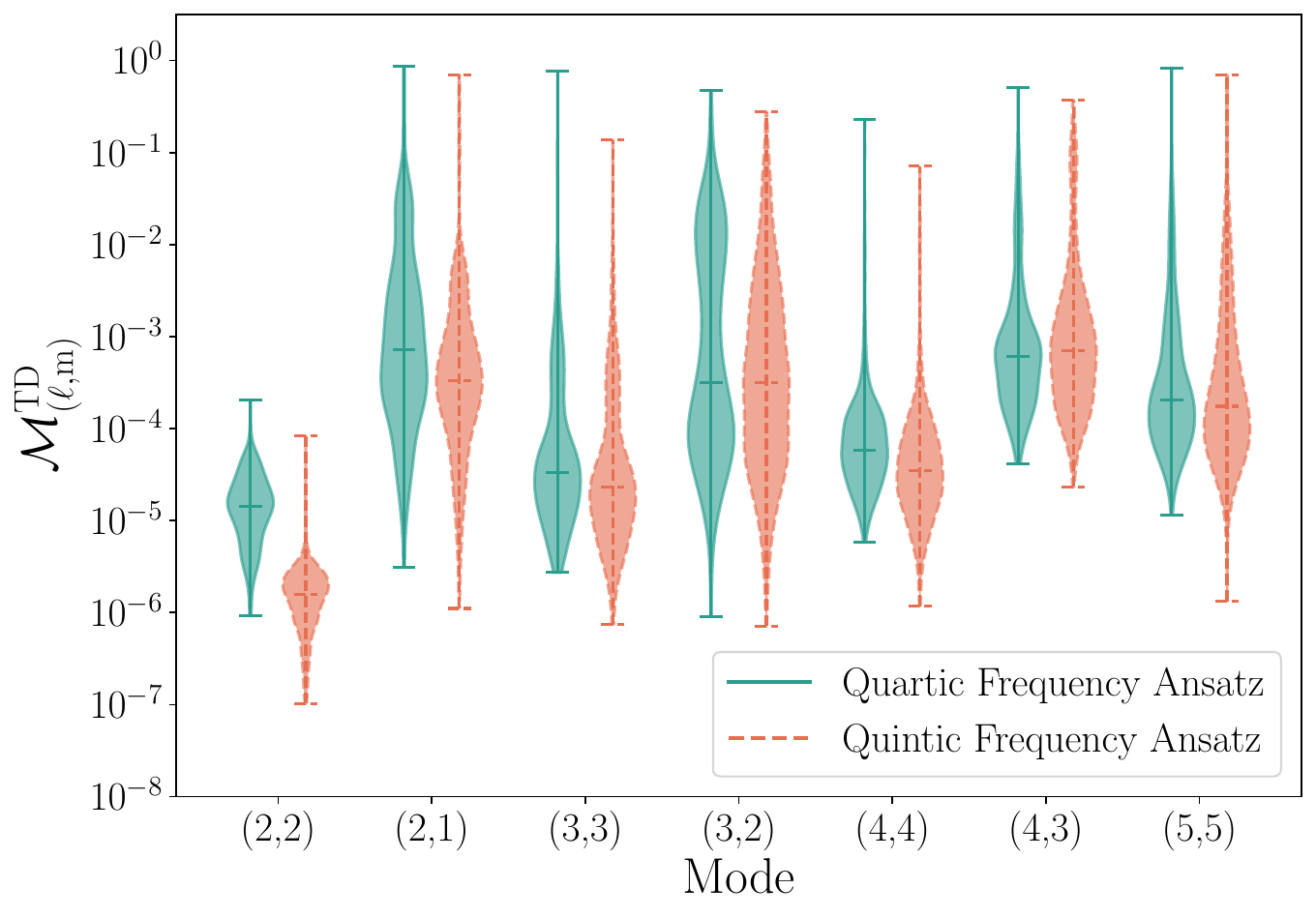}
    \caption{Evaluating the quintic frequency ansatz in Eq.~\eqref{eq:inter_freq} versus the quartic frequency ansatz used in \texttt{IMRPhenomT}. For this check, direct input values from NR waveforms are used to construct the intermediate-region waveform, which are then compared against the same NR waveforms through the time-domain unfaithfulness of Eq.~\eqref{eq:timedomain_mismatch}. 
	Top: $(2, 2)$ mode results plotted against $\chi_\mathrm{eff}$.
	We see a clear upward trend for the quartic frequency ansatz, which loses accuracy for increasing $\chi_\mathrm{eff}$. 
	The quintic ansatz instead gives consistent and more accurate results.
	Bottom: Violin plots of the mismatch distribution for all $(\ell, m)$ modes we model.}
    \label{fig:inter_reg_ansatz}
\end{figure}

We additionally control the peak time of each mode.
The peak time of the $(2, 2)$ mode is calibrated to NR as in previous \texttt{SEOBNR} models---this time shift is motivated by observations in the test-mass limit of a delay between the peak of the orbital frequency and the peak of the $(2, 2)$ mode~\cite{Barausse:2011kb,Taracchini:2014zpa,Price:2016ywk}. 
For this peak, we follow \texttt{SEOBNRv5HM} in defining a shift relative to $t_{\mathrm{ISCO}}$, the time at which $r = r_{\mathrm{ISCO}}$, where $r_{\mathrm{ISCO}}$ is the radius of the geodesic ISCO in Kerr spacetime~\cite{Bardeen:1972fi} with the same mass and spin as the remnant.
\begin{equation}
	\label{eq:t_attach}
	t^{\mathrm{peak}}_{22} = t_{\mathrm{ISCO}}+ \Delta t^{22}_{\mathrm{ISCO}}.
\end{equation}
The offset $\Delta t^{22}_{\mathrm{ISCO}}$ is determined together with the Hamiltonian calibration parameters, as detailed in Sec.~\ref{sec:Calibration_Fitting}.

In addition, since we now model the intermediate region to the peak of each $(\ell,m)$ mode, we need the difference 
\begin{equation}
	\Delta t_{22}^{\ell m} =  t^{\mathrm{peak }}_{22} - t^{\mathrm{peak }}_{\ell m}. 
\end{equation}
These time differences are treated as NR input values in the same way as the quantities above: $\Delta t_{22}^{\ell m}$ is extracted from each NR simulation, as well as from the TML waveforms, and fitted across parameter space.

The promotion of the frequency ansatz to a quintic expression, allowing us to control the frequency derivative at $t^{\mathrm{mid}}_{\ell m}$, was motivated by the observation that it improves the ansatz accuracy, especially for positive $\chi_\mathrm{eff}$ cases.
The results of this study can be seen in Fig.~\ref{fig:inter_reg_ansatz}. 
For this check, we take \emph{direct} input values from NR to construct the intermediate-region waveform.
This isolates the behavior of the ansatz itself from the accuracy of the fits across parameter space.

For each NR simulation, we extract the necessary input values and build the intermediate region waveform with both the quartic and quintic frequency ans\"atze.
We then compare this against the NR waveform in that region, from $t^{\mathrm{inter}}_{\ell m}$ to $t^{\mathrm{peak}}_{\ell m}$. 
The waveform phases are aligned by construction at $t^{\mathrm{inter}}_{\ell m}$, by taking 
\begin{equation}
	\phi_{\ell m}^\mathrm{insp}(t^{\mathrm{inter}}_{\ell m}) \equiv \phi_{\ell m}^\mathrm{NR}(t^{\mathrm{inter}}_{\ell m}).
\end{equation}
We then compute the time-domain unfaithfulness between the two waveforms, without any time or phase optimization:
\begin{equation}
	\label{eq:timedomain_mismatch}
	\mathcal{M}_{\mathrm{TD}}
	= 1 - \mathrm{Re}\frac{\displaystyle\int_{t_{\ell m}^{\mathrm{inter}}}^{t_{\ell m}^{\mathrm{peak}}} h_{\ell m}^{\mathrm{inter}}(t)\,\left(h_{\ell m}^{\mathrm{NR}}(t)\right)^*\,\mathrm{d}t}
	{\sqrt{\displaystyle\int_{t_{\ell m}^{\mathrm{inter}}}^{t_{\ell m}^{\mathrm{peak}}} \left|h_{\ell m}^{\mathrm{inter}}\right|^{2}\,\mathrm{d}t
	\;\displaystyle\int_{t_{\ell m}^{\mathrm{inter}}}^{t_{\ell m}^{\mathrm{peak}}} \left|h_{\ell m}^\mathrm{NR}\right|^{2}\,\mathrm{d}t}}
\end{equation}

In the top panel of Fig.~\ref{fig:inter_reg_ansatz}, we plot the results of the two ans\"atze for the $(2, 2)$ mode, as a function of $\chi_\mathrm{eff}$.
Whereas the quartic frequency ansatz shows a clear trend, with lower accuracy at higher values of $\chi_\mathrm{eff}$, the quintic frequency ansatz instead achieves higher accuracy, and improves in particular for large, positive values of $\chi_\mathrm{eff}$.

The bottom panel instead shows violin plots of the distribution of accuracy of the two ans\"atze evaluated for every mode. 
For the higher-order modes, the quintic ansatz does not improve the accuracy as clearly, but gives comparable results, slightly better for most modes (for the $(4,3)$ mode the median is marginally higher). 
We thus adopt this ansatz for every mode. 
Note also that, even with direct NR input values, the ansatz is markedly less accurate for the subdominant modes than for the $(2,2)$ mode.

\subsubsection{Merger-ringdown modes}
\label{sec:mr_construction}
A key difference in the construction of the merger-ringdown section between \texttt{SEOBNRv5iHM} and \texttt{SEOBNRv5HM} is that the use of the intermediate region allows us to attach this later section at an independently chosen time for each mode, rather than at $t^{\mathrm{peak}}_{22}$ for all modes. For several modes, we choose this to be the time of the peak of the mode's amplitude.
In addition to this construction difference, we use a different method to fit the free parameter quantities across parameter space, detailed in Sec.~\ref{sec:Calibration_Fitting}. 

The ansatz used is the same as that employed in previous EOB models~\cite{Damour:2014yha,Bohe:2016gbl,Cotesta:2018fcv,Gamboa:2024hli,Gamboa:2024imd,Gamboa:2026jht}, and is applicable when the amplitude and frequency of the merger-ringdown show monotonic evolution.
This is true for the modes $(2,2)$, $(3,3)$, $(2,1)$, $(4,4)$, $(5,5)$, and the ansatz reads:
\begin{equation}
h_{\ell m}^{\mathrm{merger-RD }}(t)=\nu \tilde{A}_{\ell m}(t) e^{i \tilde{\phi}_{\ell m}(t)} e^{-i \sigma_{\ell m 0}\left(t-t^{\mathrm{match }}_{\ell m}\right)}.
\label{eq:MergerRingdownAnsatz}
\end{equation}
As in the ansatz for the intermediate region, Eqs.~\eqref{eq:inter_amp}--\eqref{eq:inter_freq}, $\sigma_{\ell m 0}$ is the complex frequency of the fundamental QNM of the remnant black hole.
The amplitude $\tilde{A}_{\ell m}$ and phase $\tilde{\phi}_{\ell m}$ ans\"atze are written as~\cite{Bohe:2016gbl, Cotesta:2018fcv}
\begin{equation}
\tilde{A}_{\ell m}(t)=c_{1, c}^{\ell m} \tanh \left[c_{1, f}^{\ell m}\left(t-t^{\mathrm{match }}_{\ell m}\right)+c_{2, f}^{\ell m}\right]+c_{2, c}^{\ell m},
\label{eq:MergerRingdownAnsatzAmp}
\end{equation}
\begin{equation}
\tilde{\phi}_{\ell m}(t)=
\phi^{\mathrm{match }}_{\ell m}-d_{1, c}^{\ell m} \log \left[\frac{1+d_{2, f}^{\ell m} e^{-d_{1, f}^{\ell  m}\left(t-t^{\mathrm{match}}_{\ell m}\right)}}{1+d_{2, f}^{\ell m}}\right].
\label{eq:MergerRingdownAnsatzPhase}
\end{equation}
Here, $\phi^{\mathrm{match }}_{\ell m}$ is the phase of the intermediate-region mode at the matching time of each mode, $t^{\mathrm{match}}_{\ell m}$.
Within these ans\"atze, the coefficients $\{c_{1, c}^{\ell m},\ c_{2, c}^{\ell m},\ d_{1, c}^{\ell m}\}$ are constrained by imposing the amplitude, first derivative of the amplitude, and frequency at time $t^{\mathrm{match}}_{\ell m}$.  
These values are taken from the last point of the intermediate-region modes. 
Since the intermediate-region construction enforces these values to be equal to NR input values at $t^{\mathrm{match}}_{\ell m}$ for the modes attached at their own peak (i.e. all except the $(3,2)$ and $(4,3)$ modes, whose intermediate region is truncated at $t^\mathrm{peak}_{22}$), this enforces NR input values for the start of the merger-ringdown waveform as well.
The exact expressions of the conditions that set these coefficients can be found in Ref.~\cite{Pompili:2023tna}.

The remaining parameters $\{c_{1,f}^{\ell m},\ c_{2,f}^{\ell m},\ d_{1,f}^{\ell m},\ d_{2,f}^{\ell m}\}$ are free coefficients. These are first fitted against every NR simulation individually, and the result of this is then fitted across parameter space, with the same algorithm as that used for the input values (Sec.~\ref{sec:Calibration_Fitting}).
The exact fits are listed in Appendix~\ref{app:mrfits}.

The $(3,2)$ and $(4,3)$ modes exhibit mode mixing in the merger-ringdown, since the spherical-harmonic basis used for NR extraction differs from the spheroidal basis natural to BH perturbation theory~\cite{Buonanno:2006ui,Kelly:2012nd}.
As in \texttt{SEOBNRv5HM}~\cite{Pompili:2023tna}, we therefore apply the merger-ringdown ansatz of Eqs.~\eqref{eq:MergerRingdownAnsatz}--\eqref{eq:MergerRingdownAnsatzPhase} to the spheroidal $(\ell, m, 0)$ mode, whose input values are obtained analytically from those of the spherical $(\ell, m)$ and $(m, m)$ modes, and recover the spherical mode using the mixing coefficients of Ref.~\cite{Berti:2014fga}.

Since we find that mode mixing already sets in at $t_{22}^{\mathrm{peak}}$, we start the merger-ringdown of these modes at that time rather than at $t_{\ell m}^{\mathrm{peak}}$.
We therefore end the intermediate region of the $(3,2)$ and $(4,3)$ modes at $t_{22}^{\mathrm{peak}}$, and take the input values of the spherical $(\ell, m)$ and $(m, m)$ modes from the waveform at that time.

\section{Calibration to Numerical Relativity}
\label{sec:Calibration_Fitting}
Three ingredients of \texttt{SEOBNRv5iHM} are tuned to NR: the Hamiltonian calibration parameters $\{a_6, d_\mathrm{SO}\}$, together with $\Delta t^{22}_\mathrm{ISCO}$; the NR input values; and the free merger-ringdown coefficients.
This section describes the procedures used to produce fits across parameter space for these quantities. 

\subsection{Hamiltonian calibration}
\label{sec:hamiltoniancalibration}
As in previous EOB waveform models, we calibrate higher-order PN terms in the Hamiltonian to NR, closely following the procedure of Ref.~\cite{Pompili:2023tna}. 
However, given the removal of the NQCs, and the inclusion of matching coefficients in the factorized modes, the calibration has been redone in full, yielding substantially different results for the fitted values of $a_6$, $d_\mathrm{SO}$ and $\Delta t^{22}_\mathrm{ISCO}$.

We use the 591 NR waveforms from the third SXS catalog, discussed in Sec.~\ref{sec:Motivation}.
We calibrate using 510 of these, excluding duplicate configurations and a few outliers, compared with the 442 simulations used to calibrate \texttt{SEOBNRv5HM}.
As a check against overfitting, we verify that fits built from smaller subsets of simulations produce similar mismatches on the full set of 591 waveforms.

We consider two metrics when evaluating a given choice of calibration parameters. 
The first is the mismatch, defined in Eq.~\eqref{eq:mismatch}. 
We use the \texttt{A+} LIGO PSD~\cite{LIGO-T1500293} (the expected design sensitivity for the fifth observing run O5), with $f_\mathrm{max} = 2048\,\mathrm{Hz}$, and $f_\mathrm{min} = 1.35 f_\mathrm{start}$ as defined in Sec.~\ref{sec:Motivation}. 
The second metric is the difference in merger time $\delta t_\mathrm{merger}$, i.e., the difference between the EOB and NR times of the $(2,2)$-mode amplitude peak, after aligning the waveforms at low frequency.

Following \texttt{SEOBNRv4} and \texttt{SEOBNRv5HM}, we perform the calibration hierarchically, completing the non-spinning (noS) sector before moving to the aligned-spin (S) sector. 
We decompose the time shift as 
\begin{equation}
	\label{eq:dt_fit}
	\Delta t^{22}_{\rm{ISCO}} = \Delta t^{22}_{\rm{ISCO,\, noS}} + \Delta t^{22}_{\rm{ISCO,\, S}}.
\end{equation}
Using non-spinning simulations, we calibrate the parameters 
\begin{equation}
	\boldsymbol{\theta}_{\rm{noS}} \equiv \{a_6,\Delta t^{22}_{\rm{ISCO,\, noS}}\}.
\end{equation}
These are fitted as functions of the symmetric mass ratio only. After fixing $a_6(\nu)$ and $\Delta t^{22}_{\rm{ISCO,\, noS}}(\nu)$, we use these fits when sampling over the aligned-spin simulations to determine 
\begin{equation}
	\boldsymbol{\theta}_{\rm{S}} \equiv \{d_{\rm{SO}}, \Delta t^{22}_{\rm{ISCO,\, S}}\}.
\end{equation}

We use nested sampling~\cite{Skilling:2006gxv} to obtain posteriors for the parameters $\boldsymbol{\theta}$ for each NR simulation. 
The sampler used is \texttt{nessai}~\cite{Williams:2021qyt}, called through \texttt{Bilby}~\cite{Ashton:2018jfp}.
The likelihood function used is defined as~\cite{Pompili:2023tna}
\begin{equation}
	\label{eq:calib_likelihood}
	P(h^\mathrm{NR} | \boldsymbol{\theta}) \propto \exp \left[-\frac{1}{2}\left(\frac{{\max_M \mathcal{M}_{22}} (\boldsymbol{\theta})}{\sigma_{\mathcal{M}}}\right)^2-\frac{1}{2}\left(\frac{\delta t_{\text {merger}}(\boldsymbol{\theta})}{\sigma_t}\right)^2\right],
\end{equation}
where $\max_M \mathcal{M}_{22}$ is the maximum unfaithfulness between the EOB and NR $(2,2)$ mode waveforms over the total mass range $10\,M_\odot \leq M \leq 200\,M_\odot$.
We choose here a mass range that ends at $200\,M_\odot$ in order for the calibration to optimize mainly the inspiral and time-to-merger. 
$\sigma_{t}$ is chosen to be $5 M$, as in \texttt{SEOBNRv5HM}, and $\sigma_{\mathcal{M}} = 10^{-4}$ rather than the $10^{-3}$ used there, targeting unfaithfulnesses an order of magnitude smaller.

\begin{figure}[t]
    \centering
    \includegraphics[width=\columnwidth]{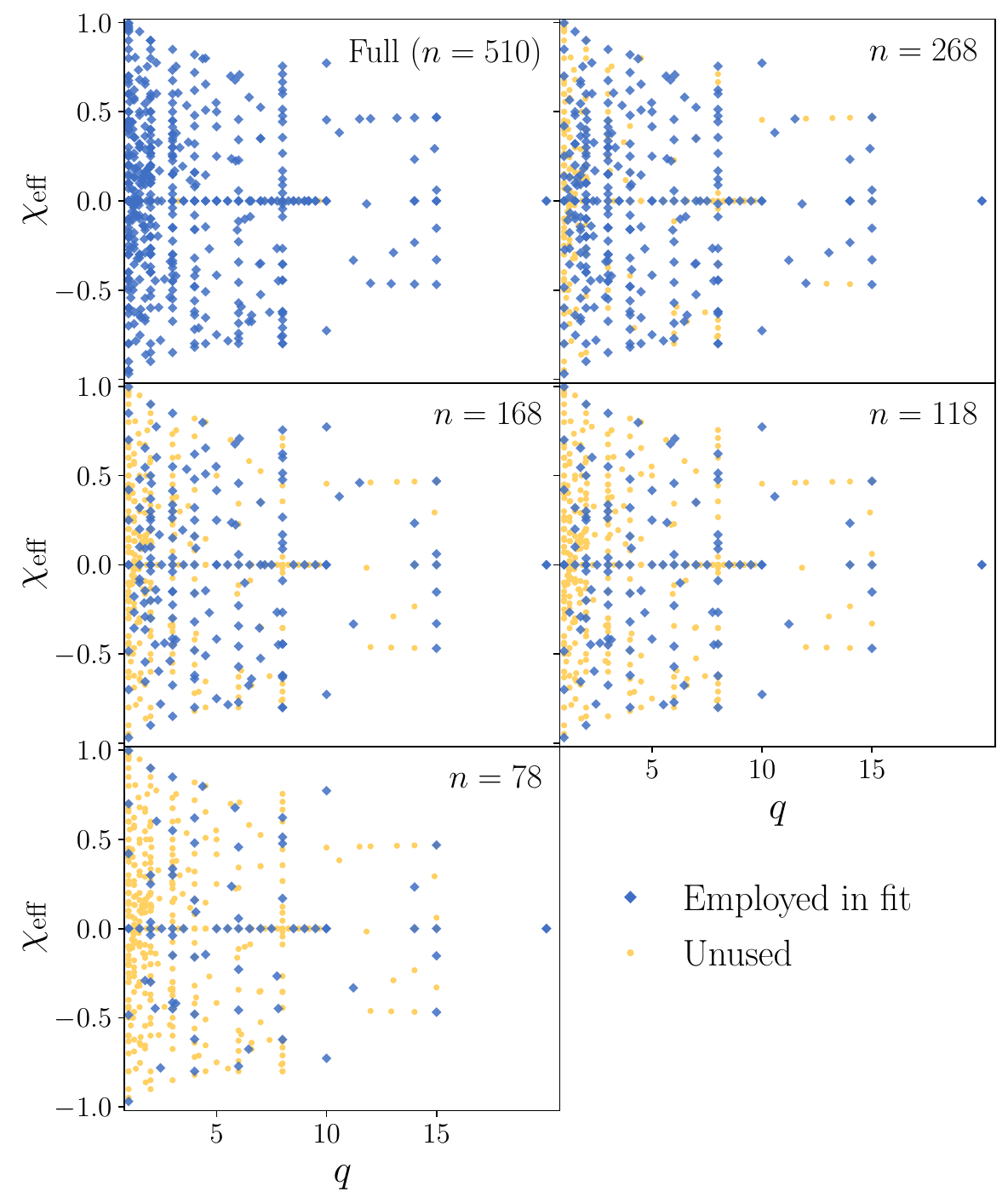}
    \caption{Parameter-space coverage of the various subsets of waveforms used in the construction of calibration parameter fits. Each nested subset contains the same 18 non-spinning simulations and an increasing number of spinning simulations, chosen by farthest-point sampling in $(\nu, a_+, \delta \cdot a_-)$.}
  	\label{fig:CalibrationSubsets}
\end{figure}

\begin{figure}[t]
    \centering
    \includegraphics[width=\columnwidth]{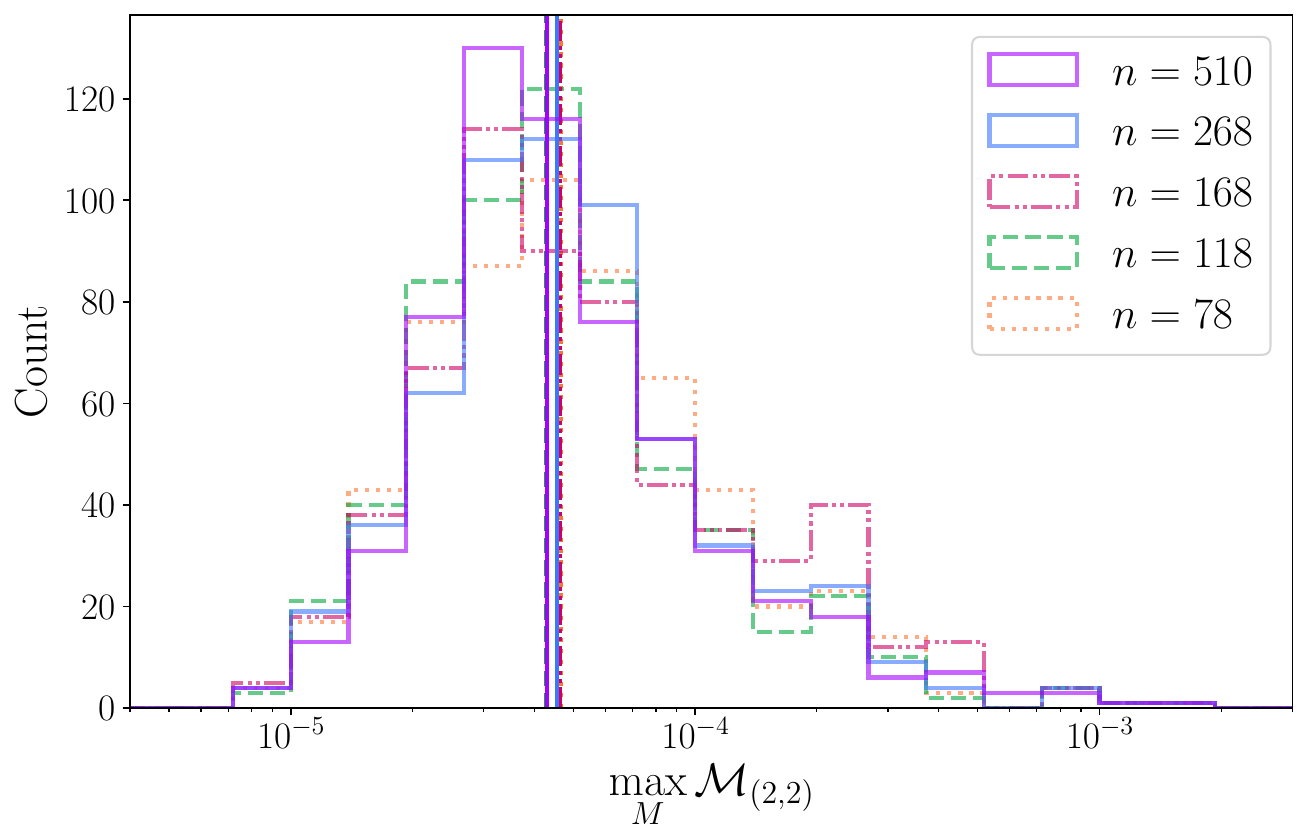}
    \caption{$(2, 2)$ mode mismatch results using calibration fits built from each of the subsets, in all cases evaluated against the full set of 591 NR simulations.}
  	\label{fig:CalibSubsetsResults}
\end{figure}

For the non-spinning sector we fit the median of each posterior rather than the maximum-likelihood point used in \texttt{SEOBNRv5HM}: the two give similar results, and the median improves the subsequent fits in the spinning sector.
In terms of fits across parameter space, we use the parametrization $\{\nu, a_+, \delta \cdot a_-\}$ rather than $\{\nu, a_+, a_-\}$ as was done in \texttt{SEOBNRv5HM}. 
This enforces symmetry under the exchange of labels $1 \leftrightarrow 2$ for the two black holes, as discussed in Ref.~\cite{Gamboa:2026jht}.
Other parts of the fitting procedure remain the same, and for additional details we refer the reader to Ref.~\cite{Pompili:2023tna}.

In order to guard against overfitting, we rebuild the calibration fits using smaller subsets of NR simulations.
We rebuild the fits using nested subsets of $n=78,118,168$, and $268$ simulations, each comprising 18 non-spinning simulations together with an increasing number of spinning simulations.
These reduced calibration sets were selected by greedy farthest-point sampling in the three-dimensional space $(\nu,\, a_+,\, \delta \cdot a_-)$.
The coordinates are rescaled to the unit cube, and each new point maximizes its Euclidean distance to the points already selected.
This prioritizes coverage, and thus samples the edges of parameter space more densely than the bulk.
The parameter space distribution of each of these subsets can be seen in Fig.~\ref{fig:CalibrationSubsets}.

The fits from the different subsets agree well.
Computing unfaithfulness against NR with each of these fits gives similar results, as shown in Fig.~\ref{fig:CalibSubsetsResults}.
Even the smallest subset, with 78 simulations, yields mismatches over the full set of 591 simulations that are comparable to those of the final fits.
This indicates that the calibration is not driven by the number of simulations used, and that the final model is not overfitted to the catalog.

\subsection{Fitting input values and merger-ringdown coefficients across parameter space}
 
The construction and accuracy goals of \texttt{SEOBNRv5iHM} demand more accurate and more flexible closed-form fits than in previous \texttt{SEOBNR} models.
The intermediate-region modes of \texttt{SEOBNRv5iHM} are determined entirely by its input values, so errors in the fitted input values propagate strongly into the accuracy of the waveform.
In particular, inaccuracies in the input values at the beginning of the intermediate region affect the entire intermediate-region waveform, where the GW amplitude approaches its maximum, as well as affecting the late inspiral slightly through the matching coefficients of Eq.~\eqref{eq:MatchingCoefficients}. 

We therefore build a new fitting pipeline for input values for the \texttt{SEOBNRv5iHM} model, which we also apply to the free merger-ringdown coefficients. 
We use a constrained symbolic regression fitting pipeline, in which the fitted expression is assembled term by term from a library of basis functions, combining a best-subset search~\cite{10.1214/19-STS733} with greedy forward selection~\cite{1993ITSP...41.3397M}.
This replaces the previous fitting method, which used a fixed polynomial ansatz pruned by recursive feature elimination~\cite{Pompili:2023tna}.

The fitted data are the input values and free merger-ringdown coefficients extracted from the 591 SXS NR simulations of Ref.~\cite{Scheel:2025jct} and TML waveforms~\cite{Nishimura:2026nse}; the TML data constrain the $\nu \to 0$ limit, improving extrapolation at more unequal mass ratios.

All fits are functions of the symmetric mass ratio $\nu$ and a small set of
aligned-spin combinations. 
For most quantities associated with even-$m$ modes we use the same variables as the Hamiltonian fits, $\{\nu,\, a_+,\, \delta \cdot a_-\}$.
The odd-$m$ mode amplitudes vanish for equal-mass, equal-spin configurations, and previous works have used spin variables that share this zero to fit these quantities, motivated by PN structure~\cite{Cotesta:2018fcv}; for example, the $(3,3)$ amplitude uses $\chi_{33} = \delta \cdot \chi_S + \chi_A$.
We continue to use these PN-motivated variables, but allow more freedom for these quantities, detailed below.

For most quantities the fit is built in two stages. The first stage determines a non-spinning backbone by best-subset selection over the $\nu$-dependent basis set
\begin{equation}
  \Bigl\{\, 1,\ \nu,\ \nu^2,\ \nu^3,\ \nu^4,\ \nu\log\nu,\ \nu^2\log\nu,\
  \sqrt{1-4\nu},\ \nu\sqrt{1-4\nu},\ \tfrac{1}{1+\nu} \,\Bigr\}.
\end{equation}
The second stage is a constrained symbolic regression, where spin dependence is added by greedy forward selection.
Candidate basis terms are formed as products of backbone functions and monomials of the spin variables.

A candidate model is specified by its active set $S$ of basis functions $B_k$ drawn from this library. 
Given $N$ training points with fit variables $v_i$, target values $y_i$, and uncertainties $\sigma_i$ (i.e. weights $1/\sigma^2_i$), its coefficients are determined by weighted
least squares,
\begin{equation}
  \hat c(S) \;=\; \operatorname*{arg\,min}_{c}\;
  \sum_{i=1}^{N} \left[ \frac{y_i - \sum_{k\in S} c_k B_k(v_i)}{\sigma_i} \right]^{2},
  \label{eq:wls}
\end{equation}
i.e.\ $\hat c(S)$ is the coefficient vector minimizing the weighted residual sum of squares for that active set, and the resulting model prediction is $\hat y(v) = \sum_{k \in S} \hat c_k(S)\, B_k(v)$. NR points enter with $\sigma_i = 1$.
TML points enter at full weight where they are consistent with the NR trend, and are down-weighted ($\sigma_i = 3$--$10^{3}$) where their extraction is in tension with it, so that they guide the $\nu \to 0$ behavior without competing with the NR data.
Candidate models are ranked by their weighted root-mean-square error.

We impose known physical constraints directly into the fit construction. 
To maintain the exchange symmetry mentioned in Sec.~\ref{sec:hamiltoniancalibration}, we exclude any basis terms $B_k$ that change sign under exchange of the two labels.
For many of the input values, it suffices to use variables which respect this symmetry.
For example, every input value used for the $(2,2)$ mode is fitted with $(\nu, a_+, \delta \cdot a_-)$.

However, for other quantities, we find that these variable combinations are too restrictive to capture the behavior accurately.
As such, we assign exchange symmetry ``odd'' or ``even'' tags to our basis variables, so that we can allow, e.g., $a^2_-$ into the fit while excluding $a_-$.  

Another physical constraint is the behavior of the odd-$m$ amplitudes, which must vanish for equal masses and equal spins.
We find that simply using variables which vanish in that limit leads to poor results for equal-mass cases with unequal spins.  
This is because terms that depend on combinations such as $\sqrt{1-4\nu}$ vanish at equal mass, and the remaining terms struggle to fit the equal-mass sector phenomenology well.
We thus first build a spin-only fit $f_{\mathrm{em}}$ to the equal-mass data, combined with a $\delta$-rescaled residual,
\begin{equation}
  f(v) \;=\; f_{\mathrm{em}}(\bm{\chi_\mathrm{em}}) \;+\; \delta\; f_{\mathrm{res}}(\nu, \bm{\chi}),
\end{equation}
where $\bm{\chi_\mathrm{em}}$ denotes the spin variables used for the equal-mass prefit, and $\bm{\chi}$ the full set of spin variables used for the mode (e.g.\ $\chi_{33}$); since $\delta= 0$ at $\nu = 1/4$, the fit reduces exactly to $f_{\mathrm{em}}$ there.
Imposing this fit structure on the amplitudes of the odd-$m$ modes dramatically reduces a tail of high mismatches.

Because the basis library is sufficiently flexible to represent many near-equivalent models, we introduce additional safeguards to prevent overfitting and instability.
The first guard is a coefficient-stability cap. 
Several basis functions are strongly correlated over the physical range $\nu \in (0, 1/4]$, and allowing an unconstrained search could lead to overfitting with near-collinear combinations, where large opposite-sign coefficients cancel inside the training range but diverge immediately outside it. 
A candidate set $S$ is therefore rejected outright if any of its fitted coefficients exceeds a cap proportional to the spread of the training data,
\begin{equation}
  \max_{k}\, |\hat c_k(S)| \;>\; R \,\bigl[\max_i y_i - \min_i y_i\bigr],
  \label{eq:coeffcap}
\end{equation}
with the dimensionless tolerance $R = 10^{3}$ in general, tightened to $R = 10^{2}$ for
the time shifts $\Delta t^{\ell m}_{22}$, whose extrapolation to small $\nu$ is the most delicate.

The second guard controls model complexity through exact leave-one-out (LOO)
cross-validation~\cite{Stone1976CrossValidatoryCA}: the error of the model at each training point when that point is excluded from the fit. For a model that is linear in its
coefficients this error need not be computed by $N$ refits; it is available in closed form~\cite{Allen1974TheRB,Hoaglin1978TheHM}. Writing the weighted
design matrix as $\tilde X_{ik} = B_k(v_i)/\sigma_i$, the hat matrix of the
least-squares problem is
$H = \tilde X (\tilde X^{\mathsf T}\tilde X)^{-1}\tilde X^{\mathsf T}$, and the
residual of the model refitted \emph{without} point $i$, evaluated at point $i$, is
\begin{equation}
  \tilde e_i^{\,(-i)} \;=\; \frac{\tilde e_i}{1 - h_i},
  \qquad
  \tilde e_i = \frac{y_i - \hat y_i}{\sigma_i},
  \qquad
  h_i = H_{ii},
  \label{eq:loo}
\end{equation}
where $\tilde e_i$ is the ordinary weighted training residual, $\hat y_i$ the model
prediction at point $i$, and $h_i \in [0,1]$ the leverage of point $i$: the diagonal
entry of $H$ measuring how strongly that point determines its own prediction, with
$h_i \to 1$ signaling that the model effectively interpolates it. 
Because the identity
is exact, a full $N$-fold cross-validation is computationally cheap, and is evaluated for every candidate model at every selection step. 

Each forward-selection step checks that (a) the training error has improved by a minimum threshold, and (b) the LOO cross-validation error has not degraded by more than a given threshold.
These thresholds are determined empirically for each quantity.
Upon ending the sequence, the model with the smallest cross-validation error is selected. 
We additionally check the smoothness of our fits across the binary parameters.
We check one-dimensional slices through parameter space, and reject any fit that oscillates beyond a given relative threshold.

\section{Evaluating Dominant Mode Accuracy}
\label{sec:Results22}
\begin{figure}[!htbp]
    \centering
    \includegraphics[width=\columnwidth]{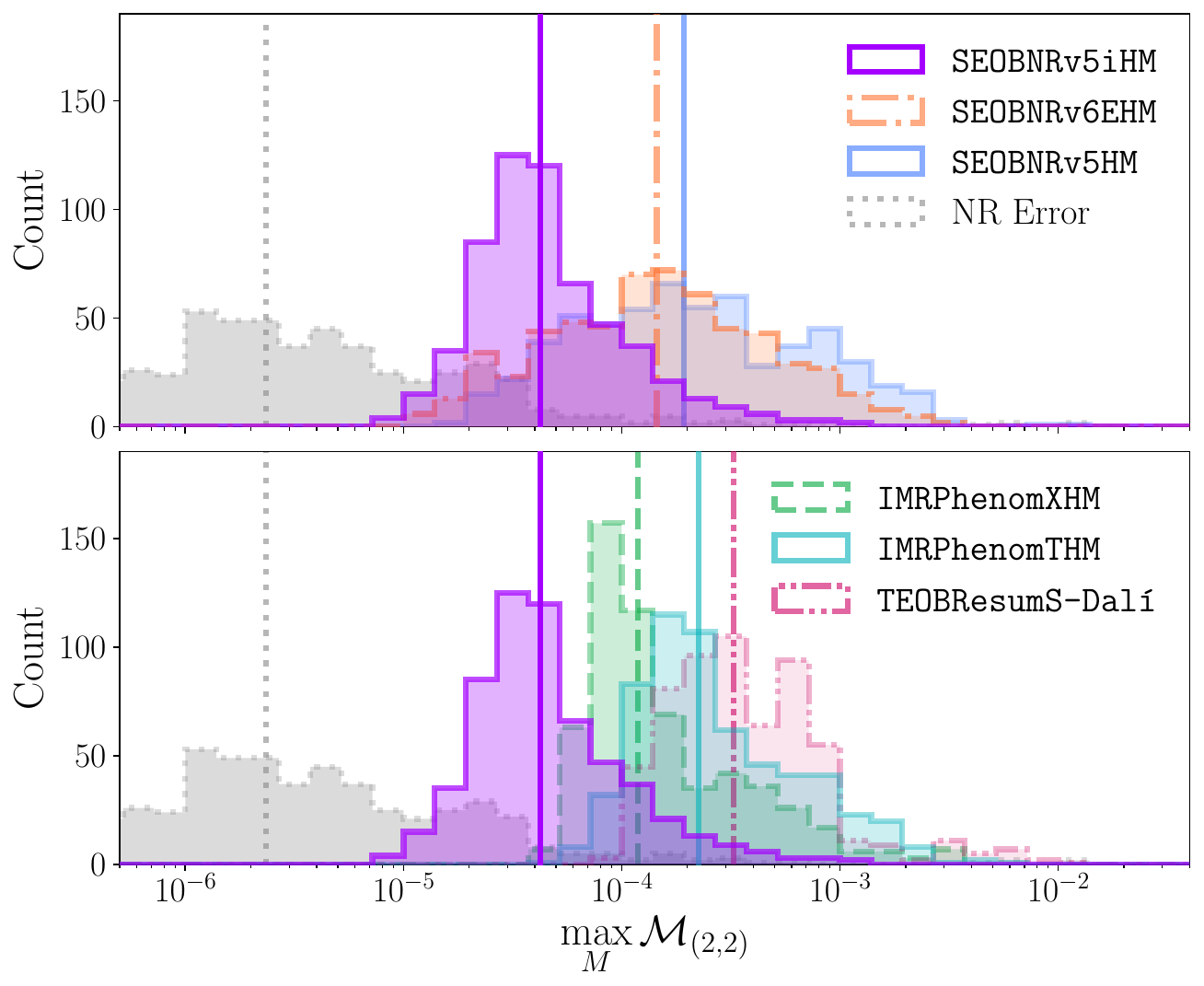}
    \includegraphics[width=\columnwidth]{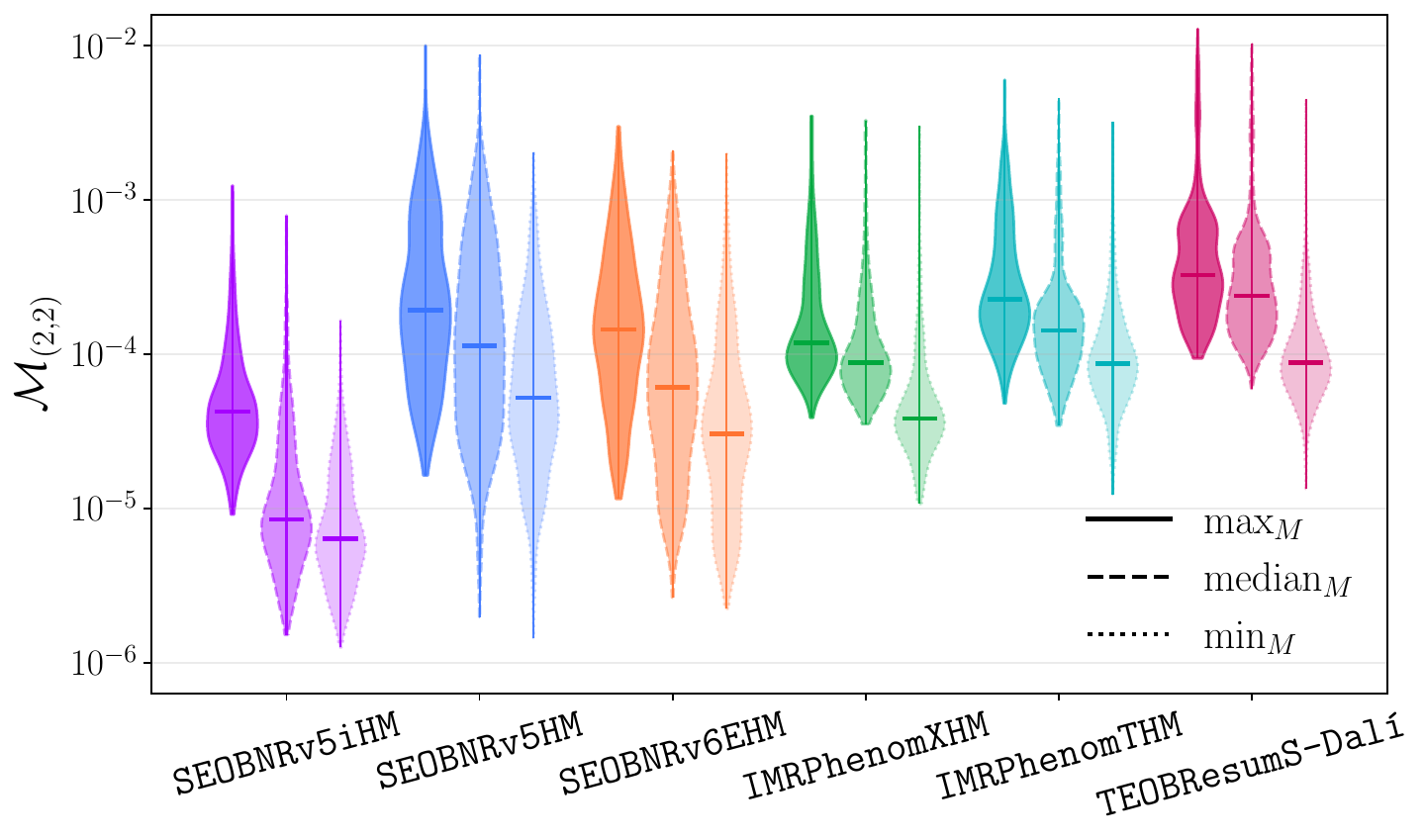}
    \caption{Top: $(2,2)$ mode mismatches versus the SXS QC aligned-spin catalog. 
    The histograms plot the maximum mismatch from $10\,M_\odot$ to $300\,M_\odot$.
    We compare the results against other state-of-the-art waveform models, 
    and find that \texttt{SEOBNRv5iHM} outperforms them.
    Bottom: $(2,2)$ mode maximum, median and minimum mismatches.}
    \label{fig:22modemismatches_models}
\end{figure}

We assess the accuracy of \texttt{SEOBNRv5iHM} against all 591 QC, aligned-spin simulations from the third SXS catalog of waveforms~\cite{Scheel:2025jct}, and compare it with that of other state-of-the-art approximants.
This comparison is performed with unfaithfulness computations, the basic form of which was detailed in Sec.~\ref{sec:Motivation}.
The set of approximants includes the semi-analytical models \texttt{SEOBNRv5HM}~\cite{Pompili:2023tna}, \texttt{IMRPhenomXHM}~\cite{Garcia-Quiros:2020qpx}, and \texttt{IMRPhenomTHM}~\cite{Estelles:2020twz,Estelles:2020osj}, as well as \texttt{SEOBNRv6EHM}~\cite{Gamboa:2026jht} and \texttt{TEOBResumS-Dal\'i} in the QC aligned-spin limit~\cite{Nagar:2024oyk}. 
The \texttt{SEOBNR} models are generated through the package \texttt{pyseobnr}~\cite{Mihaylov:2023bkc}, \texttt{IMRPhenomXHM} through \texttt{LALSimulation}~\cite{lalsuite}, \texttt{IMRPhenomTHM} through \texttt{phenomxpy}~\cite{79kn-53nt}, and \texttt{TEOBResumS-Dal\'i} from the \texttt{teobresums} package\footnote{\texttt{TEOBResumS-Dal\'i} results are computed using the commit for the \texttt{v1.1.2} release of \texttt{Dali}.}. 

We additionally compare against the accuracy of the QC aligned-spin hybridized surrogate \texttt{NRHybSur3dq8}~\cite{Varma:2018mmi}, within its allowed generation range, $q \le 10$. 
Although the generic-spin surrogate \texttt{NRSur7dq4}~\cite{Varma:2019csw} is more widely used in LVK analyses, in the aligned-spin limit \texttt{NRHybSur3dq8} has greater accuracy and greater parameter space coverage. 

\begin{figure*}[!htbp]
    \centering
    \includegraphics[width=1.0\textwidth]{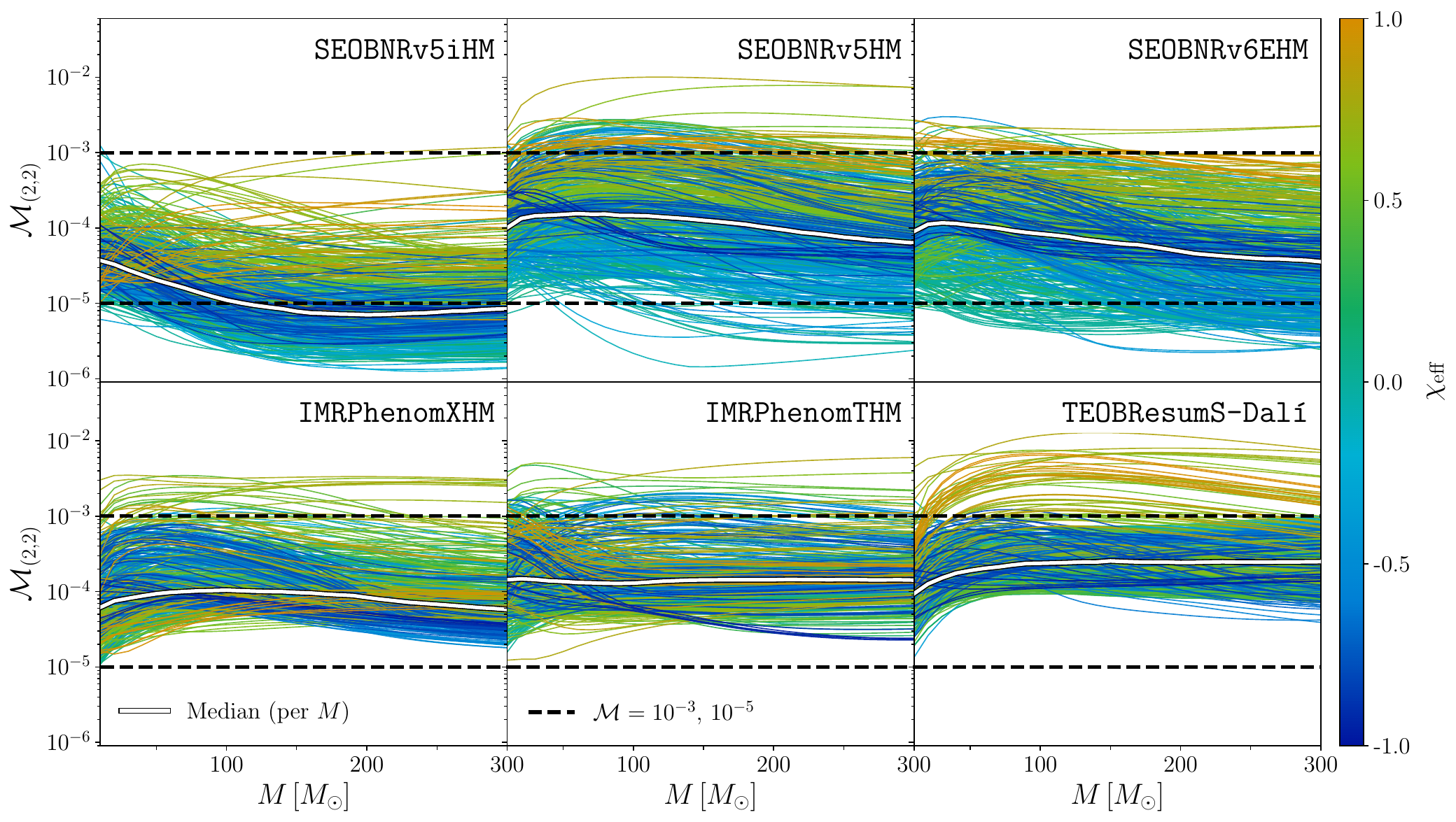}
    \caption{$(2,2)$ mode mismatches plotted against total mass $M$, colored by $\chi_\mathrm{eff}$ value. 
    We see that the \texttt{SEOBNRv5iHM} model performs better for anti-aligned systems.}
    \label{fig:22modemismatches_spaghetti_models}
\end{figure*}

The accuracy of the dominant $(2, 2)$ mode is evaluated by computing the mismatch of Eq.~\eqref{eq:mismatch} against the 591 QC aligned-spin simulations over the total mass range $10M_\odot \le M \le 300M_\odot$. 
We also employ the subsample time optimization described in Sec.~\ref{sec:Motivation}.
Other settings are the same as those described in Sec.~\ref{sec:hamiltoniancalibration}.

\subsection{Comparison against semi-analytical models}
\begin{table*}[!htbp]
    \renewcommand{\arraystretch}{1.3}
	\begin{tabular}{l | c c c c c c}
		\hline
        Approximant & \makecell[cc]{\,\,\texttt{SEOBNRv5iHM}\,\,\,\,} & \makecell[cc]{\texttt{SEOBNRv5HM}\,\,\,\,} & \makecell[cc]{\texttt{SEOBNRv6EHM}\,\,\,\,} & \makecell[cc]{\texttt{IMRPhenomXHM}\,\,\,\,} & \makecell[cc]{\texttt{IMRPhenomTHM}\,\,\,\,} & \makecell[cc]{\texttt{TEOBResumS-Dal\'i}\,\,} \\
		\hline
		$\text{median} \max_{M} \mathcal{M}_{22}$ & $4.25 \times 10^{-5}$ & $1.92 \times 10^{-4}$ & $1.45 \times 10^{-4}$ & $1.20 \times 10^{-4}$ & $2.26 \times 10^{-4}$ & $3.27 \times 10^{-4}$ \\
		$ \% \max_{M} \mathcal{M}_{22} < 10^{-3}$ & $99.7 \%$ & $88.2 \%$ & $94.9 \%$ & $96.3 \%$ & $90.7 \%$ & $91.9 \%$ \\
		$ \% \max_{M} \mathcal{M}_{22} < 10^{-4}$ & $84.1 \%$ & $29.8 \%$ & $36.2 \%$ & $38.4 \%$ & $6.9 \%$ & $0.5 \%$ \\
        $\text{median} \, \text{med}_{M} \mathcal{M}_{22}$ & $8.48 \times 10^{-6}$ & $1.13 \times 10^{-4}$ & $6.07 \times 10^{-5}$ & $8.84 \times 10^{-5}$ & $1.42 \times 10^{-4}$ & $2.39 \times 10^{-4}$ \\
        $\text{median} \min_{M} \mathcal{M}_{22}$ & $6.37 \times 10^{-6}$ & $5.21 \times 10^{-5}$ & $3.05 \times 10^{-5}$ & $3.84 \times 10^{-5}$ & $8.65 \times 10^{-5}$ & $8.78 \times 10^{-5}$ \\
		\hline
	\end{tabular}
	\caption{Summary statistics of the $(2,2)$-mode mismatches. The first three rows show statistics for mismatch maximized over total mass ($10\,M_\odot \le M \le 300\,M_\odot$), between the different waveform models and the $591$ aligned-spin SXS NR simulations. The median NR error, estimated from mismatches between the two highest NR resolutions where available, is $2.35 \times 10^{-6}$. 
    For the fourth and fifth rows, we instead consider the median across mass and the minimum across mass for each simulation, and show the median value across simulations. }
	\label{tab:mm_22}
\end{table*}

We first compute results for the semi-analytical approximants.
In Fig.~\ref{fig:22modemismatches_models}, the top panel shows histograms collecting the maximum mismatch across the mass range for each simulation, for all approximants evaluated.
We also show an estimate of the NR error computed as the mismatch between the two highest resolutions available; 27 simulations with only a single resolution available are omitted from this estimate.

Summary statistics for the semi-analytical models \texttt{SEOBNRv5iHM}, \texttt{SEOBNRv5HM}, \texttt{SEOBNRv6EHM}, \texttt{IMRPhenomXHM}, \texttt{IMRPhenomTHM}, and \texttt{TEOBResumS-Dal\'i} are presented in Table~\ref{tab:mm_22}.
Across the NR simulation set, the \texttt{SEOBNRv5iHM} $(2, 2)$ mode achieves a median maximum mismatch of $4.25 \times 10^{-5}$.
This represents an improvement by a factor of $\sim 4.5$ relative to \texttt{SEOBNRv5HM}, by a factor of $\sim 3$ relative to both \texttt{SEOBNRv6EHM} and \texttt{IMRPhenomXHM}, a factor of $\sim 5$ relative to \texttt{IMRPhenomTHM}, and by a factor of $\sim 7.5$ relative to \texttt{TEOBResumS-Dal\'i}.

The strong median performance of \texttt{SEOBNRv5iHM} is accompanied by a tight distribution of mismatches, with $99.7\%$ of cases having mismatches below $10^{-3}$, and $84.1\%$ below $10^{-4}$; the maximum mismatch value is $1.24 \times 10^{-3}$.
Only $2$ cases exceed $0.1\%$.
Note that the highest case corresponds to \texttt{SXS:BBH:1414}, which is a very long waveform of length $\sim115000M$. The error of NR simulations grows with length~\cite{Mitman:2025tmj}, and the NR error of this case is $7.14 \times 10^{-3}$, which is greater than the mismatch of \texttt{SEOBNRv5iHM}.
By comparison, \texttt{SEOBNRv5HM} has $70$ cases above $0.1\%$ mismatch (the worst being $1.0\%$ at $(q,\chi_1,\chi_2) = (4, 0.9, 0.5)$), \texttt{SEOBNRv6EHM} has $30$ cases above $0.1\%$ mismatch (worst being $0.30\%$ at $(15, -0.5, 0.0)$), \texttt{IMRPhenomXHM} has $22$ (worst $0.35\%$ at $(10, 0.8, 0.5)$), \texttt{IMRPhenomTHM} has $55$ (worst $0.60\%$ at $(4, 0.9, 0.5)$), and \texttt{TEOBResumS-Dal\'i} has $48$, one of which exceeds $1\%$ ($1.28\%$, again at $(4, 0.9, 0.5)$). 

We note that \texttt{SEOBNRv5iHM} performs better than \texttt{SEOBNRv5HM} for 91\% of the 591 NR simulations considered; in the remaining 53 cases \texttt{SEOBNRv5HM} is only marginally better. 

Looking at the mismatch across masses in Fig.~\ref{fig:22modemismatches_spaghetti_models}, we see that the accuracy of \texttt{SEOBNRv5iHM} improves with increasing total mass. 
This makes sense, since at higher masses the late-inspiral and merger-ringdown portions of the waveform dominate the signal in band, which is precisely the region we have improved most. 
Among the \texttt{SEOBNRv5iHM} results, only $8\%$ of configurations have their maximum mismatch at $M \geq 150 \, M_{\odot}$. 
Restricting the maximization to $M \geq 150 \, M_{\odot}$, the median maximum mismatch of \texttt{SEOBNRv5iHM} drops to $9.65 \times 10^{-6}$, a factor of $\sim 4.4$ below the full mass range value, with $96.1\%$ of cases below $10^{-4}$. 
The other models improve more mildly at high mass (medians of $0.65$--$2.95 \times 10^{-4}$), with none achieving median results below $10^{-5}$.

\subsection{Comparison against \texttt{NRHybSur3dq8}}

\begin{figure*}[!htbp]
    \centering
    \includegraphics[width=0.7\textwidth]{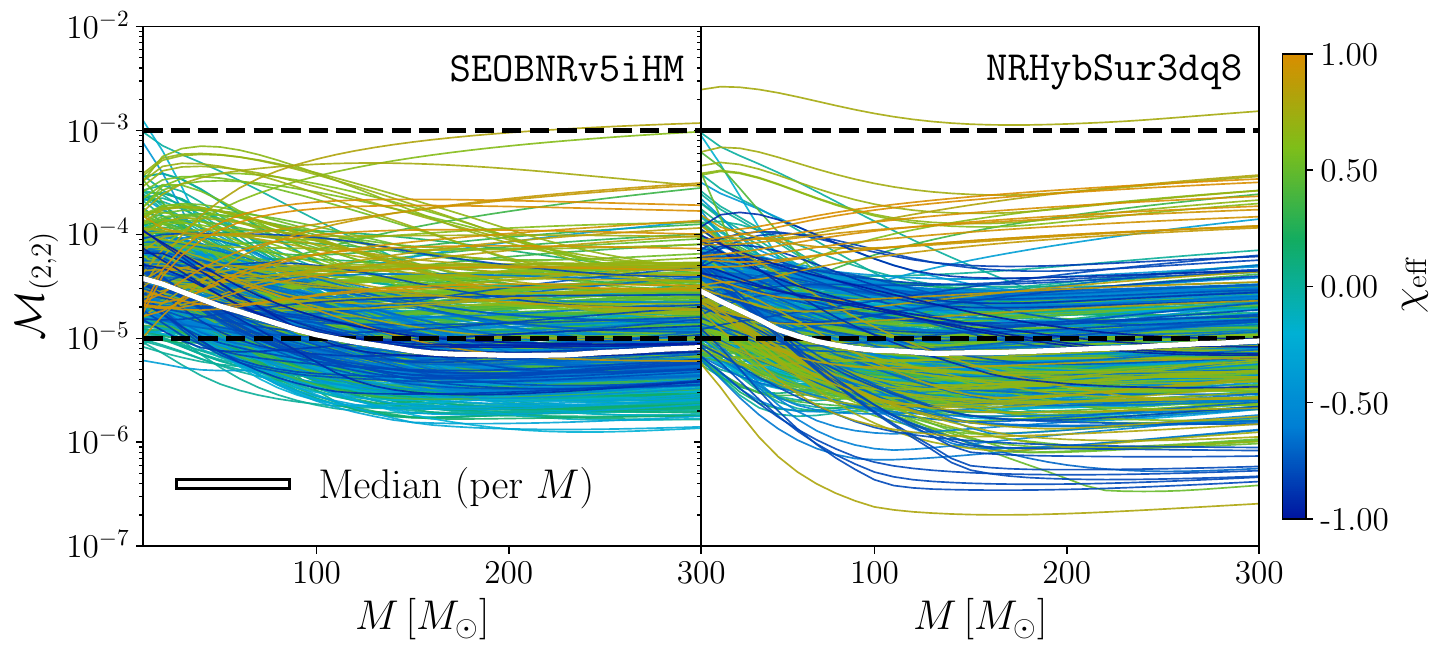}
    \caption{$(2, 2)$ mode mismatches as a function of mass, against all SXS QC aligned-spin simulations with $q \le 10$, the region where \texttt{NRHybSur3dq8} allows waveform generation.
    We see that \texttt{SEOBNRv5iHM} achieves comparable accuracy, especially at high masses. }
    \label{fig:22modemismatches_spaghetti_surrogate}
\end{figure*}

\begin{figure}[!htbp]
    \centering
    \includegraphics[width=\columnwidth]{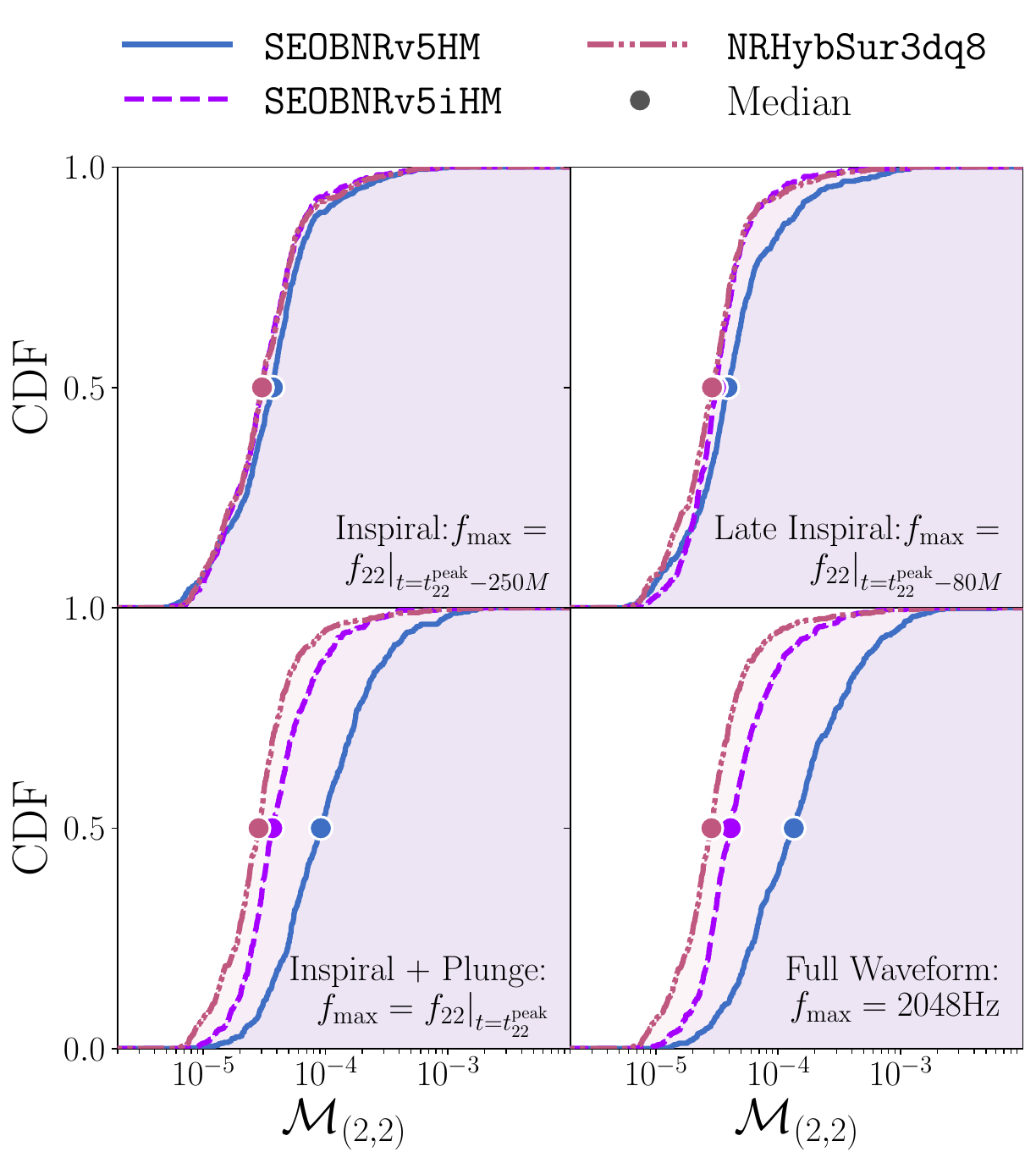}
	\caption{CDFs of the mismatch against NR for \texttt{SEOBNRv5HM} (blue-solid), \texttt{SEOBNRv5iHM} (purple-dashed), and \texttt{NRHybSur3dq8} (red-dash-dotted), computed for the $(2,2)$ mode over the 565 QC, aligned-spin SXS simulations~\cite{Scheel:2025jct} with $q \le 10$. The accuracy lost by \texttt{SEOBNRv5HM} over the final $80M$ before the peak and in the merger--ringdown is largely recovered by \texttt{SEOBNRv5iHM}.}
    \label{fig:22modeDiagnosticsPlot_With_v5_5}
\end{figure}

We also compare the performance of \texttt{SEOBNRv5iHM} with that of the aligned-spin NR surrogate \texttt{NRHybSur3dq8}, using the $565$ simulations that lie within the surrogate's generation range and are common to both comparisons.
For this set, the median maximum $(2,2)$-mode mismatch of \texttt{SEOBNRv5iHM} against NR is $4.27 \times 10^{-5}$, only a factor of $1.5$ larger than the mismatch of the surrogate itself against the same simulations ($2.93 \times 10^{-5}$).

In the high-mass regime the two are almost indistinguishable, as shown in Fig.~\ref{fig:22modemismatches_spaghetti_surrogate}: for $M \geq 150\,M_{\odot}$ the median mismatch of \texttt{SEOBNRv5iHM} is $9.23 \times
10^{-6}$, against $9.71 \times 10^{-6}$ for \texttt{NRHybSur3dq8}, with $99.8\%$ of cases below $10^{-3}$ for both. 
In this regime, the \texttt{SEOBNRv5iHM} waveform is largely determined by phenomenological fits calibrated to NR data. 
The close agreement with the surrogate \texttt{NRHybSur3dq8} therefore provides a strong validation of the accuracy of these fits. 
The model achieves surrogate-level accuracy in the merger-ringdown portion of the $(2,2)$ mode while retaining the advantages of an EOB framework: coverage of the full parameter space and the ability to generate arbitrarily long waveforms. 
By contrast, \texttt{NRHybSur3dq8}, although hybridized, remains limited in duration and in parameter-space coverage.

We also repeat the diagnostic described in Sec.~\ref{sec:Motivation} for \texttt{SEOBNRv5iHM}. 
Recall that these are flat-PSD mismatches, with the inner product truncated at different values of $f_\mathrm{max}$.
The resulting $(2,2)$-mode mismatches are shown in Fig.~\ref{fig:22modeDiagnosticsPlot_With_v5_5}. 
The first panel shows that all models have comparable accuracy in the early inspiral---this is a good check that the \texttt{SEOBNRv5iHM} early inspiral has not degraded.
Interestingly, the second panel shows that the waveform up to $80M$ before the peak of the $(2,2)$ mode improves slightly with respect to \texttt{SEOBNRv5HM}, with median mismatch values of $3.1 \times 10^{-5}$ for \texttt{SEOBNRv5iHM}, $3.9 \times 10^{-5}$ for \texttt{SEOBNRv5HM}, and $2.9 \times 10^{-5}$ for \texttt{NRHybSur3dq8}. 
The improvement could indicate that the revised treatment of the matching coefficients provides a more accurate description of the late inspiral than the NQCs used in \texttt{SEOBNRv5HM}. 
However, this interpretation cannot be established from this comparison alone: \texttt{SEOBNRv5iHM} is calibrated to a substantially larger set of NR simulations, so the improvement may instead, or in part, result from the additional calibration information.

The most indicative results are those in the third and fourth panels.
In the third, we see that the large loss of accuracy in the final $80M$ before the peak seen for \texttt{SEOBNRv5HM} is largely removed in \texttt{SEOBNRv5iHM}. 
The median value rises only to around $3.7 \times 10^{-5}$, as opposed to the $9.1 \times 10^{-5}$ value for \texttt{SEOBNRv5HM}. 
The improvement persists once the merger-ringdown is included, with the median rising only to $4.1 \times 10^{-5}$, compared with $1.3 \times 10^{-4}$ for \texttt{SEOBNRv5HM}.
These results together with those shown in Fig.~\ref{fig:22modemismatches_spaghetti_models} indicate that the late inspiral and merger-ringdown regions no longer dominate the error of the model.

\section{Evaluating Polarization and Higher Mode Accuracy}
\label{sec:ResultsPolHM}

To quantify the accuracy of the higher-order mode content, we compute sky-and-polarization-averaged, SNR-weighted mismatches for the full polarizations.

The strain measured by a detector for the signal $h(t) \equiv h_+(t) - ih_\times (t)$ of Eq.~\eqref{eq:MultiModeDecomposition} is
\begin{align}
    \label{eq:det_strain}
    h_\mathrm{det}(t) \equiv & F_+(\theta,\phi,\psi) \ h_+(t;\iota,\varphi_0, d_L, \bm{\lambda},t_{\mathrm{c}}) \nonumber \\
    &+ F_\times(\theta,\phi,\psi)\ h_\times(t;\iota,\varphi_0, d_L, \bm{\lambda},t_\mathrm{c}).
\end{align}
Here, $F_{+,\times}$ are antenna pattern functions~\cite{Sathyaprakash:1991mt,Finn:1992xs}, $(\theta,\phi)$ define the sky location, $t_\mathrm{c}$ is the time of arrival, $\psi$ the polarization angle, and $(d_L, \iota, \varphi_0)$ are as described in Eq.~\eqref{eq:MultiModeDecomposition}. 

By defining an effective polarization angle $\kappa(\theta,\phi,\psi)$, we can write the strain of Eq.~\eqref{eq:det_strain} as
\begin{align}
\label{eq:strainKappa}
h_\mathrm{det}(t) =\mathcal{A}(\theta, \phi) (h_+ \cos \kappa(\theta,\phi,\psi) + h_\times \sin \kappa(\theta,\phi,\psi)).
\end{align}
The amplitude factor $\mathcal{A}(\theta,\phi)$ is defined as in Refs.~\cite{Cotesta:2018fcv, Ossokine:2020kjp}.

To assess the accuracy of a template waveform $h_t$ relative to a signal waveform $h_s$, we use the faithfulness function $\mathcal{F} = \mathcal{F}(M_{\textrm{s}},\iota_{\textrm{s}},{\varphi_0}_{\textrm{s}},\kappa_{\textrm{s}})$~\cite{Cotesta:2018fcv,Ossokine:2020kjp,Garcia-Quiros:2020qpx},
\begin{equation}
\mathcal{F} =  \max_{t_c, {\varphi_0}_{t}, \kappa_{t}} \left[\left . \frac{ ( h_s|h_t )}{\sqrt{  ( h_s|h_s )  ( h_t|h_t )}}\right \vert_{\substack{\iota_{\mathrm{s}} = \iota_{t} \\\boldsymbol{\lambda}_\mathrm{s}(t_{\mathrm{s}} = t_{0_\mathrm{s}}) = \boldsymbol{\lambda}_{t}(t_t = t_{0_\mathrm{t}})}} \right].
\end{equation}
This is similar to Eq.~\eqref{eq:mismatch}, but includes an optimization over the effective polarization angle $\kappa_t$ of the template waveform. 
As in Eq.~\eqref{eq:mismatch}, $\bm{\lambda}$ refers to the intrinsic parameters of the signal. 
The template inclination is fixed to the signal inclination, $\iota_t = \iota_s$. 

In order to have a metric that evaluates the accuracy of the waveform, we want to reduce the dimensionality of the extrinsic parameters.
We therefore define the sky-and-polarization-averaged, SNR-weighted faithfulness~\cite{Babak:2016tgq,Ossokine:2020kjp}:
\vspace{4mm}
\begin{equation}
	\overline{\mathcal{F}}_{\mathrm{SNR}}\left(M_{\mathrm{s}}, \iota_{\mathrm{s}}\right) \equiv \sqrt[3]{\frac{\int_0^{2 \pi} d \kappa_{\mathrm{s}} \int_0^{2 \pi} d \varphi_{0 \mathrm{~s}} \mathcal{F}^3 \mathrm{SNR}^3}{\int_0^{2 \pi} d \kappa_{\mathrm{s}} \int_0^{2 \pi} d \varphi_{0 \mathrm{~s}} \operatorname{SNR}^3}}.
\end{equation}

The sky-and-polarization-averaged, SNR-weighted unfaithfulness (or mismatch) is then
\begin{equation}
	\overline{\mathcal{M}}_{\mathrm{SNR}}=1-\overline{\mathcal{F}}_{\mathrm{SNR}}.
\end{equation}

\begin{table*}[htb!]
    \renewcommand{\arraystretch}{1.3}
	\begin{tabular}{l | c c c c c c}
		\hline
        Approximant & \makecell[cc]{\,\,\texttt{SEOBNRv5iHM}\,\,\,\,} & \makecell[cc]{\texttt{SEOBNRv5HM}\,\,\,\,} & \makecell[cc]{\texttt{SEOBNRv6EHM}\,\,\,\,} & \makecell[cc]{\texttt{IMRPhenomXHM}\,\,\,\,} & \makecell[cc]{\texttt{IMRPhenomTHM}\,\,\,\,} & \makecell[cc]{\texttt{TEOBResumS-Dal\'i}\,\,} \\
        \hline
        $\text{median} \max_{M} \overline{\mathcal{M}}_{\mathrm{SNR}}$ & $4.18 \times 10^{-4}$ & $1.10 \times 10^{-3}$ & $2.41 \times 10^{-3}$ & $3.13 \times 10^{-3}$ & $8.13 \times 10^{-4}$ & $4.55 \times 10^{-3}$ \\
        $ \% \max_{M} \overline{\mathcal{M}}_{\mathrm{SNR}} < 10^{-2}$ & $100 \%$ & $95.9 \%$ & $83.1 \%$ & $77.5 \%$ & $98.0 \%$ & $74.1 \%$ \\
        $ \% \max_{M} \overline{\mathcal{M}}_{\mathrm{SNR}} < 10^{-3}$ & $80.4 \%$ & $46.7 \%$ & $25.9 \%$ & $21.8 \%$ & $56.7 \%$ & $9.0 \%$ \\
		$\text{median} \, \text{med}_{M} \overline{\mathcal{M}}_{\mathrm{SNR}}$ & $3.14 \times 10^{-4}$ & $8.04 \times 10^{-4}$ & $1.40 \times 10^{-3}$ & $1.79 \times 10^{-3}$ & $5.58 \times 10^{-4}$ & $2.95 \times 10^{-3}$ \\
 		$\text{median} \min_{M} \overline{\mathcal{M}}_{\mathrm{SNR}}$ & $1.09 \times 10^{-4}$ & $3.09 \times 10^{-4}$ & $3.01 \times 10^{-4}$ & $8.31 \times 10^{-4}$ & $2.59 \times 10^{-4}$ & $6.18 \times 10^{-4}$ \\
		\hline
	\end{tabular}
	\caption{Summary statistics of the sky-and-polarization-averaged,
	SNR-weighted mismatches of the full polarizations at inclination $\iota =
	\pi/3$, maximized over total mass ($20\,M_\odot \le M \le 300\,M_\odot$),
	against the $591$ aligned-spin SXS NR simulations: median maximum
	mismatch, and percentage of cases below $10^{-2}$ and $10^{-3}$. 
    As in Table~\ref{tab:mm_22}, the last two rows give the median across simulations of the median and of the minimum mismatch over total mass.
    The median NR resolution error with the same averaging is $3.50 \times
	10^{-5}$, and the median error from restricting NR to the model mode
	content is $1.16 \times 10^{-4}$.}
	\label{tab:mm_hm}
\end{table*}

We use inclination $\iota = \pi/3$, and a range of total masses $20\, M_{\odot} \leq M \leq 300 \, M_{\odot}$.

\subsection{Comparison against semi-analytical models}
\begin{figure}[t]
    \centering
    \includegraphics[width=\columnwidth]{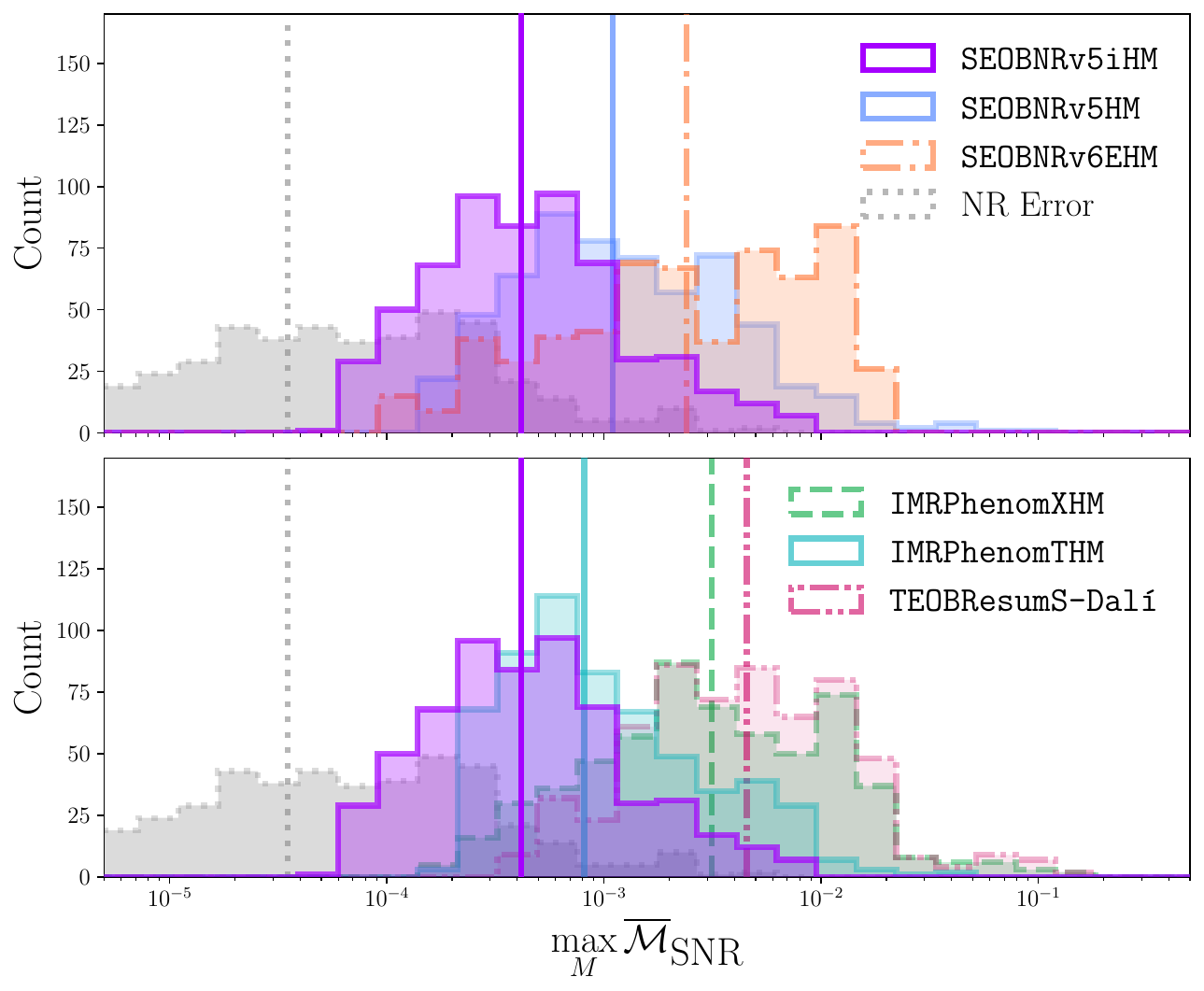}
    \includegraphics[width=\columnwidth]{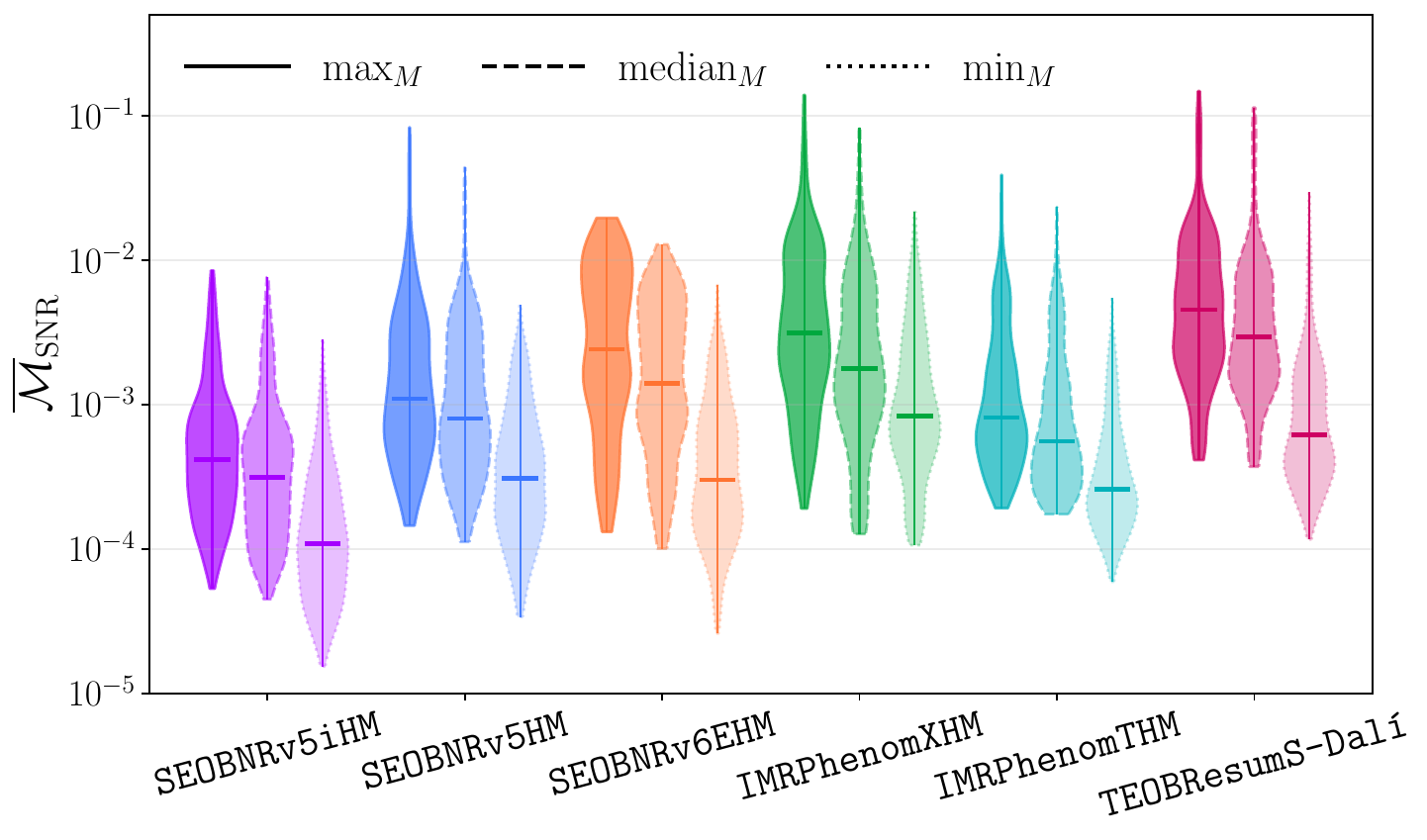}
    \caption{Top: Sky-and-polarization-averaged, SNR-weighted mismatches versus the SXS QC aligned-spin catalog. 
    The histograms plot the maximum mismatch from $20 M_\odot$ to $300M_\odot$.
    We find that \texttt{SEOBNRv5iHM} has the most accurate results.
    Bottom: Maximum, median and minimum mismatches.}
    \label{fig:polarizationmismatches_hist_models}
\end{figure}
For each approximant, we build the polarization waveform using all modes available. 
Specifically, these are: $(\ell,|m|) = (2, 2), (2,1), (3, 3), (3,2), (4, 4), (4, 3), (5, 5)$ for \texttt{SEOBNRv5iHM} and \texttt{SEOBNRv5HM}; $(\ell,|m|) = (2, 2), (2,1), (3, 3), (3,2), (4, 4), (4, 3)$ for \texttt{SEOBNRv6EHM}; $(\ell,|m|) = (2, 2), (2,1), (3, 3), (3,2), (4, 4)$ for \texttt{IMRPhenomXHM}; $(\ell,|m|) = (2, 2), (2,1), (3, 3), (4, 4), (5, 5)$ for \texttt{IMRPhenomTHM}; and $(\ell,|m|) = (2, 2), (2,1), (3, 3), (4, 4)$ for \texttt{TEOBResumS-Dal\'i}. Note that for \texttt{TEOBResumS-Dal\'i}, additional modes can be requested, but we only use the recommended modes.
For building the NR polarizations, we include all modes up to $\ell_\mathrm{max} = 5$.

\begin{figure*}[t]
    \centering
    \includegraphics[width=1.0\textwidth]{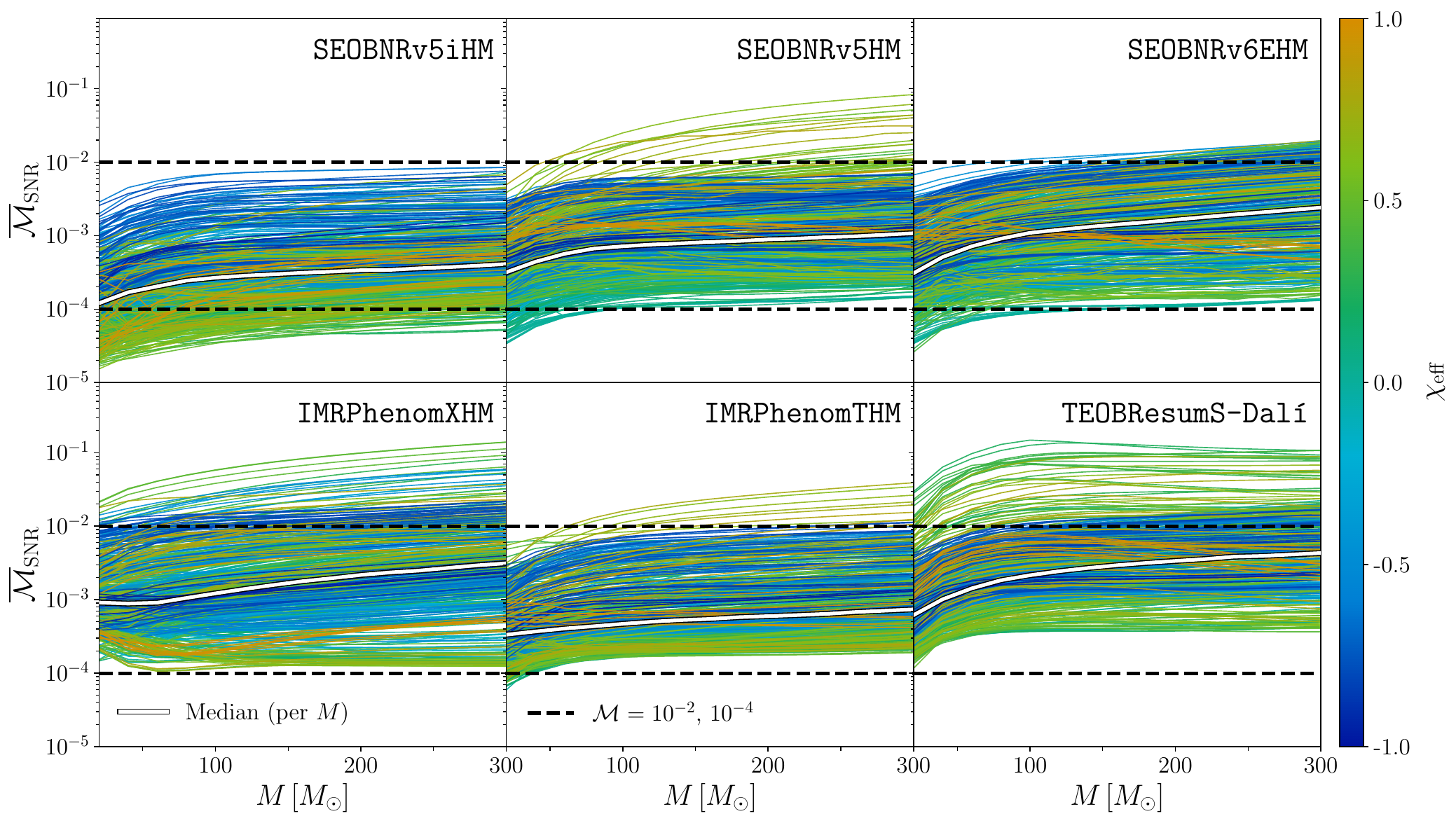}
    \caption{Sky-and-polarization-averaged, SNR-weighted mismatches plotted against total mass $M$, colored by $\chi_\mathrm{eff}$ value. 
    Here, the parameter-space trend of \texttt{SEOBNRv5iHM} is reversed, and the model now performs worse for anti-aligned systems.}
    \label{fig:polarizationmismatches_spaghetti_models}
\end{figure*}

The results are summarized in Table~\ref{tab:mm_hm}, and plotted in Fig.~\ref{fig:polarizationmismatches_hist_models}.
The dependence on total mass and $\chi_\mathrm{eff}$ is shown in Fig.~\ref{fig:polarizationmismatches_spaghetti_models}.
Overall, \texttt{SEOBNRv5iHM} provides the most accurate reconstruction of the full polarization waveform, achieving a median maximum mismatch of $4.18 \times 10^{-4}$.
This is a factor of $2.6$ below \texttt{SEOBNRv5HM} ($1.10 \times 10^{-3}$), a factor of $1.9$ below \texttt{IMRPhenomTHM} ($8.13 \times 10^{-4}$), and a factor $6$--$11$ below \texttt{TEOBResumS-Dal\'i} ($4.55 \times 10^{-3}$), \texttt{SEOBNRv6EHM} ($2.41\times 10^{-3}$) and \texttt{IMRPhenomXHM} ($3.13 \times 10^{-3}$). 

\texttt{SEOBNRv5iHM} is the only model with no case above $1\%$ mismatch: its worst case is $0.85\%$, for \texttt{SXS:BBH:2642} with $(q, \chi_1, \chi_2) = (8, -0.8, 0.8)$.
This is a challenging case, due to the large mass ratio and almost completely opposite spins.
\texttt{SEOBNRv5HM} has $24$ cases above $1\%$, reaching $8.3\%$ at $(8, 0.8,-0.8)$; \texttt{SEOBNRv6EHM} has $100$ cases above $1\%$, though none above $2\%$, mostly at high mass ratios $q \gtrsim 6$; \texttt{IMRPhenomTHM} has $12$ cases above $1\%$, with a maximum of $3.9\%$ for $(10, 0.8, 0.5)$; \texttt{IMRPhenomXHM} has $133$ cases above $1\%$ and $3$ above $10\%$, reaching $14\%$ for $(15, 0.5, 0.0)$; and \texttt{TEOBResumS-Dal\'i} has $153$ cases above $1\%$ and $5$ above $10\%$, with a highest value of $14.9\%$ for $(6, 0.4, -0.8)$. 

We note that the mode content plays a role in these results, as unmodeled modes can contribute significantly to the polarization mismatches. 
In Appendix~\ref{App:ModebyModeEllMax4}, we show results of the polarization mismatches with NR waveforms built using modes up to $\ell_\mathrm{max} = 4$, and omitting the $(5, 5)$ mode from \texttt{SEOBNRv5iHM}, \texttt{SEOBNRv5HM} and \texttt{IMRPhenomTHM}.
These $\ell_\mathrm{max} = 4$ results are comparable to those in Ref.~\cite{Gamboa:2026jht}, which also helps explain why we observe a degradation in the mismatch of \texttt{SEOBNRv6EHM} relative to \texttt{SEOBNRv5HM}, whereas Ref.~\cite{Gamboa:2026jht} shows the opposite.
The same Appendix shows that the models move closer together when removing the $(5,5)$ mode from \texttt{SEOBNRv5iHM}, \texttt{SEOBNRv5HM} and \texttt{IMRPhenomTHM}, further indicating the importance of this mode.

To isolate improvements in the modeling of individual modes, the most direct comparison is between \texttt{SEOBNRv5iHM} and \texttt{SEOBNRv5HM}, since these models share the same mode content.
When the mode content is matched, the improvement can be attributed to improved mode accuracy rather than differences in the included modes. 
We also include in Appendix~\ref{App:ModebyModeEllMax4} the mode-by-mode mismatches for the semi-analytical models. Among the semi-analytical models considered, \texttt{SEOBNRv5iHM} yields the lowest maximum mismatch over total mass for every mode it models, except for the $(3,2)$ mode, where \texttt{IMRPhenomXHM} performs better.

\subsection{Comparison against \texttt{NRHybSur3dq8} and mode-content limitations}

\begin{figure}[!tbp]
    \centering
    \includegraphics[width=\columnwidth]{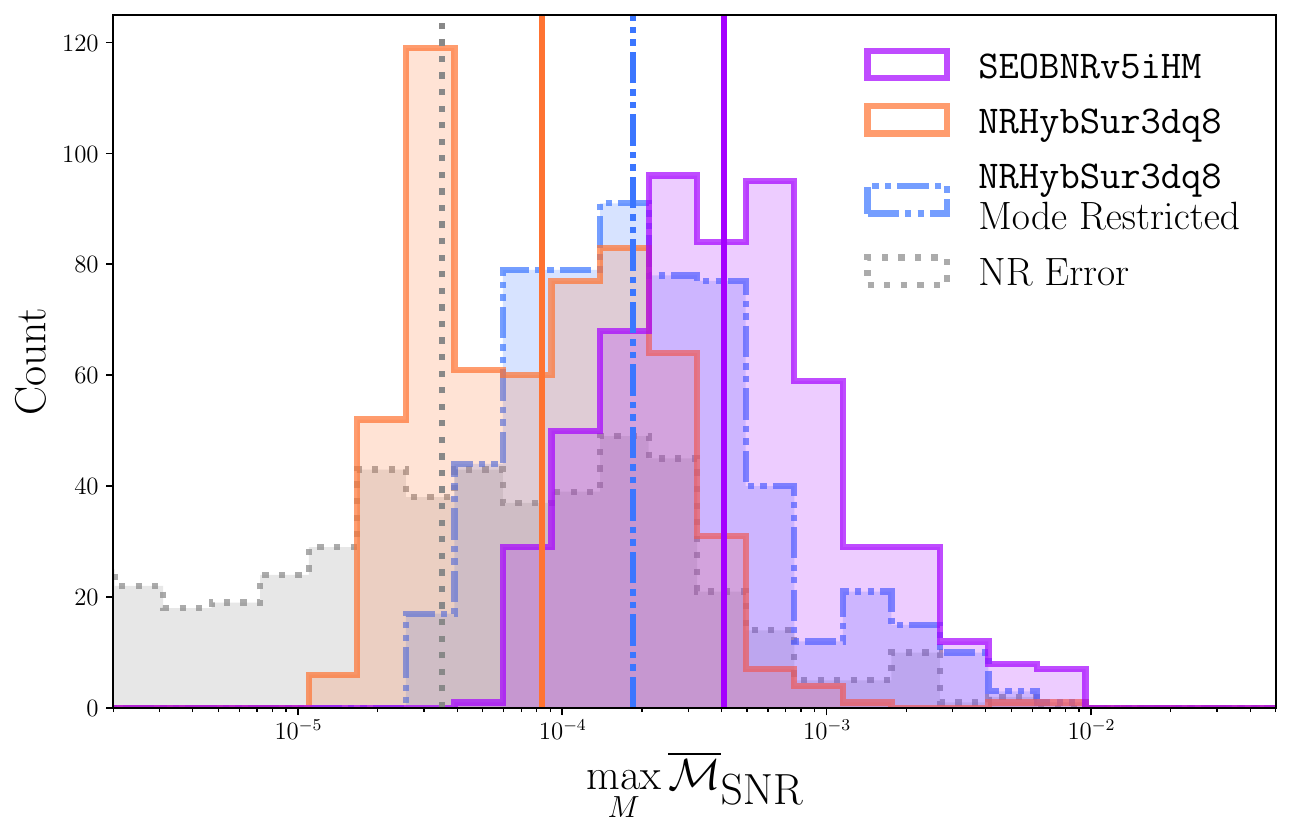}
    \caption{Sky-and-polarization-averaged, SNR-weighted mismatches versus all $q \le 10$ simulations. 
    The histograms plot the maximum mismatch from $20 M_\odot$ to $300M_\odot$, for \texttt{SEOBNRv5iHM}, for \texttt{NRHybSur3dq8} with all of its modes, and for \texttt{NRHybSur3dq8} restricted to the mode content of \texttt{SEOBNRv5iHM}. Vertical lines mark the medians, and the gray histogram shows the NR error.}
    \label{fig:polarizationmismatches_moderestricted_surrogate}
\end{figure}

\begin{figure*}[!tbp]
    \centering
    \includegraphics[width=\textwidth]{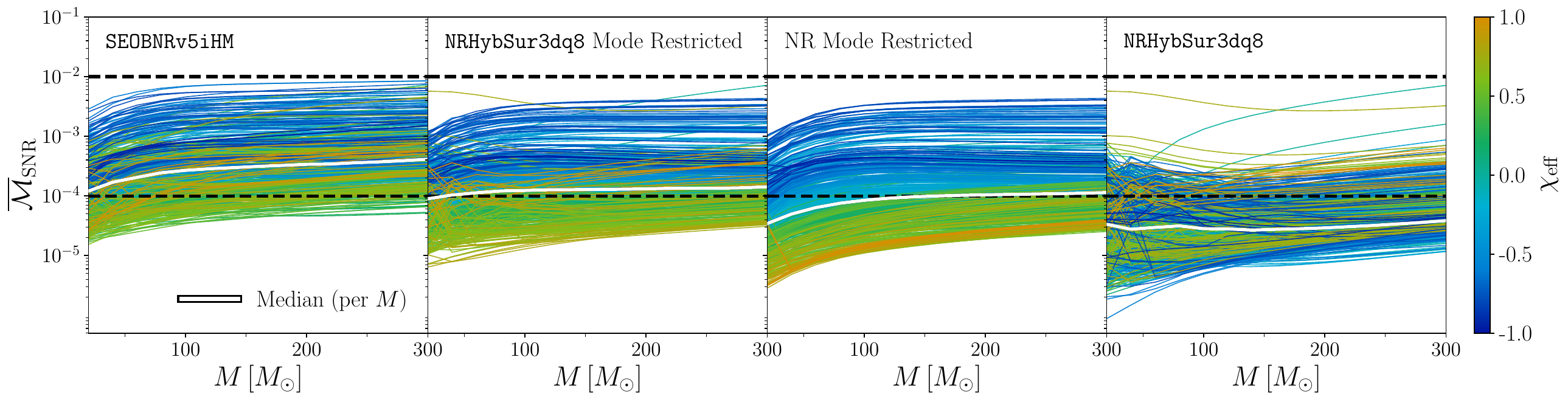}
    \caption{Sky-and-polarization-averaged, SNR-weighted mismatches plotted against total mass $M$, colored by $\chi_\mathrm{eff}$ value. From left to right: \texttt{SEOBNRv5iHM}; \texttt{NRHybSur3dq8} restricted to the mode content of \texttt{SEOBNRv5iHM}; NR restricted to the same mode content; and \texttt{NRHybSur3dq8} with all of its modes. The white line is the median at each total mass. The mode-restricted NR waveforms show that the error from the missing modes alone is largest for anti-aligned spins.}
    \label{fig:polarizationmismatches_spaghetti_moderestricted_surrogate}
\end{figure*}

\begin{figure}[!tbp]
    \centering
    \includegraphics[width=\columnwidth]{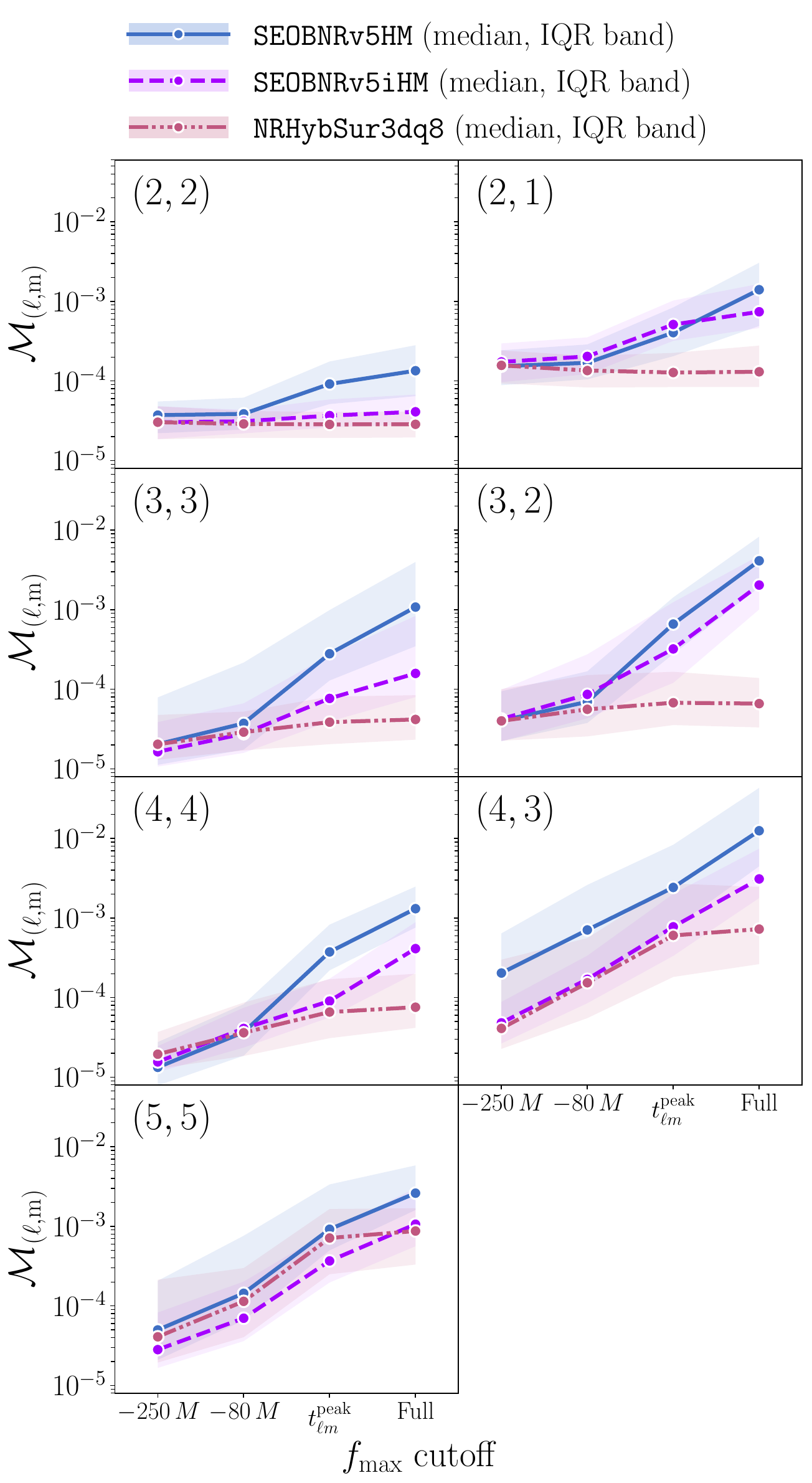}
    \caption{Mismatch against NR for each mode as a function of the upper cutoff $f_\mathrm{max}$ of the inner product in Eq.~\eqref{eq:overlap}, as in Fig.~\ref{fig:DiagnosingErrorModes}, now including \texttt{SEOBNRv5iHM} (purple, dashed).
    Lines give the median over the QC, aligned-spin SXS simulations with $q \le 10$, and shaded bands the IQR.
    For the $(2,2)$ mode, \texttt{SEOBNRv5iHM} stays nearly flat across the four cutoffs, as \texttt{NRHybSur3dq8} does.
    For the higher modes it matches the surrogate through the inspiral and improves on \texttt{SEOBNRv5HM} at the last two cutoffs (except for the $(2,1)$ mode at the third cutoff), but it still loses accuracy relative to the surrogate once the intermediate region (third cutoff) and the merger--ringdown (fourth cutoff) are included. The $(5,5)$ mode is an exception, remaining comparable or better than the surrogate.}
    \label{fig:AllModesDiagnosticsPlot}
\end{figure}

To further assess the accuracy of our model, and to understand limitations of its polarization waveform accuracy, we compare against \texttt{NRHybSur3dq8} and perform mode-restriction studies. 
The polarization waveform for \texttt{NRHybSur3dq8} is built with all modes available in the surrogate, namely all $\ell \le 4$ modes together with the $(5,5)$ mode, excluding the $(4,1)$ and $(4,0)$ modes~\cite{Varma:2018mmi}.

Fig.~\ref{fig:polarizationmismatches_moderestricted_surrogate} compares the polarization mismatches of \texttt{SEOBNRv5iHM} and \texttt{NRHybSur3dq8}. 
We restrict this comparison to simulations with $q \le 10$, corresponding to the generation range of the surrogate.
The median maximum mismatch across masses is $4.1 \times 10^{-4}$ for \texttt{SEOBNRv5iHM}, compared to $8.4 \times 10^{-5}$ for \texttt{NRHybSur3dq8}. 

To isolate the impact of missing mode content, we evaluate the error incurred by restricting the NR waveforms to have the same mode content as our model.
In our results, ``NR Mode Restricted'' refers to a polarization built directly from the highest-resolution NR simulations, but including only the $(2, 2), (2, 1), (3, 3), (3, 2), (4, 4), (4, 3), (5, 5)$ modes. 
This construction quantifies the accuracy limit imposed solely by the current model content, i.e., the minimum mismatch achievable if every included mode were modeled perfectly.

We additionally consider ``\texttt{NRHybSur3dq8} Mode Restricted'', which is constructed from the surrogate using exactly the same mode content as \texttt{SEOBNRv5iHM}. 
This comparison allows us to assess the accuracy of our higher modes versus that of \texttt{NRHybSur3dq8}, by comparing the polarization mismatch results using the same set of modes. 

We can see the parameter-space dependence of these mode-content effects in Fig.~\ref{fig:polarizationmismatches_spaghetti_moderestricted_surrogate}. 
For \texttt{SEOBNRv5iHM} and ``NR Mode Restricted'', all 591 simulations are shown, while for the \texttt{NRHybSur3dq8} results, only the $q \le 10$ subset is included. 

The ``NR Mode Restricted'' construction yields a median maximum mismatch of $1.2 \times 10^{-4}$, representing the floor set by the missing modes.
A clear parameter-space trend is visible: the impact of missing modes is substantially larger for anti-aligned than for aligned-spin systems. 
These results suggest that the missing mode content is a dominant contributor to the degradation of the polarization mismatches of \texttt{SEOBNRv5iHM} observed for anti-aligned spin systems. 

For ``\texttt{NRHybSur3dq8} Mode Restricted'', the median maximum mismatch is $1.8 \times 10^{-4}$, a factor of $2.2$ below \texttt{SEOBNRv5iHM}. 
Despite using the same mode content as \texttt{SEOBNRv5iHM}, the surrogate remains more accurate. 

This is consistent with Fig.~\ref{fig:AllModesDiagnosticsPlot}, which extends the diagnostics of Sec.~\ref{sec:Motivation} to include \texttt{SEOBNRv5iHM}. 
We see that our higher modes still degrade in accuracy compared to the surrogate.
Up to the second cutoff, i.e.\ for the inspiral up to $t^{\mathrm{peak}}_{22} - 80M$, the higher modes of \texttt{SEOBNRv5iHM} are comparable in accuracy to those of the surrogate, with only the $(2,1)$ and $(3,2)$ modes sitting slightly above it.
In particular, the loss of accuracy of the \texttt{SEOBNRv5HM} $(4,3)$ mode in the early to mid-inspiral, noted in Sec.~\ref{sec:Motivation}, is no longer present.
This seems to indicate that the matching coefficient treatment is a better way of modeling the $(4,3)$ inspiral compared to the use of NQCs. 

At the third cutoff, including the intermediate region, we now see a difference in behavior as compared to the $(2,2)$ mode. 
The median mismatch of every higher mode grows by a factor of roughly $2$--$5$ with respect to the second cutoff.
For the $(3,3)$, $(3,2)$, $(4,4)$, $(4,3)$ and $(5,5)$ modes this is still a clear improvement over \texttt{SEOBNRv5HM}, by factors of $2$--$4$.
However, in this region we see a deviation in the accuracy as compared to \texttt{NRHybSur3dq8}, with the exception of the $(5,5)$ mode, for which \texttt{SEOBNRv5iHM} remains comparable to or more accurate than the surrogate.
Including the merger--ringdown, at the fourth cutoff, widens the gap further.
The \texttt{NRHybSur3dq8} medians are essentially unchanged between the third and the fourth cutoff for all modes, while those of \texttt{SEOBNRv5iHM} grow by a further factor of $1.5$--$6$.
Unlike for the $(2,2)$ mode, then, the intermediate and merger--ringdown regions remain the dominant source of error for the higher modes.

These results suggest two clear directions for future improvements.
First, extending the model to include additional modes would directly reduce the mode-content error that currently limits the waveform accuracy, especially for anti-aligned spin systems. 
Second, the accuracy of the currently modeled higher modes can be further improved. 
For the higher modes, this likely requires improved ansätze for both the intermediate and merger-ringdown regions, tailored more towards the behavior of each subdominant mode.

\section{Conclusion}
\label{sec:Conclusion}
In this paper, we have presented \texttt{SEOBNRv5iHM}, a multipolar IMR waveform model in the EOB framework for quasi-circular, aligned-spin BBHs.
We show that this new model achieves a high level of accuracy against the third SXS catalog of NR waveforms~\cite{Scheel:2025jct}. 
The model builds upon the highly accurate analytical results of \texttt{SEOBNRv5HM}, retaining the Hamiltonian and radiation-reaction force, and using the recently corrected second-order self-force calibration of the fluxes~\cite{Leather:2025nhu}. 
To take advantage of this baseline, we have modified the construction of the waveform modes. 
We removed the NQC corrections, included a phenomenological intermediate region, and rebuilt the merger-ringdown modes, attaching at the peak of each mode, except for the $(3,2)$ and $(4,3)$ modes, which are attached at the peak of the $(2,2)$ mode.

The calibration parameters $\{a_6, d_\mathrm{SO}\}$ in the Hamiltonian, as well as $\Delta t^{22}_\mathrm{ISCO}$, have been recalibrated to optimize the new model construction using waveforms from the third SXS catalog~\cite{Scheel:2025jct}.
Nested calibration subsets of NR waveforms containing as few as 78 simulations achieve similar accuracy results, demonstrating the robustness of the calibration procedure and the performance of the mode construction choices. 

We have additionally developed a new fitting pipeline to fit the input values and the merger-ringdown free coefficients across parameter space. 
This allows us to achieve the required fitting accuracy for the model, while incorporating additional input values that determine the intermediate region waveform. 

When compared against the 591 NR simulations, the median of the maximum $(2,2)$-mode mismatch of \texttt{SEOBNRv5iHM} across the mass range considered is $4.25 \times 10^{-5}$.
This is a factor of 4.5 smaller than for \texttt{SEOBNRv5HM}, a factor of $\sim 3$ smaller than for \texttt{SEOBNRv6EHM} and \texttt{IMRPhenomXHM}, a factor of $\sim 5$ smaller than for \texttt{IMRPhenomTHM},  and a factor of $\sim 7.5$ smaller than for \texttt{TEOBResumS-Dal\'i}.
We find that the $(2,2)$-mode mismatch is comparable to that of the hybridized NR surrogate \texttt{NRHybSur3dq8}, achieving equivalent results at higher masses where the last part of the signal dominates.

For the full gravitational-wave polarizations, \texttt{SEOBNRv5iHM} achieves the highest accuracy among all semi-analytical models considered here. 
The individual modes are now accurate enough that the mode content is becoming an important limitation on the accuracy of the polarizations, particularly for anti-aligned spin systems.
The higher modes are, however, not yet at the level of the $(2,2)$ mode: a mode-by-mode diagnostic shows that they match the accuracy of \texttt{NRHybSur3dq8} through the inspiral, but still lose accuracy relative to it in the intermediate and merger--ringdown regions.

Our results identify several directions for future work. 
In particular, extending the mode content is likely to provide a significant improvement in accuracy of the full polarizations, especially for systems with spins anti-aligned to the orbital angular momentum. 
A more detailed study of the contribution of individual subdominant modes to the polarization mismatches would also be useful in identifying the modes that should be prioritized in future model development. 

A natural next step is to extend the model to include spin precession, following the approach in which \texttt{SEOBNRv5PHM} builds upon \texttt{SEOBNRv5HM}. 
The improved accuracy of \texttt{SEOBNRv5iHM} should naturally extend to the improved precessing model. 
Finally, parameter-estimation studies assessing the impact of the improved waveform accuracy provided by \texttt{SEOBNRv5iHM} are left to future work.

\section*{Acknowledgments}

We would like to warmly thank
Aurora Abbondanza,
Arnab Dhani,
Aldo Gamboa,
Marcus Haberland,
Gonzalo Morras,
Peter James Nee,
Nami Nishimura,
Harald Pfeiffer,
Lluc Planas,
and
Alessandro Santini
for useful discussions.

Computations were performed on the Hypatia cluster at the Max Planck Institute for Gravitational Physics, Potsdam.

The authors acknowledge the use of Claude (Anthropic) for code debugging and
optimization, and for grammar and language checks. The authors reviewed and
edited all output and take full responsibility for the content of this work.

This work is partly funded by the European Union (ERC grant GWSky/101167314).
Views and opinions expressed are however those of the authors only and do not necessarily reflect those of the European Union or the European Research Council Executive Agency. Neither the European Union nor the granting authority can be held responsible for them.
L.P. is supported by a UKRI Future Leaders Fellowship (grant number MR/Y018060/1).
H.E. is supported by the Spanish Ministerio de Ciencia, Innovación y Universidades (Juan de la Cierva, JDC2023-051934-I).

\appendix

\section{Mode-by-mode mismatches and assessing the importance of the $(5,5)$ mode}
\label{App:ModebyModeEllMax4}
In this appendix, we show mode-by-mode mismatches against the 591 SXS NR simulations used in this paper. 
The results here are computed with Eq.~\eqref{eq:mismatch}, using the \texttt{A+} LIGO PSD~\cite{LIGO-T1500293} and other settings detailed in Sec.~\ref{sec:hamiltoniancalibration}. 

For each mode, we compare against the same set of semi-analytical models as in Secs.~\ref{sec:Results22} and \ref{sec:ResultsPolHM}, where the mode is available.
The results are shown in Fig.~\ref{fig:modebymodemismatches}.

When considering the maximum mismatch over total mass, we find that \texttt{SEOBNRv5iHM} achieves the highest accuracy for every mode, with the exception of the $(3,2)$ mode, where \texttt{IMRPhenomXHM} achieves a higher level of accuracy.

\begin{figure}[t]
    \centering
    \includegraphics[width=\columnwidth]{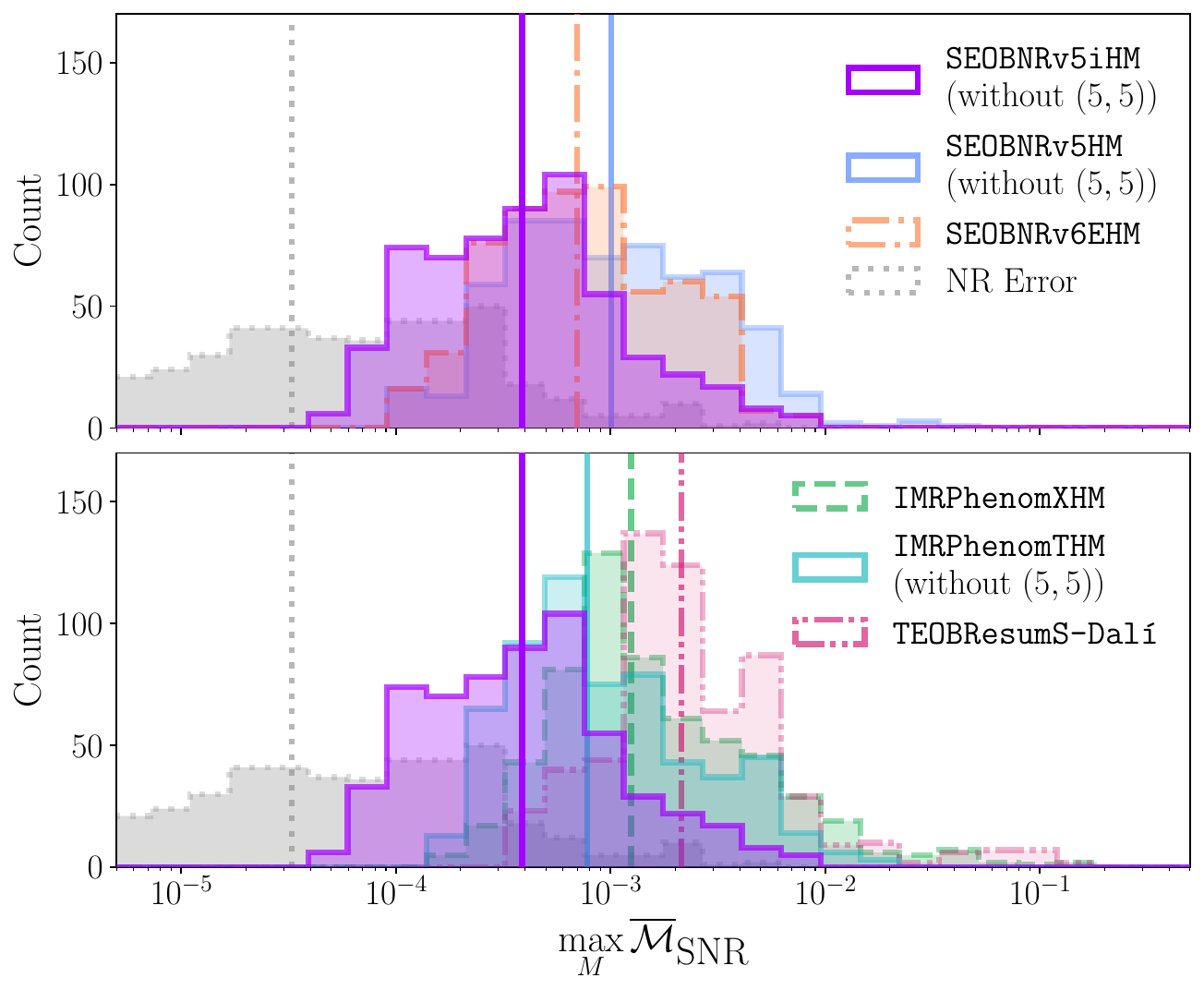}
    \caption{Polarization mismatches of the semi-analytical models when compared against NR polarizations built up to only $\ell_\mathrm{max} = 4$. For \texttt{SEOBNRv5iHM}, \texttt{SEOBNRv5HM}, and \texttt{IMRPhenomTHM}, we have removed the $(5,5)$ mode.}
    \label{fig:polarizationmismatchesEllmax4}
\end{figure}

In addition, we consider the importance of including the $(5,5)$ mode for the accuracy of the polarization waveform. 
Here, we repeat the analysis of Sec.~\ref{sec:ResultsPolHM}, but removing the $(5,5)$ mode from the three models \texttt{SEOBNRv5iHM}, \texttt{SEOBNRv5HM} and \texttt{IMRPhenomTHM} that include it. 
We then compare against an NR signal built using all modes up to $\ell_\mathrm{max} = 4$, rather than $\ell_\mathrm{max} = 5$. 
The results are shown in Fig.~\ref{fig:polarizationmismatchesEllmax4}.
The mismatches of \texttt{SEOBNRv5iHM}, \texttt{SEOBNRv5HM} and \texttt{IMRPhenomTHM} change the least with respect to Table~\ref{tab:mm_hm}, whereas those of the models without a $(5,5)$ mode decrease by a factor of $2$--$3.5$, since the missing $(5,5)$ mode is no longer counted as an error.
In particular, \texttt{SEOBNRv6EHM} now lies below \texttt{SEOBNRv5HM}, in agreement with Ref.~\cite{Gamboa:2026jht}.
\texttt{SEOBNRv5iHM} remains the most accurate model.

\begin{figure*}[t]
    \centering
    \includegraphics[width=\textwidth]{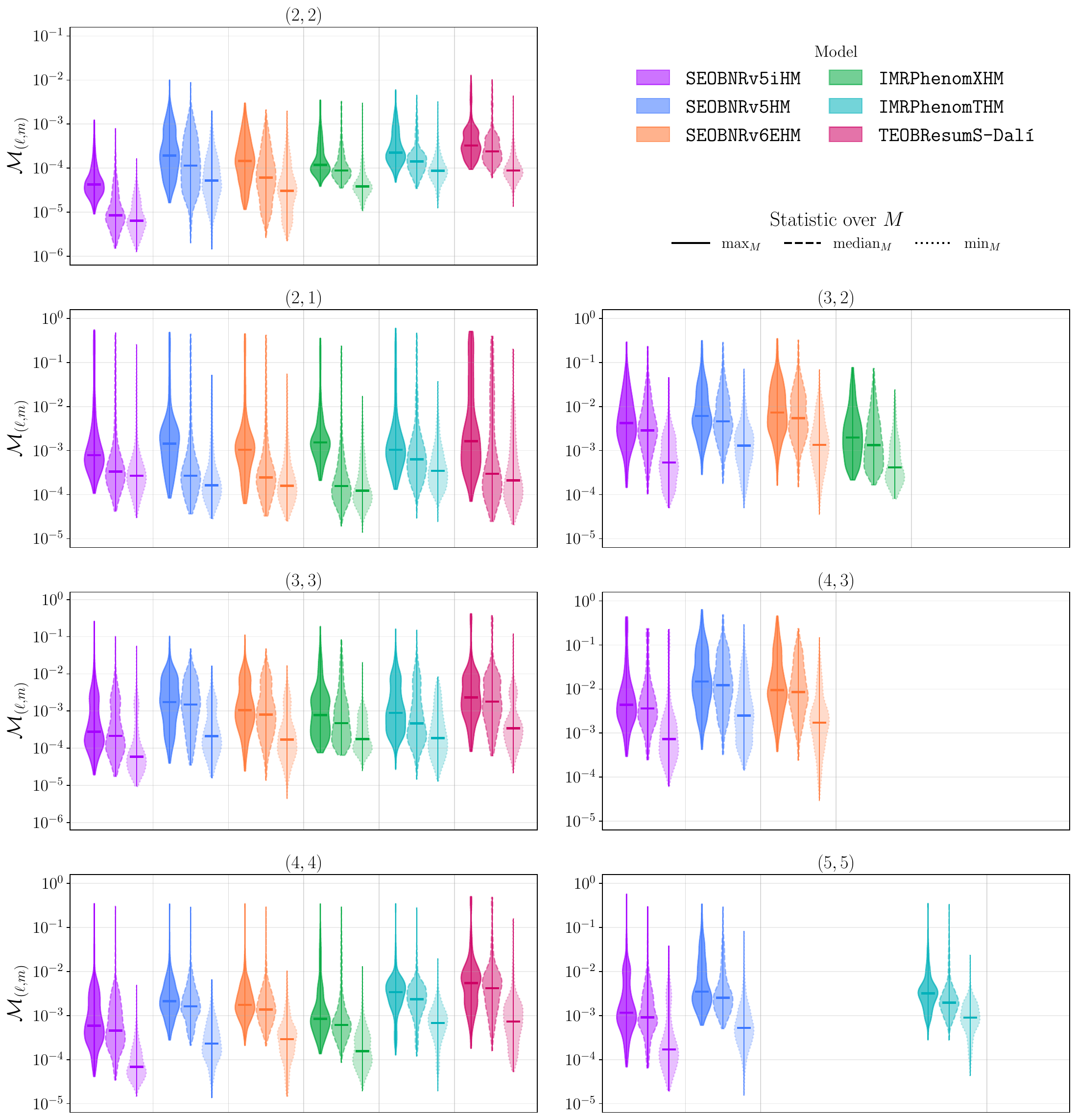}
    \caption{Mode-by-mode mismatches against the 591 SXS NR simulations for the semi-analytical models considered in this work; a model is shown only for the modes it includes. For each model, the three violins give the distributions of the maximum, median and minimum mismatch over total mass, and the horizontal lines mark their medians.}
    \label{fig:modebymodemismatches}
\end{figure*}

\section{Factorized waveform modes}
\label{app:modes}

In this appendix we specify, mode by mode, the matching terms introduced in Eq.~\eqref{eq:MatchingCoefficients}.
The factorized modes are given by Eq.~\eqref{hlmFactorised}, with $f_{\ell m}$ as in Eq.~\eqref{frholm}.
The PN expressions for $f_{\ell m}$ and $\delta_{\ell m}$ are those of \texttt{SEOBNRv5HM}, listed in the appendices of Ref.~\cite{Pompili:2023tna}, with the 2GSF calibration coefficients of Ref.~\cite{Leather:2025nhu}.

For the amplitude, the terms for each mode are
\begin{subequations}
\label{eq:MatchingAmpModes}
\begin{align}
f_{22} &\rightarrow f_{22} + \alpha_{22}\, v_\Omega^{\beta_{f,22}}, & \beta_{f,22} &= 11, \\
f_{21} &\rightarrow f_{21} + \alpha_{21}\, v_\Omega^{\beta_{f,21}}, & \beta_{f,21} &= 11, \\
f_{33} &\rightarrow f_{33} + \alpha_{33}\, v_\Omega^{\beta_{f,33}}, & \beta_{f,33} &= 11, \\
f_{32} &\rightarrow f_{32} + \alpha_{32}\, v_\Omega^{\beta_{f,32}}, & \beta_{f,32} &= 7, \\
f_{44} &\rightarrow f_{44} + \alpha_{44}\, v_\Omega^{\beta_{f,44}}, & \beta_{f,44} &= 7, \\
f_{43} &\rightarrow f_{43} + \alpha_{43}\, v_\Omega^{\beta_{f,43}}, & \beta_{f,43} &= 7, \\
f_{55} &\rightarrow f_{55} + \alpha_{55}\, v_\Omega^{\beta_{f,55}}, & \beta_{f,55} &= 7,
\end{align}
\end{subequations}

and for the phase
\begin{subequations}
\label{eq:MatchingPhaseModes}
\begin{align}
\delta_{22} &\rightarrow \delta_{22} + \gamma_{22}\, v_\Omega^{\beta_{\delta,22}}, & \beta_{\delta,22} &= 7, \\
\delta_{21} &\rightarrow \delta_{21} + \gamma_{21}\, v_\Omega^{\beta_{\delta,21}}, & \beta_{\delta,21} &= 7, \\
\delta_{33} &\rightarrow \delta_{33} + \gamma_{33}\, v_\Omega^{\beta_{\delta,33}}, & \beta_{\delta,33} &= 7, \\
\delta_{32} &\rightarrow \delta_{32} + \gamma_{32}\, v_\Omega^{\beta_{\delta,32}}, & \beta_{\delta,32} &= 7, \\
\delta_{44} &\rightarrow \delta_{44} + \gamma_{44}\, v_\Omega^{\beta_{\delta,44}}, & \beta_{\delta,44} &= 7, \\
\delta_{43} &\rightarrow \delta_{43} + \gamma_{43}\, v_\Omega^{\beta_{\delta,43}}, & \beta_{\delta,43} &= 7, \\
\delta_{55} &\rightarrow \delta_{55} + \gamma_{55}\, v_\Omega^{\beta_{\delta,55}}, & \beta_{\delta,55} &= 7.
\end{align}
\end{subequations}

The matching terms replace the calibration terms $c_{\ell m}$ that \texttt{SEOBNRv4HM} and \texttt{SEOBNRv5HM} used to cure the amplitude minima of the $(2,1)$, $(4,3)$ and $(5,5)$ modes.

\clearpage
\section{Hamiltonian Calibration Details}
\label{app:hamiltoniancalibfits}
In this appendix, we include additional details about the calibration detailed in Sec.~\ref{sec:hamiltoniancalibration}, and list the fits of the calibration parameters $a_6(\nu)$, $\Delta t^{22}_{\rm ISCO,noS}(\nu)$, $d_{\rm SO}(\nu,a_+,\delta a_-)$ and $\Delta t^{22}_{\rm ISCO,S}(\nu,a_+,\delta a_-)$. 

For nested sampling of our calibration parameters, we change our priors slightly, finding that a narrower prior gives better results when calibrating $d_\mathrm{SO}$. 

For the non-spinning fits, we fit the median points of the posterior instead of the maximum-likelihood points used in \texttt{SEOBNRv5HM}. This is because we fit the median points for the spinning calibration, and we found that for this model the two choices give similar non-spinning fits, while the median improves the subsequent fits in the spinning sector (Sec.~\ref{sec:hamiltoniancalibration}).

The final fits employed in the model are listed below.

\begin{align}
  &a_6(\nu) = 39.666 - 616.63\,\nu + 19600\,\nu^{2} \nonumber \\*
  &\qquad - 147530\,\nu^{3} + 315780\,\nu^{4}, \label{eq:a6} \\
  &\Delta t^{22}_{\mathrm{ISCO,\,noS}}(\nu) = \nu^{-1/5 + 11.229\,\nu}\,\bigl(-59.669 - 1527.6\,\nu \nonumber \\*
  &\qquad + 12963\,\nu^{2} - 31239\,\nu^{3}\bigr), \label{eq:NR_deltaT_NS} \\
  &d_{\rm SO}(\nu, a_+, \delta\,a_-) = 10.434\,a_+^{3} + 169.01\,a_+^{2}\,\nu \nonumber \\*
  &\qquad - 48.419\,a_+^{2} + 48.371\,a_+\,\delta\,a_- - 286.51\,a_+\,\nu^{2} \nonumber \\*
  &\qquad + 40.986\,a_+\,\nu + 17.198\,a_+ + 90.858\,\delta\,a_-\,\nu \nonumber \\*
  &\qquad - 64.400\,\delta\,a_- - 316.32\,\nu^{3} - 202.87\,\nu^{2} \nonumber \\*
  &\qquad + 279.05\,\nu - 79.389, \label{eq:dSO} \\
  &\Delta t^{22}_{\mathrm{ISCO,\,S}}(\nu, a_+, \delta\,a_-) = \nu^{-1/5}\,\bigl(-6.8243\,a_+^{3} \nonumber \\*
  &\qquad + 4.7476\,a_+^{2}\,\nu + 11.858\,a_+^{2} - 21.793\,a_+\,\delta\,a_- \nonumber \\*
  &\qquad + 170.54\,a_+\,\nu^{2} - 90.072\,a_+\,\nu + 19.394\,a_+ \nonumber \\*
  &\qquad + 4.0392\,\delta\,a_-\,\nu + 8.0825\,\delta\,a_-\bigr), \label{eq:NR_deltaT_spin} \\
  &\Delta t^{22}_{\mathrm{ISCO}}(\nu, a_+, \delta\,a_-) = \Delta t^{22}_{\mathrm{ISCO,\,noS}} + \Delta t^{22}_{\mathrm{ISCO,\,S}}. \label{eq:NR_deltaT}
\end{align}

\section{Input Value Fits}
\label{app:ivfits}

In this appendix, we provide fits for the input values, produced using the methodology described in Sec.~\ref{sec:Calibration_Fitting}. To produce these fits, we extracted input values from NR simulations using the highest resolution available, and extrapolation order $N=2$. As in Refs.~\cite{Cotesta:2018fcv,Pompili:2023tna}, we define the following combinations of $m_1, m_2, \chi_1, \chi_2$ to be used for the fits. 

\begin{equation}
    \delta = \frac{m_1 - m_2}{m_1 + m_2}
\end{equation}

\begin{equation}
    \chi_{33} = \chi_S \delta + \chi_A
\end{equation}

\begin{equation}
    \chi_{21A} = \frac{\chi_S}{1-1.3\nu} \delta + \chi_A
\end{equation}

\begin{equation}
    \chi_{44A} = (1-5\nu)\chi_S + \chi_A \delta
\end{equation}

\vspace{\baselineskip}

\subsection{Input Values at \texorpdfstring{$t^\mathrm{peak}_{\ell m}$}{t peak lm}}

Input value fits used to evaluate the conditions Eqs.~\eqref{eq:InterRegPeakAmpCondition},~\eqref{eq:InterRegPeakFreqCondition},~\eqref{eq:InterRegPeakDFreqCondition}, setting quantities at the peak of each mode.
The last set of fits in this subsection gives the peak-time differences $\Delta t^{\ell m}_{22}$.

\begin{widetext}
\begin{subequations}
\label{eq:h_peak}
\allowdisplaybreaks
\begin{align}
  \frac{h^\mathrm{NR}_{22}(t_{22}^\mathrm{peak})}{\nu} ={}& 1.4332 + 28.405\,\nu^{2} - 88.830\,\nu^{3} + 115.05\,\nu^{4} + 8.0097\,\nu^{2}\,\ln\nu + 0.0089553\,\sqrt{1-4\nu} \nonumber\\*
  &- 0.087793\,\delta\,a_-\,\sqrt{1-4\nu} + 0.11814\,\nu^{2}\,a_+^{2} - 0.015200\,a_+\,\sqrt{1-4\nu} + 0.62121\,\delta\,a_-\,\nu\,\ln\nu \nonumber\\*
  &- 0.18056\,a_+\,\left(\delta\,a_-\right)^{3} - 0.11108\,a_+\,\left(\delta\,a_-\right)^{2} + 0.24939\,\delta\,a_- + 0.59767\,\nu\,a_+\,\left(\delta\,a_-\right)^{2} + 0.010563\,a_+^{3} \nonumber\\*
  &+ 3.0212\,\nu^{2}\,a_+\,\delta\,a_- + 0.097684\,a_+\,\delta\,a_- - 1.2751\,\nu\,a_+\,\delta\,a_- - 0.91724\,\nu^{2}\,a_+ + 0.055974\,a_+ \nonumber\\*
  &+ 0.016290\,a_+^{4} + 0.037537\,\left(\delta\,a_-\right)^{2} \label{eq:hsign-peak-22} \\
  \frac{h^\mathrm{NR}_{33}(t_{33}^\mathrm{peak})}{\nu} ={}& 0.063903\,|\chi_{33}| - 0.0096946\,|\chi_{33}|^{2} + \delta\,\Big[83.397\,\nu - 83.212\,\nu^{2} - 311.90\,\nu^{4} + 17.916\,\sqrt{1-4\nu} \nonumber\\*
  &\quad - 64.046\,\nu\,\sqrt{1-4\nu} - \frac{17.381}{1 + \nu} - 185.20\,\nu^{2}\,\chi_{33} - 17.204\,\nu^{2}\,\chi_{33}^{2} + 106.24\,\nu^{2}\,\chi_{33}\,\delta + 155.68\,\nu^{3}\,\chi_{33} \nonumber\\*
  &\quad + 0.36970\,\chi_{33}^{4} + 0.023134\,a_+ - 8.2922\,\chi_{33}^{2}\,\delta^{2} + 7.8792\,\chi_{33}^{2}\,\delta - 17.148\,\nu\,\chi_{33}^{2}\,\delta + 0.087526\,a_+^{4} \nonumber\\*
  &\quad - 0.28608\,\chi_{33}^{2}\,a_+^{2} + 0.33335\,\chi_{33}\,\delta\,a_+ - 0.56145\,\nu\,\delta\,a_+^{2} + 9.8043\,\chi_{33}\,\delta^{3} - 9.7349\,\chi_{33}\,\delta \nonumber\\*
  &\quad + 39.200\,\nu\,\chi_{33} + 0.15642\,\nu\,\chi_{33}\,a_+^{2}\Big] \label{eq:hsign-peak-33} \\
  \frac{h^\mathrm{NR}_{21}(t_{21}^\mathrm{peak})}{\nu} ={}& 0.27909\,|\chi_{33}| - 0.17901\,|\chi_{33}|^{3}\,a_+ - 0.14612\,|\chi_{33}|^{2}\,a_+^{2} + \delta\,\Big[3.5324 - 17.795\,\nu + 8.1180\,\nu^{2} + 49.850\,\nu^{3} \nonumber\\*
  &\quad - 3.0912\,\sqrt{1-4\nu} + 12.683\,\nu\,\sqrt{1-4\nu} - 7.7411\,\nu\,\chi_{33} - 16.999\,\nu\,\chi_{33}^{2} - 229.45\,\nu^{2}\,\chi_{33}^{3} + 11.467\,\nu\,\chi_{33}^{4} \nonumber\\*
  &\quad + 50.654\,\nu^{2}\,\chi_{33}\,\delta + 0.44576\,a_+\,\nu\,\ln\nu + 181.15\,\nu^{3}\,\chi_{33}\,a_+ - 49.219\,\nu^{2}\,\chi_{33}^{2}\,\delta + 22.672\,\nu^{2}\,\chi_{33}^{2}\,a_+ \nonumber\\*
  &\quad - 47.498\,\nu\,\chi_{33}^{3}\,\delta + 69.445\,\nu\,\chi_{33}^{3} - 9.9878\,\nu\,\chi_{33}^{2}\,\delta\,a_+ - 74.371\,\nu^{3}\,a_+^{2} + 7.2649\,\nu\,a_+^{4} - 24.895\,\nu\,\chi_{33}\,a_+^{3} \nonumber\\*
  &\quad + 73.045\,\nu\,\chi_{33}^{2}\,a_+^{2} - 18.924\,\nu\,\chi_{33}^{3}\,a_+ - 587.85\,\nu^{3}\,\chi_{33}^{2} + 182.04\,\nu^{2}\,\chi_{33}^{2} - 0.93760\,\chi_{33}^{3}\,\delta^{2} \nonumber\\*
  &\quad + 13.824\,\chi_{33}^{2}\,\delta\,a_+^{2} - 14.251\,\chi_{33}^{2}\,a_+^{2} - 48.367\,\nu^{2}\,\chi_{33}\,\delta\,a_+ \nonumber\\*
  &\quad + 0.69907\,\chi_{33}^{3}\,a_+^{2} + 14.160\,\nu^{2}\,\delta\,a_+^{2}\Big] \label{eq:hsign-peak-21} \\
  \frac{h^\mathrm{NR}_{44}(t_{44}^\mathrm{peak})}{\nu} ={}& 4.9783 - 14.866\,\nu^{3} - 0.028096\,\nu\,\ln\nu + 1.1309\,\sqrt{1-4\nu} - 4.4787\,\nu\,\sqrt{1-4\nu} - \frac{5.8348}{1 + \nu} \nonumber\\*
  &+ 0.063598\,\chi_{44A}\,\sqrt{1-4\nu} + 0.19570\,\left(\delta\,a_-\right)^{4} - 0.30407\,\left(\chi_{44A}\right)^{3} + 3.9347\,\nu^{3}\,\delta\,a_- - 0.16583\,\nu\,\delta\,a_- \nonumber\\*
  &+ 2.9155\,\nu^{2}\,\left(\chi_{44A}\right)^{2} + 0.026325\,\left(\delta\,a_-\right)^{3} + 0.061985\,\nu\,\chi_{44A} - 0.19301\,\chi_{44A}\,\left(\delta\,a_-\right)^{3} \nonumber\\*
  &- 1.3028\,\nu^{2}\,\chi_{44A}\,\delta\,a_- + 0.041915\,\left(\delta\,a_-\right)^{2} - 0.41045\,\nu\,\left(\chi_{44A}\right)^{2} \nonumber\\*
  &- 0.21701\,\nu\,\left(\chi_{44A}\right)^{3} + 0.30078\,\left(\chi_{44A}\right)^{2}\,\delta\,a_- \label{eq:hsign-peak-44} \\
  \frac{h^\mathrm{NR}_{55}(t_{55}^\mathrm{peak})}{\nu} ={}& 0.010871\,|\chi_{33}| + 0.0047946\,|\chi_{33}|\,a_+^{2} + \delta\,\Big[7.0122 + 79.950\,\nu^{4} + 5.6927\,\nu^{2}\,\ln\nu + 1.5619\,\sqrt{1-4\nu} \nonumber\\*
  &\quad - 5.4769\,\nu\,\sqrt{1-4\nu} - \frac{8.4239}{1 + \nu} + 0.31708\,\nu\,\chi_{33} - 3.0387\,\nu^{2}\,\chi_{33}^{2} - 2.5973\,\nu^{2}\,\chi_{33}\,\delta + 0.040976\,\chi_{33}\,\delta^{2} \nonumber\\*
  &\quad + 0.075514\,\chi_{33}^{4} - 1.5676\,\chi_{33}^{2}\,\delta^{2} + 1.4781\,\chi_{33}^{2}\,\delta - 3.2556\,\nu\,\chi_{33}^{2}\,\delta - 0.013320\,\delta\,a_+ - 0.033643\,\chi_{33}^{3}\,a_+ \nonumber\\*
  &\quad + 0.031674\,a_+^{4} + 0.0056999\,\chi_{33}^{3} + 17.785\,\nu^{2}\,\delta^{2} - 0.047164\,\chi_{33}\,a_+^{3} \nonumber\\*
  &\quad + 0.13840\,\chi_{33}\,\delta^{2}\,a_+ - 0.039105\,\delta\,a_+^{2}\Big] \label{eq:hsign-peak-55} \\
  \frac{h^\mathrm{NR}_{32}(t_{32}^\mathrm{peak})}{\nu} ={}& 1.8852 - 7.6565\,\nu + 115.45\,\nu^{4} + 0.93019\,\nu\,\ln\nu - 1.6844\,\sqrt{1-4\nu} + 6.4494\,\nu\,\sqrt{1-4\nu} + 3.7757\,\nu^{2}\,a_+ \nonumber\\*
  &+ 0.043618\,\left(\delta\,a_-\right)^{3} + 40.533\,\nu^{3}\,\delta\,a_- + 0.021551\,a_+^{2} - 0.055889\,\delta\,a_-\,\sqrt{1-4\nu} - 0.044697\,a_+\,\sqrt{1-4\nu} \nonumber\\*
  &+ 3.6843\,\nu^{2}\,\left(\delta\,a_-\right)^{2} - 16.990\,\nu^{3}\,a_+ - 6.4864\,\nu^{2}\,\delta\,a_- - 0.16418\,a_+\,\nu\,\ln\nu \label{eq:hsign-peak-32} \\
  \frac{h^\mathrm{NR}_{43}(t_{43}^\mathrm{peak})}{\nu} ={}& 0.025449\,|\chi_{33}| - 0.0070159\,|\chi_{33}|\,a_+ + 0.0011663\,|\chi_{33}|^{4} + 0.0023518\,|\chi_{33}|\,a_+^{2} + \delta\,\Big[-3.0254 - 13.675\,\nu^{2} \nonumber\\*
  &\quad - 4.9395\,\nu\,\ln\nu - 25.530\,\nu^{2}\,\ln\nu + 2.9871\,\sqrt{1-4\nu} - 10.533\,\nu\,\sqrt{1-4\nu} - 3.3612\,\nu^{2}\,\chi_{33}^{2} \nonumber\\*
  &\quad + 1.6194\,\nu^{2}\,\delta\,a_+ - 94.225\,\nu^{3}\,\chi_{33} + 0.14104\,\chi_{33}^{4} - 0.69111\,\chi_{33}^{2}\,\delta^{2} + 0.19110\,\nu\,\chi_{33}^{3} + 2.4598\,\chi_{33}\,\delta^{3} \nonumber\\*
  &\quad - 40.239\,\nu^{2}\,\chi_{33}\,\delta + 60.880\,\nu^{2}\,\chi_{33} + 0.34715\,\chi_{33}^{2}\,\delta + 0.010735\,a_+^{3} - 0.067295\,\delta^{2}\,a_+^{2} - 0.044528\,\chi_{33}^{3}\,a_+ \nonumber\\*
  &\quad + 14.339\,\nu\,\chi_{33}\,\delta - 2.4838\,\chi_{33} + 0.40494\,\chi_{33}\,\delta^{2}\,a_+ + 0.033746\,a_+^{4} - 0.11429\,\chi_{33}\,\delta\,a_+ \nonumber\\*
  &\quad - 0.055546\,\chi_{33}\,a_+^{3} + 0.024021\,\delta^{3}\,a_+\Big] \label{eq:hsign-peak-43}
\end{align}
\end{subequations}
\end{widetext}

\begin{widetext}
\begin{subequations}
\label{eq:omega_peak}
\allowdisplaybreaks
\begin{align}
  \omega^\mathrm{NR}_{22}(t_{22}^\mathrm{peak}) ={}& -0.66045 - 2.7574\,\nu^{3} + 5.3768\,\nu^{4} - 0.052308\,\nu\,\ln\nu + 0.0033571\,\sqrt{1-4\nu} + \frac{0.38165}{1 + \nu} - 0.047266\,a_+ \nonumber\\*
  &- 0.056733\,a_+^{3}\,\delta\,a_- - 0.041240\,\delta\,a_- - 0.44913\,\nu^{2}\,a_+^{2} - 0.012838\,a_+^{3} - 2.9810\,\nu^{3}\,\delta\,a_- \nonumber\\*
  &- 0.056525\,a_+\,\left(\delta\,a_-\right)^{2} - 0.045738\,a_+\,\delta\,a_- - 0.10273\,\nu\,a_+ \label{eq:omega-peak-22} \\
  \omega^\mathrm{NR}_{33}(t_{33}^\mathrm{peak}) ={}& -65.068 + 438.51\,\nu^{4} - 6.4252\,\nu^{2}\,\ln\nu - 13.078\,\sqrt{1-4\nu} + 49.718\,\nu\,\sqrt{1-4\nu} + \frac{77.709}{1 + \nu} - 0.14065\,a_+ \nonumber\\*
  &- 0.066106\,a_+^{2} + 0.24408\,\delta\,a_-\,\nu\,\ln\nu - 0.12696\,a_+\,\delta\,a_-\,|\chi_{33}| \nonumber\\*
  &- 11.760\,\nu^{2}\,a_+\,\left(\delta\,a_-\right)^{2} - 0.79462\,\nu^{2}\,a_+^{3} \label{eq:omega-peak-33} \\
  \omega^\mathrm{NR}_{21}(t_{21}^\mathrm{peak}) ={}& -43.368 + 52.702\,\nu - 51.088\,\nu^{3} + 2.4529\,\nu\,\ln\nu + 34.369\,\nu^{2}\,\ln\nu + \frac{43.136}{1 + \nu} + 0.12428\,a_+^{4} \nonumber\\*
  &- 1.6581\,\nu\,a_+\,a_-^{2} + 0.70527\,\nu\,a_+\,a_-\,\chi_{21A} - 0.11875\,\nu\,a_-^{2} - 8.4047\,\nu^{2}\,a_+^{2} - 23.483\,\nu^{3}\,a_+ + 5.7241\,\nu^{2}\,a_+ \nonumber\\*
  &+ 2.2420\,\nu\,a_+^{2} - 0.054852\,a_+\,\sqrt{1-4\nu} + 0.10376\,a_+\,a_-\,\chi_{21A} + 0.013942\,a_-^{4} - 3.1290\,a_+^{2}\,\left(\chi_{21A}\right)^{2} \nonumber\\*
  &+ 5.8506\,a_+^{2}\,a_-\,\chi_{21A} - 2.7924\,a_+^{2}\,a_-^{2} - 0.087917\,a_+^{2} \label{eq:omega-peak-21} \\
  \omega^\mathrm{NR}_{44}(t_{44}^\mathrm{peak}) ={}& -20.992\,\nu + 162.38\,\nu^{3} - 216.26\,\nu^{4} - 4.1027\,\sqrt{1-4\nu} + 16.719\,\nu\,\sqrt{1-4\nu} + \frac{3.4527}{1 + \nu} - 0.61901\,a_+ \nonumber\\*
  &- 0.021996\,a_+^{4} - 0.26726\,a_+\,\left(\delta\,a_-\right)^{2} + 0.030762\,\left(\delta\,a_-\right)^{2} - 1.1062\,\nu\,\delta\,a_- + 0.46140\,a_+\,\sqrt{1-4\nu} \nonumber\\*
  &+ 2.2262\,\nu\,a_+ - 1.3988\,\nu^{2}\,a_+^{2} - 0.15251\,\nu\,a_+^{3} - 0.31693\,a_+^{2}\,\left(\delta\,a_-\right)^{2} \nonumber\\*
  &- 1.1546\,\nu\,\left(\delta\,a_-\right)^{2} + 0.44748\,a_+\,\nu\,\ln\nu \label{eq:omega-peak-44} \\
  \omega^\mathrm{NR}_{55}(t_{55}^\mathrm{peak}) ={}& 8.3880 - 55.744\,\nu + 59.136\,\nu^{2} + 186.84\,\nu^{4} - 9.1955\,\sqrt{1-4\nu} + 36.257\,\nu\,\sqrt{1-4\nu} - 0.42202\,a_+ \nonumber\\*
  &- 0.11993\,a_+^{2} - 0.049745\,a_-^{2}\,\delta\,a_- + 12.418\,\nu^{4}\,a_+ - 0.36392\,\delta\,a_-\,\sqrt{1-4\nu} - 0.15564\,a_-^{2}\,\left(\delta\,a_-\right)^{2} \nonumber\\*
  &- 12.071\,\nu^{2}\,a_+\,\left(\delta\,a_-\right)^{2} + 0.42439\,a_+\,\sqrt{1-4\nu} - 0.059850\,a_+^{5} - 33.378\,\nu^{3}\,a_+^{2} - 0.025034\,|\chi_{33}| \nonumber\\*
  &+ 7.2626\,\nu^{2}\,a_+^{2} - 0.11558\,a_+^{2}\,a_-^{2} \label{eq:omega-peak-55} \\
  \omega^\mathrm{NR}_{32}(t_{32}^\mathrm{peak}) ={}& -4.3966 + 94.443\,\nu^{2} - 273.25\,\nu^{3} - 6.4370\,\nu\,\ln\nu + 3.9435\,\sqrt{1-4\nu} - 15.924\,\nu\,\sqrt{1-4\nu} \nonumber\\*
  &- 7.3712\,\nu^{2}\,a_+\,\delta\,a_- + 0.054388\,\nu^{2}\,a_+ - 0.30148\,\left(\delta\,a_-\right)^{3}\,a_-^{2} - 0.056714\,\nu\,a_+^{4} - 36.012\,\nu^{3}\,\left(\delta\,a_-\right)^{2} \nonumber\\*
  &+ 0.63664\,\left(\delta\,a_-\right)^{3} + 274.25\,\nu^{4}\,\delta\,a_- - 3.2386\,\delta\,a_-\,\sqrt{1-4\nu} - 0.037629\,a_+^{3} - 1.6978\,\nu^{3}\,a_-^{2} \nonumber\\*
  &- 0.13001\,\nu\,a_+^{2} + 1.6441\,\delta\,a_-\,\nu\,\ln\nu - 235.71\,\nu^{3}\,\delta\,a_- + 3.2113\,\delta\,a_- - 4.6809\,\nu\,\left(\delta\,a_-\right)^{3} \nonumber\\*
  &+ 0.34560\,\delta\,a_-\,a_-^{4} - 0.69998\,a_+\,\left(\delta\,a_-\right)^{2}\,a_-^{2} + 0.21997\,a_+^{3}\,a_-^{2} - 0.30829\,a_+\,\left(\delta\,a_-\right)^{2} \label{eq:omega-peak-32} \\
  \omega^\mathrm{NR}_{43}(t_{43}^\mathrm{peak}) ={}& 180.81\,\nu^{2} - 491.75\,\nu^{3} + 400.02\,\nu^{4} + 69.245\,\nu^{2}\,\ln\nu - 0.62575\,\sqrt{1-4\nu} + 1.4974\,\nu\,\sqrt{1-4\nu} \nonumber\\*
  &- 84.937\,\nu^{3}\,\delta\,a_- + 6.4477\,\nu^{2}\,a_+ - 0.68241\,\nu^{2}\,a_+\,\delta\,a_-\,|\chi_{33}| - 1.0440\,\nu^{3}\,a_+^{2} + 11.788\,\nu^{2}\,a_+^{2}\,\delta\,a_- \nonumber\\*
  &- 111.96\,\nu^{4}\,a_+ + 0.12512\,\delta\,a_-\,\sqrt{1-4\nu} - 15.041\,\nu^{2}\,\left(\delta\,a_-\right)^{2} - 3.5729\,\nu^{2}\,a_+^{2}\,|\chi_{33}| - 0.13537\,a_+ \nonumber\\*
  &+ 0.20635\,a_+^{5} - 4.1406\,\nu\,a_+^{3}\,|\chi_{33}| + 41.620\,\nu^{3}\,\delta\,a_-\,|\chi_{33}| + 24.449\,\nu^{3}\,a_+\,|\chi_{33}| \nonumber\\*
  &- 0.043808\,a_+^{4} - 8.1274\,\nu^{2}\,a_+\,a_-^{2} \label{eq:omega-peak-43}
\end{align}
\end{subequations}
\end{widetext}

\begin{widetext}
\begin{subequations}
\label{eq:omegadot_peak}
\allowdisplaybreaks
\begin{align}
  \dot{\omega}^\mathrm{NR}_{22}(t_{22}^\mathrm{peak}) ={}& -0.17801\,\nu^{2} + 0.40896\,\nu^{3} - 0.52406\,\nu^{4} + 0.0058633\,\nu\,\ln\nu - 0.030603\,\nu^{2}\,\ln\nu - \frac{0.0062976}{1 + \nu} \nonumber\\*
  &- 0.0029528\,\nu^{2}\,\delta\,a_- + 0.0028783\,\left(\delta\,a_-\right)^{3} + 0.0024053\,\left(\delta\,a_-\right)^{2} - 0.020779\,\nu\,\left(\delta\,a_-\right)^{2} \nonumber\\*
  &- 0.0076115\,a_+\,\nu\,\ln\nu + 0.0075092\,\delta\,a_-\,\nu\,\ln\nu - 0.011870\,\nu\,a_+ + 0.00086225\,a_+\,\delta\,a_- \label{eq:omegadot-peak-22} \\
  \dot{\omega}^\mathrm{NR}_{33}(t_{33}^\mathrm{peak}) ={}& 1.7674\,\nu^{2} + 0.30020\,\nu\,\ln\nu + 2.5869\,\nu^{2}\,\ln\nu - 0.26003\,\sqrt{1-4\nu} + 0.91955\,\nu\,\sqrt{1-4\nu} + \frac{0.25008}{1 + \nu} \nonumber\\*
  &- 0.11936\,\nu^{2}\,\delta\,a_- + 0.0051544\,\left(\delta\,a_-\right)^{4} + 0.0080458\,\left(\delta\,a_-\right)^{3}\,|\chi_{33}| - 0.00075322\,a_+ + 0.10122\,\nu^{3}\,|\chi_{33}| \nonumber\\*
  &- 0.12803\,\nu^{2}\,\left(\delta\,a_-\right)^{2} - 0.0022192\,\delta\,a_-\,|\chi_{33}| \label{eq:omegadot-peak-33} \\
  \dot{\omega}^\mathrm{NR}_{21}(t_{21}^\mathrm{peak}) ={}& 0.79656\,\nu^{2} - 2.3565\,\nu^{3} - 0.044106\,\nu\,\ln\nu + 0.049255\,\sqrt{1-4\nu} - 0.21074\,\nu\,\sqrt{1-4\nu} - \frac{0.051637}{1 + \nu} \nonumber\\*
  &- 0.0096316\,a_+\,\sqrt{1-4\nu} - 0.019764\,a_+^{2} + 0.092872\,\nu\,a_+^{2} + 0.0085974\,a_+^{3} - 0.20472\,\left(\chi_{21A}\right)^{2} \nonumber\\*
  &+ 3.1658\,\nu\,a_-^{2} - 0.011371\,a_+ + 0.67737\,\nu\,a_+ + 0.63885\,a_-\,\chi_{21A} - 5.7083\,\nu\,a_-\,\chi_{21A} \nonumber\\*
  &- 0.024600\,a_+\,\left(\chi_{21A}\right)^{2} + 2.4049\,\nu\,\left(\chi_{21A}\right)^{2} - 0.40073\,a_-^{2} + 0.21345\,a_+\,\nu\,\ln\nu \nonumber\\*
  &- 1.4196\,\nu^{2}\,a_+ + 0.026058\,a_+\,a_-\,\chi_{21A} \label{eq:omegadot-peak-21} \\
  \dot{\omega}^\mathrm{NR}_{44}(t_{44}^\mathrm{peak}) ={}& 5.5106\,\nu^{3} + 0.39744\,\nu\,\ln\nu + 2.2559\,\nu^{2}\,\ln\nu - 0.28977\,\sqrt{1-4\nu} + 1.1646\,\nu\,\sqrt{1-4\nu} + \frac{0.27816}{1 + \nu} \nonumber\\*
  &- 0.038012\,a_+\,\nu\,\ln\nu + 0.023939\,\nu\,\delta\,a_- + 0.011635\,a_+\,\sqrt{1-4\nu} + 0.047132\,\delta\,a_-\,\nu\,\ln\nu \nonumber\\*
  &- 0.016427\,a_+ + 0.011759\,\left(\delta\,a_-\right)^{3} \label{eq:omegadot-peak-44} \\
  \dot{\omega}^\mathrm{NR}_{55}(t_{55}^\mathrm{peak}) ={}& -0.030620 - 0.71662\,\nu^{3} - 0.010432\,\nu\,\ln\nu - 0.095041\,\nu^{2}\,\ln\nu + 0.011792\,\sqrt{1-4\nu} \nonumber\\*
  &- 0.087180\,\nu\,\sqrt{1-4\nu} + 0.62409\,\nu\,\delta\,a_- - 0.21530\,\nu^{3}\,a_+ + 0.0080579\,\left(\delta\,a_-\right)^{4} \nonumber\\*
  &+ 0.011487\,a_+\,\left(\delta\,a_-\right)^{2}\,|\chi_{33}| - 0.28893\,\delta\,a_- + 0.28410\,\delta\,a_-\,\sqrt{1-4\nu} + 8.5592\,\nu^{3}\,\delta\,a_- - 0.99085\,\nu^{2}\,\delta\,a_- \nonumber\\*
  &+ 0.084436\,\nu^{3}\,|\chi_{33}| - 0.16855\,\nu^{2}\,\left(\delta\,a_-\right)^{2} \label{eq:omegadot-peak-55} \\
  \dot{\omega}^\mathrm{NR}_{32}(t_{32}^\mathrm{peak}) ={}& -2.1044\,\nu^{3} - 0.016616\,\nu\,\ln\nu - 0.91258\,\nu^{2}\,\ln\nu + 0.066962\,\sqrt{1-4\nu} - 0.23232\,\nu\,\sqrt{1-4\nu} - \frac{0.081871}{1 + \nu} \nonumber\\*
  &- 0.0052273\,a_+\,\delta\,a_- - 0.16767\,\nu^{2}\,a_+\,\delta\,a_- + 0.017397\,\left(\delta\,a_-\right)^{3} - 0.072302\,\delta\,a_-\,\nu\,\ln\nu - 5.8474\,\nu^{3}\,a_+ \nonumber\\*
  &+ 1.4153\,\nu^{2}\,a_+ - 0.019167\,a_+^{2}\,\delta\,a_- - 0.51257\,\nu\,a_+^{2} - 3.9560\,\nu^{3}\,\delta\,a_- + 0.26569\,\nu^{2}\,\delta\,a_- \nonumber\\*
  &- 0.66573\,\nu\,\left(\delta\,a_-\right)^{3} - 0.011817\,a_+\,\sqrt{1-4\nu} - 0.031986\,\left(\delta\,a_-\right)^{4} + 0.14219\,\nu\,\delta\,a_-\,a_-^{2} \nonumber\\*
  &- 3.8836\,\nu^{2}\,\left(\delta\,a_-\right)^{2} + 0.66165\,\nu\,\left(\delta\,a_-\right)^{2} + 1.3600\,\nu^{2}\,a_+^{2} + 0.019101\,a_+^{3}\,\delta\,a_- + 0.046343\,a_+^{2} \label{eq:omegadot-peak-32} \\
  \dot{\omega}^\mathrm{NR}_{43}(t_{43}^\mathrm{peak}) ={}& -1.4478 - 11.885\,\nu^{2} - 2.0969\,\nu\,\ln\nu - 16.623\,\nu^{2}\,\ln\nu + 1.3987\,\sqrt{1-4\nu} - 5.2271\,\nu\,\sqrt{1-4\nu} \nonumber\\*
  &+ 2.2498\,\nu^{3}\,\delta\,a_- + 0.011299\,|\chi_{33}|\,\sqrt{1-4\nu} - 0.70822\,\nu^{2}\,\left(\delta\,a_-\right)^{2} + 0.062499\,a_+\,\left(\delta\,a_-\right)^{2} \nonumber\\*
  &- 0.24660\,\nu\,a_+\,\left(\delta\,a_-\right)^{2} - 0.0046832\,a_+\,\sqrt{1-4\nu} - 0.019606\,\delta\,a_-\,|\chi_{33}|^{3} \label{eq:omegadot-peak-43}
\end{align}
\end{subequations}
\end{widetext}

\begin{widetext}
\begin{subequations}
\label{eq:nrdeltaTlm}
\allowdisplaybreaks
\begin{align}
  \Delta t_{22}^{22} ={}& 0 \label{eq:nrdeltaTlm-22} \\
  \Delta t_{22}^{33} ={}& -442.46 + 1150.7\,\nu^{2} + 912.44\,\nu^{2}\,\ln\nu - 119.65\,\sqrt{1-4\nu} + 467.62\,\nu\,\sqrt{1-4\nu} + \frac{555.33}{1 + \nu} - 2.0093\,a_+ \nonumber\\*
  &- 1.0466\,a_+^{2} - 54.385\,\nu\,a_+^{2}\,\delta\,a_- - 3.6999\,a_+^{2}\,a_-^{2} + 1256.4\,\nu^{3}\,\delta\,a_- - 242.64\,\nu^{2}\,\delta\,a_- + 4.2304\,a_+^{2}\,\delta\,a_- \label{eq:nrdeltaTlm-33} \\
  \Delta t_{22}^{21} ={}& -512.87 - 530.03\,\nu^{3} - 139.12\,\nu\,\ln\nu - 763.00\,\nu^{2}\,\ln\nu + 2.4928\,\nu\,\sqrt{1-4\nu} + \frac{500.70}{1 + \nu} + 1.1774\,a_+\,\sqrt{1-4\nu} \nonumber\\*
  &- 82.118\,a_+^{4} + 8.4774\,a_+^{5} + 193.75\,\nu\,\chi_A^{2} - 307.54\,\nu^{2}\,\chi_A^{2}\,a_+ - 994.26\,\nu\,\chi_A^{2}\,a_+^{2} + 114.73\,\nu^{2}\,a_-^{2}\,a_+ + 111.04\,\chi_S\,a_+^{3} \nonumber\\*
  &- 2.4189\,\chi_S\,\sqrt{1-4\nu} - 18.964\,a_+\,\nu\,\ln\nu - 1291.4\,\nu^{4}\,a_+ + 141.78\,\chi_A^{2}\,a_+^{2} + 121.12\,\nu\,a_-\,\chi_A\,a_+^{2} - 4.7743\,\chi_S\,a_+^{2} \nonumber\\*
  &+ 302.07\,\nu\,\chi_S\,\chi_A^{2}\,a_+ - 2.3291\,\chi_S\,a_+ - 5.8620\,\chi_A^{4} - 672.17\,\nu^{2}\,\chi_A^{2} + 44.260\,\nu\,a_-^{4} \nonumber\\*
  &- 21.749\,\chi_S^{4} - 11.110\,\chi_A^{2} \label{eq:nrdeltaTlm-21} \\
  \Delta t_{22}^{44} ={}& 155.42 - 1039.9\,\nu + 1610.5\,\nu^{2} - 156.52\,\sqrt{1-4\nu} + 639.47\,\nu\,\sqrt{1-4\nu} - 18.304\,a_+ - 1.0816\,a_+^{2} \nonumber\\*
  &- 6.2908\,\left(\delta\,a_-\right)^{3} + 19.962\,a_+\,\sqrt{1-4\nu} + 248.45\,\nu^{2}\,a_+ - 3.4804\,\left(\delta\,a_-\right)^{2} - 2.0237\,\delta\,a_- \label{eq:nrdeltaTlm-44} \\
  \Delta t_{22}^{55} ={}& -51.312 + 160.27\,\nu + 429.92\,\nu^{2} - 1562.0\,\nu^{3} + 46.545\,\sqrt{1-4\nu} - 111.63\,\nu\,\sqrt{1-4\nu} - 657.69\,\nu^{3}\,a_+ \nonumber\\*
  &- 9.7754\,\nu\,a_+^{2} - 19.968\,a_+\,\nu\,\ln\nu - 1.0474\,a_-^{2} - 4.0725\,a_+\,a_-^{2} - 5.7767\,a_+\,\sqrt{1-4\nu} \label{eq:nrdeltaTlm-55} \\
  \Delta t_{22}^{32} ={}& -2013.2 - 153.43\,\nu\,\ln\nu - 1163.5\,\nu^{2}\,\ln\nu - 305.69\,\sqrt{1-4\nu} + 1082.6\,\nu\,\sqrt{1-4\nu} + \frac{2309.9}{1 + \nu} - 522.29\,\nu^{2}\,a_+\,\delta\,a_- \nonumber\\*
  &+ 31.057\,a_+\,\sqrt{1-4\nu} - 2712.7\,\nu^{2}\,\delta\,a_- - 801.61\,\nu^{2}\,a_+^{2}\,\delta\,a_- + 297.82\,\nu^{2}\,\chi_S\,\delta\,a_- + 2387.8\,\nu^{4}\,a_+ - 8.8833\,\chi_S \nonumber\\*
  &+ 73.847\,\left(\delta\,a_-\right)^{5} + 2228.0\,\nu^{3}\,\delta\,a_- + 86.331\,\chi_S\,a_+^{3}\,\delta\,a_- + 11.937\,\nu\,\chi_S\,a_+ - 2.6269\,a_+^{4} - 1109.6\,\nu\,\chi_S^{2}\,a_+ \nonumber\\*
  &- 216.69\,\chi_S\,\left(\delta\,a_-\right)^{4} + 117.74\,\nu\,\chi_S\,a_+^{2} - 168.53\,\delta\,a_-\,\sqrt{1-4\nu} + 17.331\,\chi_A^{2}\,\delta\,a_- + 83.525\,\chi_S^{2}\,a_+ \nonumber\\*
  &- 129.19\,\chi_S^{3}\,\left(\delta\,a_-\right)^{2} - 1816.9\,\nu^{3}\,a_+\,\delta\,a_- - 82.442\,\chi_A^{2}\,\left(\delta\,a_-\right)^{3} + 14.548\,\chi_S^{4}\,\delta\,a_- \nonumber\\*
  &+ 137.01\,\delta\,a_- + 2627.1\,\nu^{2}\,\chi_S^{2}\,a_+ \label{eq:nrdeltaTlm-32} \\
  \Delta t_{22}^{43} ={}& 106.43 - 797.22\,\nu^{2} + 189.40\,\nu\,\ln\nu - 109.12\,\sqrt{1-4\nu} + 316.82\,\nu\,\sqrt{1-4\nu} - 455.37\,\nu^{2}\,\delta\,a_- \nonumber\\*
  &- 66.857\,\nu\,a_+\,\delta\,a_-\,|\chi_{33}| + 96.060\,\delta\,a_-\,\sqrt{1-4\nu} - 2.5391\,a_+\,|\chi_{33}|^{2} + 5.7172\,a_+^{2}\,\left(\delta\,a_-\right)^{2} + 36.755\,a_+\,\sqrt{1-4\nu} \nonumber\\*
  &- 35.366\,a_+ + 63.235\,\nu\,a_+ + 559.89\,\nu^{2}\,\delta\,a_-\,|\chi_{33}| - 91.617\,\nu^{2}\,a_+^{2} - 665.13\,\nu^{2}\,\left(\delta\,a_-\right)^{2} - 194.86\,\nu\,\left(\delta\,a_-\right)^{3} \nonumber\\*
  &+ 1.2228\,a_-^{2} + 4.2547\,\nu\,a_+^{2}\,\delta\,a_- - 1.6300\,a_+^{2}\,\delta\,a_- + 37.145\,a_+\,\left(\delta\,a_-\right)^{2} - 207.63\,\delta\,a_-\,\nu\,\ln\nu - 115.29\,\delta\,a_- \nonumber\\*
  &- 377.23\,\nu\,a_+\,\left(\delta\,a_-\right)^{2} + 980.29\,\nu^{3}\,a_+ + 3.6298\,a_+^{2} - 9.3498\,a_+^{3}\,\delta\,a_- \label{eq:nrdeltaTlm-43}
\end{align}
\end{subequations}
\end{widetext}

\subsection{Input Values at \texorpdfstring{$t^\mathrm{mid}_{\ell m}$}{t mid lm}}

\begin{widetext}
\begin{subequations}
\label{eq:h_mid}
\allowdisplaybreaks
\begin{align}
  \frac{h^\mathrm{NR}_{22}(t_{22}^\mathrm{mid})}{\nu} ={}& 7.2910\,\nu - 27.138\,\nu^{3} + 1.9472\,\nu^{4} + 0.40508\,\nu\,\ln\nu + 1.2921\,\sqrt{1-4\nu} - 4.9748\,\nu\,\sqrt{1-4\nu} + 0.28423\,a_+ \nonumber\\*
  &+ 0.53851\,a_+\,\nu\,\ln\nu + 0.029664\,a_+^{2} + 1.0132\,\nu^{2}\,\delta\,a_- + 0.23119\,\nu\,\left(\delta\,a_-\right)^{2} + 0.66862\,\nu\,a_+^{3} \nonumber\\*
  &- 0.14928\,a_+^{3} + 0.093887\,a_+^{2}\,\delta\,a_- \label{eq:hsign-mid-22} \\
  \frac{h^\mathrm{NR}_{33}(t_{33}^\mathrm{mid})}{\nu} ={}& 0.019912\,|\chi_{33}| + 0.015028\,|\chi_{33}|\,a_+^{2} + \delta\,\Big[-193.79 + 145.86\,\nu - 188.17\,\nu^{4} - 3.8160\,\sqrt{1-4\nu} \nonumber\\*
  &\quad + 17.142\,\nu\,\sqrt{1-4\nu} + \frac{198.05}{1 + \nu} - 0.66494\,\chi_{33}\,\nu\,\ln\nu - 37.530\,\nu^{2}\,\chi_{33}^{2} + 0.14534\,\nu\,a_+ - 10.023\,\nu^{2}\,\chi_{33}\,\delta \nonumber\\*
  &\quad - 24.399\,\nu\,\chi_{33}^{2}\,\delta + 7.3524\,\nu\,\chi_{33}^{2} + 0.14702\,\chi_{33}^{4} + 0.059565\,\chi_{33}\,a_+ + 1.8447\,\nu^{2}\,\delta\,a_+ - 0.094563\,\chi_{33}^{3}\,a_+ \nonumber\\*
  &\quad + 0.086738\,\nu\,\chi_{33}^{3} + 8.4909\,\chi_{33}^{2}\,\delta - 8.5468\,\chi_{33}^{2}\,\delta^{2} + 0.011034\,a_+^{3} + 25.964\,\nu\,\delta^{3}\Big] \label{eq:hsign-mid-33} \\
  \frac{h^\mathrm{NR}_{21}(t_{21}^\mathrm{mid})}{\nu} ={}& 0.11702\,|\chi_{33}| + 0.019639\,|\chi_{33}|\,a_+ + \delta\,\Big[-19.375 - 5.0674\,\nu^{2} + 70.680\,\nu^{3} - 3.7709\,\sqrt{1-4\nu} \nonumber\\*
  &\quad + 16.016\,\nu\,\sqrt{1-4\nu} + \frac{23.395}{1 + \nu} - 12.293\,\nu^{2}\,\chi_{33} + 4.0027\,\nu\,\chi_{33}^{2} - 17.479\,\nu\,\chi_{33}^{4} - 54.341\,\nu^{2}\,\chi_{33}^{3} \nonumber\\*
  &\quad + 1.8461\,\chi_{33}^{3}\,\delta^{2} + 127.82\,\nu^{3}\,\chi_{33}\,\delta - 7.2975\,\chi_{33}^{3}\,\delta + 5.0757\,\chi_{33}^{3} - 6.7213\,\chi_{33}^{4}\,\delta - 1.0971\,\nu\,\chi_{33} \nonumber\\*
  &\quad + 6.5822\,\chi_{33}^{4} - 98.361\,\nu^{3}\,\chi_{33}^{2} + 0.28720\,\chi_{33}\,\delta\,a_+ + 1.3709\,\chi_{33}^{3}\,\delta\,a_+ - 1.0148\,\chi_{33}^{2}\,a_+^{2} - 0.61643\,\chi_{33}^{2} \nonumber\\*
  &\quad + 0.34966\,\delta\,a_+^{4} - 1.1556\,\nu^{2}\,\chi_{33}\,a_+^{2} + 3.3545\,\nu\,\chi_{33}^{3}\,a_+ - 0.19836\,\chi_{33}\,a_+ - 0.13727\,\delta\,a_+^{2}\Big] \label{eq:hsign-mid-21} \\
  \frac{h^\mathrm{NR}_{44}(t_{44}^\mathrm{mid})}{\nu} ={}& 1.5442\,\nu - 4.3490\,\nu^{3} + 0.31562\,\nu\,\ln\nu + 1.8386\,\nu^{2}\,\ln\nu + 0.17686\,\sqrt{1-4\nu} - 0.69378\,\nu\,\sqrt{1-4\nu} \nonumber\\*
  &+ 0.13137\,\delta\,a_-\,\sqrt{1-4\nu} - 0.0035343\,\left(\delta\,a_-\right)^{4} + 0.27721\,\chi_{44A}\,\nu\,\ln\nu + 1.7533\,\nu^{3}\,\delta\,a_- + 0.46870\,\nu\,\chi_{44A} \nonumber\\*
  &+ 0.071349\,\left(\delta\,a_-\right)^{2} - 0.59422\,\nu\,\chi_{44A}\,\delta\,a_- + 0.029605\,\left(\delta\,a_-\right)^{3} - 0.047650\,\chi_{44A} \nonumber\\*
  &+ 1.2260\,\nu^{2}\,\left(\delta\,a_-\right)^{2} + 0.58433\,\nu^{2}\,\left(\chi_{44A}\right)^{2} \label{eq:hsign-mid-44} \\
  \frac{h^\mathrm{NR}_{55}(t_{55}^\mathrm{mid})}{\nu} ={}& 0.0023583\,|\chi_{33}| + 0.0016022\,|\chi_{33}|\,a_+^{2} + 0.0026012\,|\chi_{33}|\,a_+^{3} + \delta\,\Big[-7.0309\,\nu^{2} - 0.21366\,\nu\,\ln\nu \nonumber\\*
  &\quad - 3.9659\,\nu^{2}\,\ln\nu + 0.79952\,\sqrt{1-4\nu} - 2.8373\,\nu\,\sqrt{1-4\nu} + \frac{0.075376}{1 + \nu} + 0.12420\,\chi_{33} + 0.022596\,a_+ \nonumber\\*
  &\quad + 1.5997\,\nu^{2}\,\chi_{33}\,\delta - 0.00063851\,a_+^{2} + 2.6222\,\nu^{2}\,\chi_{33}^{2} + 0.59148\,\nu\,\chi_{33}^{2}\,\delta + 0.014423\,\chi_{33}^{4} - 0.84760\,\nu\,\chi_{33}^{2} \nonumber\\*
  &\quad + 0.054889\,\chi_{33}\,\delta^{2}\,a_+ - 0.79236\,\delta\,\sqrt{1-4\nu} - 0.00092756\,\delta\,a_+^{3} - 0.21465\,\nu\,\chi_{33}^{3} + 0.016648\,\nu^{2}\,\delta\,a_+ \nonumber\\*
  &\quad + 0.15183\,\nu^{2}\,\chi_{33}\,a_+ - 0.022112\,\chi_{33}\,\delta\,a_+ - 0.023163\,\nu\,\chi_{33}\,a_+^{2} + 0.0056903\,\chi_{33}\,a_+^{2} - 0.050806\,\chi_{33}^{3}\,\delta \nonumber\\*
  &\quad + 0.066313\,\chi_{33}^{3} + 0.043555\,a_+\,\nu\,\ln\nu + 0.14430\,\chi_{33}\,\delta^{3} - 5.9427\,\nu^{3}\,\chi_{33} - 0.24122\,\chi_{33}\,\delta \nonumber\\*
  &\quad + 0.0036551\,a_+^{4} - 0.0068243\,\chi_{33}^{2}\,a_+^{2}\Big] \label{eq:hsign-mid-55} \\
  \frac{h^\mathrm{NR}_{32}(t_{32}^\mathrm{mid})}{\nu} ={}& 0.45204 - 1.8407\,\nu^{3} + 0.090544\,\nu\,\ln\nu + 0.096909\,\sqrt{1-4\nu} - 0.38654\,\nu\,\sqrt{1-4\nu} - \frac{0.45772}{1 + \nu} \nonumber\\*
  &+ 0.036248\,a_+ + 0.013122\,a_+^{2} + 0.086866\,\delta\,a_-\,\nu\,\ln\nu + 0.010999\,a_+^{2}\,\delta\,a_- + 0.35003\,\nu^{2}\,\delta\,a_- \nonumber\\*
  &+ 0.0099644\,\left(\delta\,a_-\right)^{4} + 0.0029165\,a_+^{4} + 0.0074248\,\nu\,a_+^{3} + 0.0024094\,\delta\,a_- \label{eq:hsign-mid-32} \\
  \frac{h^\mathrm{NR}_{43}(t_{43}^\mathrm{mid})}{\nu} ={}& 0.0083064\,|\chi_{33}| + 0.0017365\,|\chi_{33}|\,a_+ - 0.0023654\,|\chi_{33}|^{4} - 0.0030227\,|\chi_{33}|^{2} + 0.0048895\,|\chi_{33}|^{3} \nonumber\\*
  &+ \delta\,\Big[0.84381 + 6.4325\,\nu^{2} + 1.3245\,\nu\,\ln\nu + 8.9104\,\nu^{2}\,\ln\nu - 0.78075\,\sqrt{1-4\nu} + 3.0203\,\nu\,\sqrt{1-4\nu} \nonumber\\*
  &\quad + 14.165\,\nu^{3}\,\chi_{33} - 2.1342\,a_+\,\nu\,\ln\nu - 1.0632\,\nu^{2}\,\chi_{33}^{2} + 0.0090981\,a_+^{4} - 0.34691\,\nu^{2}\,\chi_{33}\,a_+ \nonumber\\*
  &\quad - 13.903\,\nu\,\delta\,a_+ + 0.064429\,\chi_{33}^{4} - 0.011845\,\chi_{33}^{2}\,\delta^{2} + 0.30730\,\nu\,\chi_{33}\,\delta\,a_+ - 0.048388\,\chi_{33}^{3}\,a_+ \nonumber\\*
  &\quad - 2.4952\,\nu\,\chi_{33}^{3} + 2.5723\,\nu\,\chi_{33}^{2}\,a_+ + 0.0055504\,\nu\,a_+^{3} + 4.4946\,\nu^{2}\,\chi_{33}\,\delta - 5.0395\,\nu^{2}\,\chi_{33} \nonumber\\*
  &\quad + 0.016018\,\chi_{33}\,\delta\,a_+^{2} - 0.60988\,\chi_{33}^{3}\,\delta + 0.76002\,\chi_{33}^{3} + 0.62549\,\chi_{33}^{2}\,\delta\,a_+ - 0.78843\,\chi_{33}^{2}\,a_+ + 3.9011\,\delta\,a_+ \nonumber\\*
  &\quad - 3.9388\,\delta^{2}\,a_+ - 48.098\,\nu^{3}\,a_+ - 0.021425\,\chi_{33}\,\delta^{2} - 0.026322\,\chi_{33}^{2}\,\delta\Big] \label{eq:hsign-mid-43}
\end{align}
\end{subequations}
\end{widetext}

\begin{widetext}
\begin{subequations}
\label{eq:omega_mid}
\allowdisplaybreaks
\begin{align}
  \omega^\mathrm{NR}_{22}(t_{22}^\mathrm{mid}) ={}& -0.31179 + 4.1510\,\nu^{3} - 14.508\,\nu^{4} - 0.29110\,\nu\,\ln\nu + 0.11023\,\sqrt{1-4\nu} - 0.43191\,\nu\,\sqrt{1-4\nu} \nonumber\\*
  &- 0.057455\,a_+ - 0.020164\,a_+^{2} - 0.062636\,\delta\,a_- - 0.044717\,a_+\,\delta\,a_- - 0.0069018\,a_+^{3} \nonumber\\*
  &- 0.066793\,a_+\,\left(\delta\,a_-\right)^{2} - 0.039363\,a_+\,\left(\delta\,a_-\right)^{3} - 2.0999\,\nu^{3}\,\delta\,a_- + 0.27308\,\nu\,a_+\,\left(\delta\,a_-\right)^{2} + 3.1761\,\nu^{2}\,\delta\,a_- \nonumber\\*
  &- 0.65932\,\delta\,a_-\,\nu\,\ln\nu - 0.0020387\,a_+^{4} - 1.1322\,\nu^{2}\,a_+\,\delta\,a_- - 1.4815\,\nu\,\delta\,a_- + 0.31152\,\nu\,a_+\,\delta\,a_- \nonumber\\*
  &- 0.12087\,\nu\,a_+^{2}\,\delta\,a_- - 0.23984\,\nu^{3}\,a_+ - 0.013780\,a_+^{3}\,\delta\,a_- - 0.32911\,\nu^{2}\,\left(\delta\,a_-\right)^{2} \label{eq:omega-mid-22} \\
  \omega^\mathrm{NR}_{33}(t_{33}^\mathrm{mid}) ={}& -1.8684\,\nu + 6.6004\,\nu^{3} + 0.015143\,\nu\,\ln\nu - 0.36436\,\sqrt{1-4\nu} + 1.3963\,\nu\,\sqrt{1-4\nu} + \frac{0.077347}{1 + \nu} \nonumber\\*
  &- 0.071745\,a_+ - 0.043009\,a_+^{3}\,\delta\,a_- - 0.15385\,a_+\,\left(\delta\,a_-\right)^{2} - 0.029612\,a_+^{2} + 0.18596\,\delta\,a_- \nonumber\\*
  &- 0.076745\,\left(\delta\,a_-\right)^{4} + 0.71121\,\nu\,\left(\delta\,a_-\right)^{3} + 0.055573\,a_+\,\nu\,\ln\nu - 0.069037\,a_+\,\delta\,a_- - 0.063481\,\nu\,a_+^{3} \nonumber\\*
  &- 3.9445\,\nu^{2}\,\delta\,a_- - 0.25352\,\delta\,a_-\,\sqrt{1-4\nu} - 2.6219\,\nu^{2}\,a_+\,\delta\,a_- \nonumber\\*
  &+ 0.61420\,\nu\,a_+\,\delta\,a_- + 3.1671\,\nu^{3}\,\delta\,a_- \label{eq:omega-mid-33} \\
  \omega^\mathrm{NR}_{21}(t_{21}^\mathrm{mid}) ={}& -1.0561\,\nu + 9.1371\,\nu^{4} - 0.11303\,\nu\,\ln\nu - 0.91829\,\nu^{2}\,\ln\nu - 0.10397\,\sqrt{1-4\nu} + 0.46778\,\nu\,\sqrt{1-4\nu} \nonumber\\*
  &- 0.033399\,a_+ - 0.020776\,a_+^{3}\,\delta\,a_- - 0.037016\,a_+\,\left(\delta\,a_-\right)^{2} - 0.012293\,a_+^{2} \nonumber\\*
  &- 0.0099109\,\delta\,a_- - 0.044019\,\left(\delta\,a_-\right)^{4} \label{eq:omega-mid-21} \\
  \omega^\mathrm{NR}_{44}(t_{44}^\mathrm{mid}) ={}& -0.61339 + 7.4227\,\nu^{3} - 24.710\,\nu^{4} - 0.55626\,\nu\,\ln\nu + 0.20879\,\sqrt{1-4\nu} - 0.82290\,\nu\,\sqrt{1-4\nu} \nonumber\\*
  &- 0.11439\,a_+ - 0.063545\,a_+^{2} + 0.050931\,\delta\,a_- - 0.061182\,a_+\,\delta\,a_- + 0.0016931\,a_+^{2}\,\delta\,a_- + 0.083023\,\nu\,a_+^{3} \nonumber\\*
  &- 0.11795\,a_+\,\left(\delta\,a_-\right)^{3} - 0.12969\,\nu^{2}\,a_+ - 0.0062830\,a_+^{4} - 0.90573\,\nu\,a_+\,\left(\delta\,a_-\right)^{2} - 0.15409\,\left(\delta\,a_-\right)^{3} \nonumber\\*
  &- 1.9152\,\nu^{2}\,\delta\,a_- + 1.4756\,\nu\,\left(\delta\,a_-\right)^{3} - 2.3142\,\nu^{2}\,a_+\,\delta\,a_- + 0.10166\,\nu\,a_+^{2} - 0.11997\,\delta\,a_-\,\sqrt{1-4\nu} \nonumber\\*
  &+ 0.47960\,\nu\,a_+\,\delta\,a_- - 0.56249\,\nu^{2}\,\left(\delta\,a_-\right)^{2} - 0.034904\,a_+^{3} \label{eq:omega-mid-44} \\
  \omega^\mathrm{NR}_{55}(t_{55}^\mathrm{mid}) ={}& -3.9743\,\nu + 14.930\,\nu^{3} + 0.071991\,\nu\,\ln\nu - 0.82984\,\sqrt{1-4\nu} + 3.1583\,\nu\,\sqrt{1-4\nu} + \frac{0.35132}{1 + \nu} \nonumber\\*
  &- 0.15034\,a_+ - 0.065628\,a_+^{3}\,\delta\,a_- - 0.19515\,a_+\,\left(\delta\,a_-\right)^{2} - 0.049561\,a_+^{2} + 0.16943\,\delta\,a_- - 0.13147\,\left(\delta\,a_-\right)^{4} \nonumber\\*
  &+ 1.2443\,\nu\,\left(\delta\,a_-\right)^{3} - 5.7767\,\nu^{2}\,\delta\,a_- - 0.11374\,a_+\,\delta\,a_- - 0.11043\,\nu\,a_+^{3} - 0.24538\,\delta\,a_-\,\sqrt{1-4\nu} \nonumber\\*
  &- 4.1008\,\nu^{2}\,a_+\,\delta\,a_- + 0.95907\,\nu\,a_+\,\delta\,a_- + 8.9787\,\nu^{3}\,\delta\,a_- - 0.075357\,\left(\delta\,a_-\right)^{3} \label{eq:omega-mid-55} \\
  \omega^\mathrm{NR}_{32}(t_{32}^\mathrm{mid}) ={}& -2.0847\,\nu^{3} + 6.3536\,\nu^{4} - 0.080432\,\nu\,\ln\nu + 0.073183\,\sqrt{1-4\nu} - 0.26546\,\nu\,\sqrt{1-4\nu} - \frac{0.27353}{1 + \nu} \nonumber\\*
  &+ 0.018948\,a_+ - 0.051055\,a_+^{2} + 0.15904\,\delta\,a_-\,\sqrt{1-4\nu} + 0.014306\,a_+\,\delta\,a_- + 0.034352\,a_+\,\left(\delta\,a_-\right)^{2}\,a_-^{2} \nonumber\\*
  &+ 0.045053\,a_+^{4} + 1.7813\,\nu^{2}\,a_+ - 0.076572\,a_+^{4}\,\delta\,a_- + 61.330\,\nu^{4}\,\delta\,a_- + 0.29357\,a_+\,\nu\,\ln\nu - 0.26020\,\delta\,a_- \nonumber\\*
  &+ 0.028829\,a_+^{5} - 0.11813\,\nu\,a_+^{3} - 0.066396\,\left(\delta\,a_-\right)^{5} - 0.0085945\,\delta\,a_-\,a_-^{2} - 11.863\,\nu^{3}\,\delta\,a_- \nonumber\\*
  &- 0.077381\,\left(\delta\,a_-\right)^{4} + 0.76524\,\nu\,\left(\delta\,a_-\right)^{4} - 0.29478\,\nu\,a_+\,\delta\,a_- - 0.030728\,a_+\,\sqrt{1-4\nu} - 24.466\,\nu^{4}\,a_+ \nonumber\\*
  &- 0.49556\,\delta\,a_-\,\nu\,\ln\nu + 0.18903\,\nu\,a_+^{2} - 0.26587\,\nu\,a_+^{4} - 0.070963\,a_+^{2}\,\left(\delta\,a_-\right)^{2} \label{eq:omega-mid-32} \\
  \omega^\mathrm{NR}_{43}(t_{43}^\mathrm{mid}) ={}& 1.6847 + 26.985\,\nu + 11.276\,\nu\,\ln\nu + 55.768\,\nu^{2}\,\ln\nu - 1.6904\,\sqrt{1-4\nu} + 6.6562\,\nu\,\sqrt{1-4\nu} - 0.093256\,a_+ \nonumber\\*
  &- 0.028529\,a_+^{2} - 0.044172\,\delta\,a_- - 0.043568\,a_+\,\delta\,a_- - 0.043179\,a_+^{3}\,a_-^{2} - 0.25903\,\nu^{2}\,a_+^{3} \nonumber\\*
  &- 0.31507\,\nu\,a_+^{3}\,\delta\,a_- - 0.048473\,\left(\delta\,a_-\right)^{4} - 0.031433\,a_+\,\left(\delta\,a_-\right)^{2} \label{eq:omega-mid-43}
\end{align}
\end{subequations}
\end{widetext}

\begin{widetext}
\begin{subequations}
\label{eq:omegadot_mid}
\allowdisplaybreaks
\begin{align}
  \dot{\omega}^\mathrm{NR}_{22}(t_{22}^\mathrm{mid}) ={}& 0.021774\,\nu - 0.016224\,\nu^{3} + 0.011163\,\nu\,\ln\nu + 0.049571\,\nu^{2}\,\ln\nu - 0.0013725\,\sqrt{1-4\nu} \nonumber\\*
  &+ 0.0053724\,\nu\,\sqrt{1-4\nu} - 0.00011190\,a_+ - 0.000088542\,a_+^{2} + 0.0024647\,\nu^{2}\,\delta\,a_- \nonumber\\*
  &+ 0.000018005\,a_+\,\sqrt{1-4\nu} - 0.00017298\,a_+^{3} - 0.0022238\,\nu\,a_+\,\delta\,a_- - 0.0039494\,\nu\,\delta\,a_- - 0.0021447\,\nu\,a_+ \nonumber\\*
  &- 0.00094949\,\nu\,a_+^{2} - 0.0015348\,\nu\,\left(\delta\,a_-\right)^{2} - 0.00028716\,\left(\delta\,a_-\right)^{3} \label{eq:omegadot-mid-22} \\
  \dot{\omega}^\mathrm{NR}_{33}(t_{33}^\mathrm{mid}) ={}& -0.48161\,\nu - 0.12521\,\nu^{3} + 1.2986\,\nu^{4} - 0.13430\,\nu\,\ln\nu - 0.81276\,\nu^{2}\,\ln\nu - \frac{0.0054005}{1 + \nu} + 0.00089285\,a_+ \nonumber\\*
  &- 0.000024754\,a_+^{2} - 0.11163\,\nu^{2}\,\delta\,a_- - 0.00036478\,a_+^{3} - 0.0056149\,\nu\,a_+\,\delta\,a_- + 0.0053203\,a_+\,\nu\,\ln\nu \nonumber\\*
  &- 0.010821\,\delta\,a_-\,\sqrt{1-4\nu} + 0.0096968\,\delta\,a_- - 0.0017202\,\nu\,a_+^{2} + 0.0044745\,\delta\,a_-\,\nu\,\ln\nu \label{eq:omegadot-mid-33} \\
  \dot{\omega}^\mathrm{NR}_{21}(t_{21}^\mathrm{mid}) ={}& 0.031398 - 0.28023\,\nu^{2} + 2.3905\,\nu^{4} + 0.072759\,\nu\,\ln\nu - 0.029679\,\sqrt{1-4\nu} \nonumber\\*
  &+ 0.12164\,\nu\,\sqrt{1-4\nu} + 0.0011719\,a_+\,\nu\,\ln\nu \label{eq:omegadot-mid-21} \\
  \dot{\omega}^\mathrm{NR}_{44}(t_{44}^\mathrm{mid}) ={}& 0.0095804\,\nu + 0.027390\,\nu^{3} + 0.012555\,\nu\,\ln\nu + 0.048173\,\nu^{2}\,\ln\nu - 0.0029933\,\sqrt{1-4\nu} \nonumber\\*
  &+ 0.0099116\,\nu\,\sqrt{1-4\nu} - 0.0012966\,a_+ - 0.00074615\,a_+^{2} - 0.037331\,\nu^{2}\,\delta\,a_- \nonumber\\*
  &- 0.00037059\,a_+^{3} + 0.00030141\,a_+\,\sqrt{1-4\nu} \label{eq:omegadot-mid-44} \\
  \dot{\omega}^\mathrm{NR}_{55}(t_{55}^\mathrm{mid}) ={}& 0.90479 - 0.50811\,\nu + 0.098982\,\nu\,\ln\nu + 0.033613\,\sqrt{1-4\nu} - 0.12329\,\nu\,\sqrt{1-4\nu} - \frac{0.93873}{1 + \nu} \nonumber\\*
  &- 0.0017041\,a_+ - 0.00078887\,a_+^{2} - 0.035314\,\nu^{2}\,\delta\,a_- \label{eq:omegadot-mid-55} \\
  \dot{\omega}^\mathrm{NR}_{32}(t_{32}^\mathrm{mid}) ={}& 0.26200 + 0.076032\,\nu\,\ln\nu + 0.27346\,\nu^{2}\,\ln\nu + 0.0064829\,\sqrt{1-4\nu} - 0.021912\,\nu\,\sqrt{1-4\nu} - \frac{0.26846}{1 + \nu} \nonumber\\*
  &- 0.019833\,\nu^{2}\,a_+ - 0.00096346\,a_+^{4} - 0.0014422\,a_+\,\left(\delta\,a_-\right)^{2} + 0.0015487\,\nu\,a_+^{3} - 0.0061772\,\nu^{2}\,\left(\delta\,a_-\right)^{2} \nonumber\\*
  &+ 0.0021892\,\nu\,\delta\,a_-\,a_-^{2} + 0.00038244\,a_+^{2} + 0.00013456\,a_-^{4} + 0.031593\,\nu^{2}\,a_+\,\delta\,a_- \nonumber\\*
  &+ 0.0014532\,a_+^{3}\,\delta\,a_- - 0.013666\,\nu\,a_+\,\delta\,a_- \label{eq:omegadot-mid-32} \\
  \dot{\omega}^\mathrm{NR}_{43}(t_{43}^\mathrm{mid}) ={}& 0.073933 + 1.0973\,\nu^{3} + 0.13452\,\nu\,\ln\nu + 0.56909\,\nu^{2}\,\ln\nu - 0.073807\,\sqrt{1-4\nu} + 0.29096\,\nu\,\sqrt{1-4\nu} \nonumber\\*
  &+ 0.0029848\,a_+\,\nu\,\ln\nu - 0.00043113\,a_+^{4} - 0.0020758\,\nu\,\delta\,a_- - 0.046851\,\nu^{2}\,a_+\,\delta\,a_- \nonumber\\*
  &- 0.0016437\,\nu\,a_+^{3} + 0.025637\,\nu^{2}\,\left(\delta\,a_-\right)^{2} \label{eq:omegadot-mid-43}
\end{align}
\end{subequations}
\end{widetext}

\subsection{Input Values at \texorpdfstring{$t^\mathrm{inter}_{\ell m}$}{t inter lm}}

\begin{widetext}
\begin{subequations}
\label{eq:h_inter}
\allowdisplaybreaks
\begin{align}
  \frac{h^\mathrm{NR}_{22}(t_{22}^\mathrm{inter})}{\nu} ={}& 18.890\,\nu - 82.886\,\nu^{4} + 3.6864\,\nu\,\ln\nu + 24.978\,\nu^{2}\,\ln\nu + 1.0971\,\sqrt{1-4\nu} - 4.2901\,\nu\,\sqrt{1-4\nu} \nonumber\\*
  &+ 0.35244\,a_+ + 1.1639\,a_+\,\nu\,\ln\nu + 0.16147\,a_+\,\delta\,a_- + 0.022942\,a_+^{2} - 14.124\,\nu^{3}\,\delta\,a_- + 0.66414\,\nu\,a_+ \nonumber\\*
  &+ 4.1749\,\nu^{2}\,\delta\,a_- - 2.0972\,\nu\,a_+\,\delta\,a_- + 0.080763\,a_+^{2}\,\left(\delta\,a_-\right)^{2} + 6.7271\,\nu^{2}\,a_+\,\delta\,a_- + 0.31914\,\nu\,a_+^{2}\,\delta\,a_- \nonumber\\*
  &- 0.23147\,\left(\delta\,a_-\right)^{4} + 0.072146\,\left(\delta\,a_-\right)^{2} \label{eq:hsign-inter-22} \\
  \frac{h^\mathrm{NR}_{33}(t_{33}^\mathrm{inter})}{\nu} ={}& 0.0079323\,|\chi_{33}| + 0.0093326\,|\chi_{33}|\,a_+^{2} + \delta\,\Big[-119.13 + 87.448\,\nu - 94.917\,\nu^{4} - 2.8993\,\sqrt{1-4\nu} \nonumber\\*
  &\quad + 27.818\,\nu\,\sqrt{1-4\nu} + \frac{122.37}{1 + \nu} + 0.049108\,a_+ + 0.090744\,\chi_{33} - 18.138\,\nu^{2}\,\chi_{33}^{2} - 6.1508\,\nu\,\chi_{33}\,\delta \nonumber\\*
  &\quad - 10.698\,\nu\,\chi_{33}^{2}\,\delta + 0.042038\,\chi_{33}\,\delta\,a_+^{2} + 0.26898\,\nu^{2}\,\chi_{33}\,a_+ + 18.601\,\nu\,\chi_{33}^{2} + 0.10423\,\chi_{33}\,\delta^{2}\,a_+ \nonumber\\*
  &\quad + 0.038679\,\chi_{33}^{4} + 0.020378\,\nu\,\delta\,a_+ + 0.044905\,\chi_{33}^{3}\,a_+ + 4.6730\,\nu\,\chi_{33}\,\delta^{2} + 14.897\,\nu^{2}\,\chi_{33}\,\delta \nonumber\\*
  &\quad - 0.089580\,\chi_{33}\,\nu\,\ln\nu + 0.048359\,\nu\,\chi_{33}^{3} - 61.164\,\nu^{2}\,\delta + 0.018052\,a_+^{4} - 0.087643\,\chi_{33}^{2}\,a_+^{2} \nonumber\\*
  &\quad - 0.11389\,\nu^{2}\,a_+^{2} - 3.7100\,\chi_{33}^{2} + 3.6247\,\chi_{33}^{2}\,\delta - 0.012392\,\chi_{33}^{2}\,a_+\Big] \label{eq:hsign-inter-33} \\
  \frac{h^\mathrm{NR}_{21}(t_{21}^\mathrm{inter})}{\nu} ={}& 0.066888\,|\chi_{33}| + 0.013578\,|\chi_{33}|\,a_+ + \delta\,\Big[-8.0454 - 2.2835\,\nu + 36.635\,\nu^{3} - 1.9331\,\sqrt{1-4\nu} \nonumber\\*
  &\quad + 8.3664\,\nu\,\sqrt{1-4\nu} + \frac{10.165}{1 + \nu} + 2.3987\,\nu^{2}\,\chi_{33} - 3.1798\,\nu\,\chi_{33}^{2} - 15.076\,\nu\,\chi_{33}^{4} - 74.885\,\nu^{2}\,\chi_{33}^{3} \nonumber\\*
  &\quad + 4.7309\,\nu^{3}\,\delta\,a_+ + 0.13221\,\chi_{33}^{4}\,a_+ + 1.5396\,\nu\,\chi_{33}^{3}\,a_+ + 49.603\,\nu^{3}\,\chi_{33}\,\delta - 16.647\,\nu\,\chi_{33}^{3}\,\delta \nonumber\\*
  &\quad + 22.594\,\nu\,\chi_{33}^{3} - 0.80887\,\chi_{33} - 6.5792\,\nu^{3}\,a_+^{2} - 4.7959\,\chi_{33}^{4}\,\delta - 5.1428\,\nu\,\chi_{33}^{2}\,\delta^{2} + 5.2800\,\chi_{33}^{4} \nonumber\\*
  &\quad - 5.3413\,\chi_{33}^{2}\,\delta^{3} - 0.040710\,\chi_{33}\,a_+ + 4.8822\,\chi_{33}^{2}\,\delta^{2} - 0.52096\,\chi_{33}^{3}\,\delta^{2} + 0.72164\,\nu\,a_+^{4} - 1.2691\,\nu\,\chi_{33}\,a_+^{3} \nonumber\\*
  &\quad - 1.3365\,\nu^{2}\,\chi_{33}\,a_+^{2} + 0.88881\,\chi_{33}\,\sqrt{1-4\nu}\Big] \label{eq:hsign-inter-21} \\
  \frac{h^\mathrm{NR}_{44}(t_{44}^\mathrm{inter})}{\nu} ={}& 2.8355\,\nu - 14.970\,\nu^{4} + 0.77538\,\nu\,\ln\nu + 4.0808\,\nu^{2}\,\ln\nu + 0.14269\,\sqrt{1-4\nu} - 0.57582\,\nu\,\sqrt{1-4\nu} \nonumber\\*
  &+ 0.052021\,\left(\delta\,a_-\right)^{3} + 0.086936\,\left(\delta\,a_-\right)^{4} + 0.10412\,\delta\,a_-\,\sqrt{1-4\nu} - 0.22682\,\nu\,\left(\chi_{44A}\right)^{3} \nonumber\\*
  &- 0.025239\,\chi_{44A}\,\sqrt{1-4\nu} + 0.27284\,\chi_{44A}\,\nu\,\ln\nu + 1.4876\,\nu^{3}\,\delta\,a_- + 0.29851\,\nu\,\chi_{44A} \label{eq:hsign-inter-44} \\
  \frac{h^\mathrm{NR}_{55}(t_{55}^\mathrm{inter})}{\nu} ={}& 0.00068397\,|\chi_{33}| + 0.0015136\,|\chi_{33}|\,a_+^{2} + \delta\,\Big[-37.173 + 42.240\,\nu - 26.272\,\nu^{2} + 0.84793\,\nu\,\ln\nu \nonumber\\*
  &\quad + 10.161\,\nu^{2}\,\ln\nu + \frac{36.810}{1 + \nu} + 0.75193\,\nu\,\delta\,a_+ + 0.0080581\,\nu\,\chi_{33}^{3} + 0.14056\,a_+ + 0.0074395\,\delta^{2}\,a_+^{2} \nonumber\\*
  &\quad + 0.64984\,a_+\,\nu\,\ln\nu + 0.17658\,\nu\,\chi_{33} + 4.2552\,\nu^{2}\,\chi_{33}^{2} + 0.0051107\,\chi_{33}^{4} - 0.043893\,\chi_{33}^{2}\,\delta\,a_+ \nonumber\\*
  &\quad - 0.079576\,\delta\,a_+ - 0.13859\,\nu^{2}\,\chi_{33}\,\delta - 2.2099\,\nu^{3}\,\chi_{33} - 1.5251\,\nu^{2}\,\delta\,a_+ + 0.043300\,\delta + 1.5573\,\nu\,\chi_{33}^{2}\,\delta \nonumber\\*
  &\quad - 1.1121\,\nu\,\chi_{33}^{2} + 0.41260\,\chi_{33}^{2}\,\delta^{2} + 0.0015193\,\chi_{33}\,a_+ - 0.22273\,\nu\,\chi_{33}^{2}\,a_+ + 0.060220\,\chi_{33}^{2}\,a_+ \nonumber\\*
  &\quad + 0.0033920\,\chi_{33}\,a_+^{2} - 0.38735\,\chi_{33}^{2}\,\delta + 0.38452\,\delta^{3} + 0.35677\,\nu\,a_+ - 0.12466\,\nu\,\chi_{33}\,\delta\Big] \label{eq:hsign-inter-55} \\
  \frac{h^\mathrm{NR}_{32}(t_{32}^\mathrm{inter})}{\nu} ={}& 0.087270 - 0.30372\,\nu^{3} + 0.15275\,\nu\,\ln\nu + 0.18978\,\nu^{2}\,\ln\nu - 0.024661\,\sqrt{1-4\nu} + 0.086763\,\nu\,\sqrt{1-4\nu} \nonumber\\*
  &+ 0.052289\,a_+ + 0.012365\,\left(\delta\,a_-\right)^{3} + 0.021402\,a_+^{3}\,\delta\,a_- - 0.079004\,\nu\,\left(\delta\,a_-\right)^{3} + 0.73433\,\nu^{2}\,a_+^{2} \nonumber\\*
  &+ 0.27281\,a_+\,\nu\,\ln\nu - 0.0094307\,\delta\,a_-\,\sqrt{1-4\nu} + 0.072797\,\left(\delta\,a_-\right)^{4} + 0.030960\,a_+^{2} - 0.27517\,\nu\,a_+^{2} \nonumber\\*
  &- 0.032086\,\nu\,a_+^{3} - 0.085612\,a_+\,\left(\delta\,a_-\right)^{3} + 0.41826\,\nu\,a_+ - 0.69132\,\nu^{2}\,a_+ \nonumber\\*
  &- 0.41443\,\nu^{3}\,\delta\,a_- + 0.0094267\,a_+^{3} \label{eq:hsign-inter-32} \\
  \frac{h^\mathrm{NR}_{43}(t_{43}^\mathrm{inter})}{\nu} ={}& 0.0034107\,|\chi_{33}| + 0.0012474\,|\chi_{33}|\,a_+ + 0.00027535\,|\chi_{33}|\,a_+^{2} + 0.000042530\,|\chi_{33}|^{3} + \delta\,\Big[0.73179 \nonumber\\*
  &\quad + 7.3071\,\nu^{3} + 1.2331\,\nu\,\ln\nu + 4.6951\,\nu^{2}\,\ln\nu - 0.67804\,\sqrt{1-4\nu} + 2.4900\,\nu\,\sqrt{1-4\nu} + 0.032451\,a_+ \nonumber\\*
  &\quad + 0.42066\,\nu^{2}\,\chi_{33} - 0.72141\,\nu^{2}\,\chi_{33}^{2} + 0.0068302\,a_+^{4} + 0.015522\,\chi_{33}^{4} - 0.061616\,\chi_{33}^{2}\,\delta - 0.48166\,\nu\,\delta^{2}\,a_+ \nonumber\\*
  &\quad + 2.4265\,\nu^{2}\,\chi_{33}\,a_+ + 0.37366\,\nu\,\chi_{33}\,\delta\,a_+ + 0.00038300\,\chi_{33}^{2}\,a_+^{2} - 0.088691\,\nu\,\chi_{33}^{2}\,a_+ + 0.0048665\,a_+^{3} \nonumber\\*
  &\quad + 0.028123\,\nu\,\chi_{33}^{3} - 0.16477\,\nu^{2}\,\chi_{33}\,\delta + 0.14564\,\nu\,\delta\,a_+ - 1.9608\,\nu^{3}\,a_+ + 0.021591\,\chi_{33}^{2} \nonumber\\*
  &\quad + 0.019621\,\chi_{33}\,\delta\,a_+^{2} + 0.044222\,\chi_{33}\,\delta\,a_+ - 0.72883\,\nu\,\chi_{33}\,a_+ - 0.014020\,\chi_{33}\,\delta^{3} - 0.013663\,\delta\,a_+^{3} \nonumber\\*
  &\quad - 2.6695\,\nu^{3}\,\chi_{33} + 0.14768\,\nu\,\chi_{33}^{2}\,\delta - 0.010471\,\chi_{33}\,a_+^{3}\Big] \label{eq:hsign-inter-43}
\end{align}
\end{subequations}
\end{widetext}

\begin{widetext}
\begin{subequations}
\label{eq:omega_inter}
\allowdisplaybreaks
\begin{align}
  \omega^\mathrm{NR}_{22}(t_{22}^\mathrm{inter}) ={}& -0.19326\,\nu + 1.4901\,\nu^{3} - 2.8806\,\nu^{4} - 0.10073\,\nu\,\ln\nu + 0.088827\,\nu^{2}\,\ln\nu - \frac{0.15072}{1 + \nu} - 0.054263\,a_+ \nonumber\\*
  &+ 0.0051682\,\delta\,a_-\,\sqrt{1-4\nu} - 0.013237\,a_+^{2} - 0.038800\,a_+\,\delta\,a_- - 0.013617\,a_+^{2}\,\delta\,a_- + 2.2600\,\nu^{3}\,\delta\,a_- \nonumber\\*
  &- 0.0087682\,a_+^{2}\,\left(\delta\,a_-\right)^{2} + 0.22225\,\nu\,\delta\,a_- + 0.23197\,\nu\,a_+\,\delta\,a_- - 0.0071763\,a_+^{3} - 0.043757\,a_+\,\nu\,\ln\nu \nonumber\\*
  &- 0.018862\,a_+\,\left(\delta\,a_-\right)^{2} - 0.0080416\,a_+^{3}\,\delta\,a_- + 0.027427\,\nu\,a_+^{2}\,\delta\,a_- - 0.52796\,\nu^{2}\,a_+\,\delta\,a_- \nonumber\\*
  &- 0.16173\,\nu^{2}\,\left(\delta\,a_-\right)^{2} + 0.038978\,\nu^{2}\,a_+^{2} - 0.0057754\,\left(\delta\,a_-\right)^{4} + 0.019695\,\nu\,a_+^{3} - 1.1903\,\nu^{2}\,\delta\,a_- \nonumber\\*
  &- 0.036675\,\delta\,a_- + 0.038403\,\nu\,a_+\,\left(\delta\,a_-\right)^{2} \label{eq:omega-inter-22} \\
  \omega^\mathrm{NR}_{33}(t_{33}^\mathrm{inter}) ={}& (-19.276\,\nu^{2} + 47.069\,\nu^{3} - 42.702\,\nu^{4} - 0.28656\,\nu\,\ln\nu - 9.6886\,\nu^{2}\,\ln\nu \nonumber\\*
  &\quad + \frac{1.5009}{1 + \nu} + 0.034386\,\nu^{2}\,a_+)\,\omega^\mathrm{NR}_{22}(t_{22}^\mathrm{inter}) \label{eq:omega-inter-33} \\
  \omega^\mathrm{NR}_{21}(t_{21}^\mathrm{inter}) ={}& -1.3063\,\nu + 7.4667\,\nu^{4} - 0.27020\,\nu\,\ln\nu - 1.5893\,\nu^{2}\,\ln\nu - 0.080299\,\sqrt{1-4\nu} + 0.33048\,\nu\,\sqrt{1-4\nu} \nonumber\\*
  &- 0.033877\,a_+ - 0.067042\,\left(\delta\,a_-\right)^{3} - 0.12548\,a_+^{2}\,\left(\delta\,a_-\right)^{2} - 0.62250\,\nu^{2}\,a_+^{2} - 0.039290\,a_+\,\nu\,\ln\nu \nonumber\\*
  &+ 3.1434\,\nu^{3}\,\delta\,a_- + 0.49949\,\nu\,\left(\delta\,a_-\right)^{3} - 0.89743\,\nu^{2}\,\delta\,a_- - 0.0080491\,a_+^{2}\,\delta\,a_- - 0.028903\,a_+^{2} \nonumber\\*
  &+ 0.24199\,\nu\,a_+^{2} - 0.19871\,\left(\delta\,a_-\right)^{4} + 0.28122\,a_+\,\left(\delta\,a_-\right)^{3} \label{eq:omega-inter-21} \\
  \omega^\mathrm{NR}_{44}(t_{44}^\mathrm{inter}) ={}& (2.1830 - 1.0192\,\nu + 0.76571\,\nu^{2} + 4.3651\,\nu^{4} - 0.18174\,\sqrt{1-4\nu} + 0.71288\,\nu\,\sqrt{1-4\nu} \nonumber\\*
  &\quad + 0.0016974\,a_+ - 0.010905\,\delta\,a_-\,\nu\,\ln\nu)\,\omega^\mathrm{NR}_{22}(t_{22}^\mathrm{inter}) \label{eq:omega-inter-44} \\
  \omega^\mathrm{NR}_{55}(t_{55}^\mathrm{inter}) ={}& (0.40102\,\nu - 0.38953\,\nu^{2} + 15.981\,\nu^{4} - 0.44682\,\sqrt{1-4\nu} + 1.7064\,\nu\,\sqrt{1-4\nu} \nonumber\\*
  &\quad + \frac{2.9481}{1 + \nu} + 0.0041209\,a_+)\,\omega^\mathrm{NR}_{22}(t_{22}^\mathrm{inter}) \label{eq:omega-inter-55} \\
  \omega^\mathrm{NR}_{32}(t_{32}^\mathrm{inter}) ={}& 1.1016\,\nu - 0.43372\,\nu^{3} - 0.49125\,\nu^{4} + 0.28894\,\nu\,\ln\nu + 2.0934\,\nu^{2}\,\ln\nu - \frac{0.14254}{1 + \nu} - 0.060570\,a_+ \nonumber\\*
  &- 0.024546\,\delta\,a_-\,\sqrt{1-4\nu} - 0.021057\,a_+\,\delta\,a_- - 0.017652\,a_+^{2} - 0.034997\,a_+^{2}\,\delta\,a_- - 0.033252\,a_+^{2}\,\left(\delta\,a_-\right)^{2} \nonumber\\*
  &- 0.12067\,\nu^{2}\,a_+^{3} + 2.6507\,\nu^{4}\,\delta\,a_- + 0.12327\,\nu\,a_+^{2}\,\delta\,a_- + 0.037412\,\nu\,a_+^{2} - 0.059815\,a_+\,\nu\,\ln\nu \nonumber\\*
  &+ 0.0029123\,a_+\,\sqrt{1-4\nu} + 0.13795\,\nu\,a_+^{2}\,\left(\delta\,a_-\right)^{2} - 0.0015795\,a_+^{4} + 0.0045590\,a_+^{5} \nonumber\\*
  &- 1.4839\,\nu^{3}\,\delta\,a_- - 0.011831\,a_+^{4}\,\delta\,a_- \label{eq:omega-inter-32} \\
  \omega^\mathrm{NR}_{43}(t_{43}^\mathrm{inter}) ={}& 0.63290\,\nu + 0.43898\,\nu^{3} + 0.14576\,\nu\,\ln\nu + 1.5509\,\nu^{2}\,\ln\nu + 0.042963\,\nu\,\sqrt{1-4\nu} - \frac{0.21950}{1 + \nu} \nonumber\\*
  &- 0.082836\,a_+ - 0.040756\,\delta\,a_-\,\sqrt{1-4\nu} - 0.058127\,a_+\,\delta\,a_- - 0.016566\,a_+^{2} - 0.025055\,a_+^{2}\,\delta\,a_- \nonumber\\*
  &- 5.4559\,\nu^{4}\,\delta\,a_- - 0.021336\,a_+^{2}\,\left(\delta\,a_-\right)^{2} - 0.067039\,a_+\,\nu\,\ln\nu + 0.26616\,\nu\,a_+\,\delta\,a_- - 0.030124\,a_+^{2}\,\left(\delta\,a_-\right)^{3} \nonumber\\*
  &- 0.0030618\,a_+^{3} - 0.010088\,a_+^{3}\,\delta\,a_- - 0.54636\,\nu^{2}\,a_+\,\delta\,a_- \label{eq:omega-inter-43}
\end{align}
\end{subequations}
\end{widetext}

\section{Merger-Ringdown Coefficient Fits}
\label{app:mrfits}

In this appendix, we provide fits for the free merger-ringdown coefficients $\{c_{1,f}^{\ell m},\ c_{2,f}^{\ell m},\ d_{1,f}^{\ell m},\ d_{2,f}^{\ell m}\}$ of Eqs.~\eqref{eq:MergerRingdownAnsatzAmp} and \eqref{eq:MergerRingdownAnsatzPhase}, produced using the methodology described in Sec.~\ref{sec:Calibration_Fitting}. 
For the $(\ell, m) = (3,2), (4,3)$ modes, these coefficients describe the spheroidal $(\ell,m,0)$ mode.

\begin{widetext}
\begin{subequations}
\label{eq:c1f}
\allowdisplaybreaks
\begin{align}
  c^{22}_{1,f} ={}& 0.90730\,\nu - 1.2181\,\nu^{3} + 0.24716\,\nu\,\ln\nu + 1.2375\,\nu^{2}\,\ln\nu + 0.0033015\,\nu\,\sqrt{1-4\nu} + \frac{0.093048}{1 + \nu} - 0.0048502\,a_+ \nonumber\\*
  &- 0.0057660\,a_+^{2} - 1.2428\,\nu^{3}\,\delta\,a_- - 0.081966\,\nu\,\left(\delta\,a_-\right)^{2} - 0.0034666\,a_+^{3} - 0.090849\,\nu\,\left(\delta\,a_-\right)^{3} - 0.051188\,\nu\,a_+ \nonumber\\*
  &- 0.0088797\,a_+^{2}\,a_-^{2} - 0.013842\,a_+^{2}\,\delta\,a_- - 0.0048042\,\delta\,a_- + 0.021369\,\nu^{2}\,a_-^{2} \nonumber\\*
  &- 0.039916\,\nu\,a_+\,\delta\,a_- - 0.0017796\,a_+^{4} \label{eq:c1f-22} \\
  c^{33}_{1,f} ={}& -11.367 - 141.37\,\nu^{3} - 19.335\,\nu\,\ln\nu - 80.159\,\nu^{2}\,\ln\nu + 11.003\,\sqrt{1-4\nu} - 41.540\,\nu\,\sqrt{1-4\nu} \nonumber\\*
  &- 0.016868\,a_+ - 0.23172\,\nu\,a_+\,\delta\,a_- \label{eq:c1f-33} \\
  c^{21}_{1,f} ={}& -4.7569\,\nu + 44.286\,\nu^{3} - 65.085\,\nu^{4} - 0.87403\,\sqrt{1-4\nu} + 3.4757\,\nu\,\sqrt{1-4\nu} + \frac{1.0532}{1 + \nu} - 0.19384\,a_+\,\left(\delta\,a_-\right)^{2} \nonumber\\*
  &+ 0.32441\,a_+\,\delta\,a_- + 0.059166\,a_+\,\sqrt{1-4\nu} + 0.60186\,\nu^{2}\,\delta\,a_- - 0.43661\,a_+ - 1.1282\,a_+\,\nu\,\ln\nu \nonumber\\*
  &+ 0.038426\,a_+^{2} - 1.1471\,\nu\,a_+\,\delta\,a_- \label{eq:c1f-21} \\
  c^{44}_{1,f} ={}& 340.37\,\nu^{2} + 46.430\,\nu\,\ln\nu + 407.86\,\nu^{2}\,\ln\nu - 37.196\,\sqrt{1-4\nu} + 144.29\,\nu\,\sqrt{1-4\nu} + \frac{37.868}{1 + \nu} + 0.13621\,a_+\,\nu\,\ln\nu \nonumber\\*
  &- 2.4241\,\nu^{2}\,\delta\,a_- - 4.1921\,\nu^{2}\,a_+\,\delta\,a_- \label{eq:c1f-44} \\
  c^{55}_{1,f} ={}& -24.744 - 420.23\,\nu - 162.91\,\nu\,\ln\nu - 847.96\,\nu^{2}\,\ln\nu + 21.018\,\sqrt{1-4\nu} - 82.840\,\nu\,\sqrt{1-4\nu} \label{eq:c1f-55} \\
  c^{32}_{1,f} ={}& \exp\Big[2426.8\,\nu^{3} + 223.41\,\nu\,\ln\nu + 1019.0\,\nu^{2}\,\ln\nu - 155.17\,\sqrt{1-4\nu} + 605.29\,\nu\,\sqrt{1-4\nu} + \frac{157.87}{1 + \nu} + 90.314\,\nu^{2}\,a_+ \nonumber\\*
  &\quad - 2.3500\,a_+^{3} - 5.2904\,\nu\,a_-^{2} + 36.323\,\nu\,\left(\delta\,a_-\right)^{2} - 4.3137\,a_+\,\left(\delta\,a_-\right)^{2} - 11.372\,\nu\,a_+ - 3.3346\,a_+\,\delta\,a_- \nonumber\\*
  &\quad + 6.5583\,a_+^{2}\,\delta\,a_- + 1.6897\,\delta\,a_-\,\sqrt{1-4\nu} + 15.142\,\nu\,a_+\,\delta\,a_- - 1.1385\,a_+\,\sqrt{1-4\nu}\Big] \label{eq:c1f-32} \\
  c^{43}_{1,f} ={}& -180.68\,\nu^{2} + 395.96\,\nu^{4} - 158.07\,\nu^{2}\,\ln\nu + 5.4783\,\sqrt{1-4\nu} - 21.465\,\nu\,\sqrt{1-4\nu} - \frac{4.7725}{1 + \nu} \label{eq:c1f-43}
\end{align}
\end{subequations}
\end{widetext}

\begin{widetext}
\begin{subequations}
\label{eq:c2f}
\allowdisplaybreaks
\begin{align}
  c^{22}_{2,f} ={}& 0.31472 - 4.9486\,\nu^{2} + 49.095\,\nu^{4} + 1.6850\,\nu\,\ln\nu - 0.74617\,\sqrt{1-4\nu} + 2.8554\,\nu\,\sqrt{1-4\nu} - 0.25049\,a_+ \nonumber\\*
  &- 0.32137\,a_+\,\nu\,\ln\nu - 0.16023\,a_+\,\delta\,a_- + 0.23780\,\nu\,a_+^{3} + 3.2331\,\nu^{2}\,\left(\delta\,a_-\right)^{2} + 0.056908\,a_+^{4} - 0.19950\,a_+\,\left(\delta\,a_-\right)^{2} \nonumber\\*
  &+ 0.39021\,\nu\,a_+\,a_-^{2} - 0.036629\,a_+^{2} + 2.3050\,\nu^{2}\,a_+\,\delta\,a_- \label{eq:c2f-22} \\
  c^{33}_{2,f} ={}& 29.016 + 354.63\,\nu^{3} + 49.500\,\nu\,\ln\nu + 202.48\,\nu^{2}\,\ln\nu - 28.178\,\sqrt{1-4\nu} + 105.78\,\nu\,\sqrt{1-4\nu} - 0.70264\,\nu\,a_+ \nonumber\\*
  &+ 20.881\,\nu^{2}\,\delta\,a_- - 2.6264\,\nu\,\delta\,a_- + 0.39691\,\nu\,a_-^{2} \label{eq:c2f-33} \\
  c^{21}_{2,f} ={}& 108.65\,\nu - 132.96\,\nu^{3} + 31.637\,\nu\,\ln\nu + 162.93\,\nu^{2}\,\ln\nu + 0.77611\,\sqrt{1-4\nu} - 4.1625\,\nu\,\sqrt{1-4\nu} \nonumber\\*
  &- 0.22668\,a_+\,\sqrt{1-4\nu} - 3.5186\,\nu\,a_+\,\delta\,a_- - 8.2382\,\nu^{2}\,\delta\,a_- - 0.14705\,a_+^{2} - 0.16508\,a_+^{3} \label{eq:c2f-21} \\
  c^{44}_{2,f} ={}& -374.75\,\nu^{2} - 52.137\,\nu\,\ln\nu - 447.51\,\nu^{2}\,\ln\nu + 41.131\,\sqrt{1-4\nu} - 159.69\,\nu\,\sqrt{1-4\nu} - \frac{42.136}{1 + \nu} - 0.093844\,a_+ \nonumber\\*
  &+ 23.365\,\nu^{3}\,\delta\,a_- + 0.33545\,\nu\,a_+^{3} \label{eq:c2f-44} \\
  c^{55}_{2,f} ={}& 568.04\,\nu - 377.11\,\nu^{3} + 163.71\,\nu\,\ln\nu + 919.90\,\nu^{2}\,\ln\nu + 3.5450\,\sqrt{1-4\nu} \nonumber\\*
  &- 6.6595\,\nu\,\sqrt{1-4\nu} - 0.16989\,a_+^{3} \label{eq:c2f-55} \\
  c^{32}_{2,f} ={}& 46.813 + 941.26\,\nu^{3} + 75.119\,\nu\,\ln\nu + 412.51\,\nu^{2}\,\ln\nu - 46.632\,\sqrt{1-4\nu} + 184.33\,\nu\,\sqrt{1-4\nu} - 0.0049727\,\nu\,a_-^{2} \nonumber\\*
  &+ 93.570\,\nu^{2}\,\delta\,a_- - 15.876\,\nu\,\delta\,a_- + 1.7895\,\left(\delta\,a_-\right)^{3} - 1.2811\,\nu\,a_+^{2} + 23.032\,\nu\,a_+\,\delta\,a_- - 2.4198\,a_+\,\delta\,a_- \nonumber\\*
  &+ 0.85044\,a_+\,\nu\,\ln\nu + 0.43915\,a_+^{3} \label{eq:c2f-32} \\
  c^{43}_{2,f} ={}& -35.285 + 189.08\,\nu - 748.35\,\nu^{3} - 234.35\,\nu^{4} + 35.098\,\sqrt{1-4\nu} - 139.70\,\nu\,\sqrt{1-4\nu} + 0.81857\,\nu\,a_+ \nonumber\\*
  &+ 20.880\,\nu\,a_+\,\delta\,a_- - 2.6380\,a_+\,\delta\,a_- + 15.175\,\nu^{2}\,\delta\,a_- - 0.56203\,\delta\,a_-\,\sqrt{1-4\nu} \label{eq:c2f-43}
\end{align}
\end{subequations}
\end{widetext}

\begin{widetext}
\begin{subequations}
\label{eq:d1f}
\allowdisplaybreaks
\begin{align}
  d^{22}_{1,f} ={}& -0.084907\,\nu + 5.7849\,\nu^{4} + 0.78558\,\nu^{2}\,\ln\nu - 0.14100\,\sqrt{1-4\nu} + 0.55666\,\nu\,\sqrt{1-4\nu} + \frac{0.28816}{1 + \nu} \nonumber\\*
  &- 0.096929\,\nu\,a_+ - 0.0089591\,a_+^{2} - 0.17405\,\nu\,a_+\,\delta\,a_- - 0.13407\,\nu\,a_+^{2}\,\delta\,a_- - 0.056397\,\nu\,a_+^{3} - 1.3772\,\nu^{3}\,\delta\,a_- \nonumber\\*
  &- 0.034965\,a_+\,a_-^{2}\,\delta\,a_- - 0.017544\,a_+\,a_-^{2} + 0.0093562\,a_+ - 0.0099542\,a_+^{4} \label{eq:d1f-22} \\
  d^{33}_{1,f} ={}& 5.2563 + 44.175\,\nu^{3} + 8.7048\,\nu\,\ln\nu + 31.277\,\nu^{2}\,\ln\nu - 4.8202\,\sqrt{1-4\nu} \nonumber\\*
  &+ 17.199\,\nu\,\sqrt{1-4\nu} - 0.067781\,a_+^{3} \label{eq:d1f-33} \\
  d^{21}_{1,f} ={}& -0.18491 - 52.285\,\nu + 131.37\,\nu^{4} - 15.433\,\nu\,\ln\nu - 87.410\,\nu^{2}\,\ln\nu + 0.019010\,\sqrt{1-4\nu} \nonumber\\*
  &- 0.040452\,a_+ - 0.11587\,a_+\,\left(\delta\,a_-\right)^{2} \label{eq:d1f-21} \\
  d^{44}_{1,f} ={}& -8.7383 - 137.83\,\nu^{2} - 10.203\,\nu\,\ln\nu - 161.02\,\nu^{2}\,\ln\nu + 9.1925\,\sqrt{1-4\nu} - 36.764\,\nu\,\sqrt{1-4\nu} - 0.36279\,\nu\,\delta\,a_- \nonumber\\*
  &- 1.0509\,\nu\,a_+\,\delta\,a_- - 0.028716\,a_+^{3} \label{eq:d1f-44} \\
  d^{55}_{1,f} ={}& 158.15 - 172.12\,\nu + 345.76\,\nu^{4} - 112.35\,\nu^{2}\,\ln\nu - 0.13839\,\sqrt{1-4\nu} - \frac{157.55}{1 + \nu} \nonumber\\*
  &- 0.11507\,a_+\,\delta\,a_- + 0.15201\,\nu\,a_-^{2} \label{eq:d1f-55} \\
  d^{32}_{1,f} ={}& \exp\Big[909.79\,\nu + 838.66\,\nu^{2} - 4737.4\,\nu^{4} + 223.21\,\nu\,\ln\nu + 2138.2\,\nu^{2}\,\ln\nu + 1.0429\,\nu\,\sqrt{1-4\nu} \nonumber\\*
  &\quad - 5.3814\,\nu\,a_+^{2} - 0.94027\,a_+^{3}\Big] \label{eq:d1f-32} \\
  d^{43}_{1,f} ={}& 197.22 + 12.471\,\nu^{2} + 62.522\,\nu\,\ln\nu + 229.67\,\nu^{2}\,\ln\nu + 1.4473\,\nu\,\sqrt{1-4\nu} - \frac{195.33}{1 + \nu} + 13.694\,\nu^{2}\,\delta\,a_- \nonumber\\*
  &- 1.8652\,\nu\,\delta\,a_- - 0.058366\,a_+^{2} \label{eq:d1f-43}
\end{align}
\end{subequations}
\end{widetext}

\begin{widetext}
\begin{subequations}
\label{eq:d2f}
\allowdisplaybreaks
\begin{align}
  d^{22}_{2,f} ={}& \exp\Big[-154.07\,\nu + 170.19\,\nu^{3} - 49.845\,\nu\,\ln\nu - 232.76\,\nu^{2}\,\ln\nu + 0.14061\,\nu\,\sqrt{1-4\nu} - \frac{1.1267}{1 + \nu} + 0.42536\,a_+ \nonumber\\*
  &\quad + 0.43051\,\delta\,a_- + 0.41116\,a_+\,\delta\,a_- + 0.25533\,a_+^{2}\,\delta\,a_- + 0.79444\,\nu^{2}\,a_+^{2} \Big] \label{eq:d2f-22} \\
  d^{33}_{2,f} ={}& \exp\Big[314.85 + 3412.2\,\nu^{3} + 529.69\,\nu\,\ln\nu + 2111.1\,\nu^{2}\,\ln\nu - 301.11\,\sqrt{1-4\nu} \nonumber\\*
  &\quad + 1107.4\,\nu\,\sqrt{1-4\nu} + 0.28242\,a_+ \Big] \label{eq:d2f-33} \\
  d^{21}_{2,f} ={}& \exp\Big[-463.35\,\nu^{2} + 4273.9\,\nu^{4} + 140.63\,\nu\,\ln\nu - 69.881\,\sqrt{1-4\nu} + 276.17\,\nu\,\sqrt{1-4\nu} + \frac{76.145}{1 + \nu} + 1.6687\,a_+^{3} \nonumber\\*
  &\quad + 0.68439\,a_+^{2} + 1.6146\,a_+\,a_-^{2} + 14.723\,\nu\,a_+\,\delta\,a_- - 0.33271\,a_+\Big] \label{eq:d2f-21} \\
  d^{44}_{2,f} ={}& \exp\Big[169.72\,\nu - 1881.5\,\nu^{3} - 366.68\,\nu^{2}\,\ln\nu + 59.924\,\sqrt{1-4\nu} - 241.25\,\nu\,\sqrt{1-4\nu} - \frac{55.828}{1 + \nu} \nonumber\\*
  &\quad + 1.1556\,\nu\,a_+ - 0.54641\,a_+\,\delta\,a_- \Big] \label{eq:d2f-44} \\
  d^{55}_{2,f} ={}& \exp\Big[109.36 - 832.41\,\nu^{2} + 5470.4\,\nu^{4} + 225.25\,\nu\,\ln\nu - 100.23\,\sqrt{1-4\nu} \nonumber\\*
  &\quad + 366.49\,\nu\,\sqrt{1-4\nu} - 3.9089\,\delta\,a_-\,\nu\,\ln\nu \Big] \label{eq:d2f-55} \\
  d^{32}_{2,f} ={}& \exp\Big[260.80 + 2734.2\,\nu^{2} + 370.77\,\nu\,\ln\nu + 3473.0\,\nu^{2}\,\ln\nu - 257.05\,\sqrt{1-4\nu} + 1002.9\,\nu\,\sqrt{1-4\nu} - 41.416\,\nu^{2}\,a_+^{2} \nonumber\\*
  &\quad + 36.206\,\nu\,a_+\,a_-^{2} - 69.784\,\nu\,a_+\,\left(\delta\,a_-\right)^{2} \Big] \label{eq:d2f-32} \\
  d^{43}_{2,f} ={}& \exp\Big[568.74\,\nu - 1452.8\,\nu^{2} + 5013.3\,\nu^{4} + 225.02\,\nu\,\ln\nu + 14.200\,\nu\,\sqrt{1-4\nu} + \frac{9.9809}{1 + \nu} + 65.240\,\nu^{2}\,\delta\,a_- \nonumber\\*
  &\quad + 19.020\,\nu^{2}\,a_+ - 1.2962\,a_+^{3}\Big] \label{eq:d2f-43}
\end{align}
\end{subequations}
\end{widetext}

\bibliography{SEOBNRv5iHM}

\end{document}